\documentclass[hyper]{JHEP3} 

\usepackage{epsfig}
\usepackage{amsmath}
\usepackage{cite}
\usepackage{axodraw}
\usepackage{lscape}
\usepackage{multirow}

\newcommand{\beq}{\begin{equation}}
\newcommand{\eeq}{\end{equation}}
\newcommand{\hm}{\hat{m}^2}

\title{A General Method for Model-Independent Measurements of Particle Spins, 
Couplings and Mixing Angles in Cascade Decays with Missing Energy at Hadron Colliders}

\author{Michael Burns\\
        Physics Department, University of Florida,
        Gainesville, FL 32611, USA
        }

\author{Kyoungchul Kong\\
        Theoretical Physics Department, Fermilab,
        Batavia, IL 60510, USA
        }

\author{Konstantin T.~Matchev, Myeonghun Park\\
        Physics Department, University of Florida,
        Gainesville, FL 32611, USA
        }

\received{\today}               
\accepted{\today}               

\preprint{FERMILAB-PUB-08-255-T\\
          UFIFT-HEP-08-11 \\
          August 19, 2008
          } 

\abstract{We outline a general strategy for measuring spins, couplings and 
mixing angles in the case of a heavy partner decay chain terminating in an invisible
particle. We consider the common example of a heavy scalar or fermion $D$ 
decaying sequentially to other heavy particles $C$, $B$ and $A$ by emitting
a quark jet $j$ and two leptons $\ell^\pm_n$ and $\ell^\mp_f$. We derive 
analytic formulas for the dilepton ($\{\ell^+\ell^-\}$) and the
two jet-lepton ($\{j\ell_n\}$ and $\{j\ell_f\}$) invariant mass distributions 
for the case of most general couplings and mixing angles of the heavy partners. 
We then consider various spin assignments for the heavy particles 
$A$, $B$, $C$ and $D$, and for each case, derive the relevant
functional basis for the invariant mass distributions
which contains the intrinsic spin information and 
does not depend on the couplings and mixing angles. 
We propose a new method for determining the 
spins of the heavy partners, using the three experimentally observable
distributions $\{\ell^+\ell^-\}$, 
$\{j\ell^+\}+\{j\ell^-\}$ and $\{j\ell^+\}-\{j\ell^-\}$.
We show that the former two only depend on {\em a single} 
model-dependent parameter $\alpha$, while the latter may
depend on two other parameters $\beta$ and $\gamma$.
By fitting these distributions to our set of basis functions, 
we are able to do a pure measurement of the spins per se. 
Our method is also applicable at a $p\bar{p}$ collider 
such as the Tevatron, for which the previously proposed 
lepton charge asymmetry is identically zero and does not 
contain any spin information. In the process of determining 
the spins, we also end up with an independent {\em measurement}
of the parameters $\alpha$, $\beta$ and $\gamma$, which
represent certain combinations of the couplings and the mixing 
angles of the heavy partners $A$, $B$, $C$ and $D$.
}

\begin{document} 

\section{Introduction}
\label{sec:intro}

The ongoing Run II of the Fermilab Tevatron and the imminent 
turn-on of the Large Hadron Collider (LHC) at CERN are
beginning to explore the physics of the Terascale. 
There are sound theoretical reasons to believe that some
new physics beyond the Standard Model (BSM) is going to be 
revealed in those experiments. Perhaps the most compelling 
{\em phenomenological} evidence for BSM particles and 
interactions at the TeV scale is provided by the 
dark matter problem \cite{Bertone:2004pz}. It is a tantalizing coincidence that
a neutral, weakly interacting massive particle (WIMP) in the TeV 
range can explain all of the observed dark matter in the Universe.
A typical WIMP does not interact in the detector and 
can only manifest itself as missing energy. The WIMP idea 
therefore greatly motivates the study of missing energy 
signatures at the Tevatron and the LHC \cite{Hubisz:2008gg}.

The long lifetime of the dark matter WIMPs is typically ensured by 
some new exact symmetry, e.g. $R$-parity in supersymmetry \cite{Jungman:1995df}, 
KK parity in models with extra dimensions \cite{Hooper:2007qk}, 
$T$-parity in Little Higgs models \cite{Cheng:2003ju,Birkedal:2006fz}, 
$U$-parity \cite{Hur:2007ur,Lee:2008pc} etc. The particles of the 
Standard Model (SM) are not charged under this new symmetry, 
but the new particles are, and the lightest among them is the 
dark matter WIMP. This setup guarantees that the WIMP cannot decay, 
and more importantly, that WIMPs are always 
pair-produced at colliders. The cross-sections for direct 
production of WIMPs (tagged with a jet or a photon from initial 
state radiation) at hadron colliders are typically too small
to allow observation above the SM backgrounds \cite{Birkedal:2004xn}.
Therefore one typically concentrates on the pair production of the
other, heavier particles (e.g. superpartners, KK-partners, or $T$-partners), 
which also carry nontrivial new quantum numbers just like the WIMPs.
Once produced, those heavier partners will cascade decay 
down, emitting SM particles which are in principle 
observable in the detector. However, each such cascade also
inevitably ends up with an invisible WIMP, whose energy and momentum 
are unknown. Since the heavy partners are being pair-produced, 
there are two such cascades in each event, and therefore, two unknown 
WIMP momenta. In addition, at hadron colliders the total
parton level energy and momentum in the center of mass frame 
are also unknown, and thus the exact 
reconstruction of the decay chains 
on an event by event basis is a very challenging 
task \cite{Cheng:2007xv,Nojiri:2007pq,Cheng:2008mg}.

The lack of fully reconstructed events makes the mass and spin 
determination of the heavy partners rather difficult. Due to 
the escaping WIMPs, the heavy partners cannot be reconstructed as 
resonances in the invariant mass distributions of their decay products.
Their masses therefore must be measured from (a sufficient number of)
kinematic endpoints 
\cite{Hinchliffe:1996iu,Allanach:2000kt,Gjelsten:2004ki,Gjelsten:2005aw,Miller:2005zp}. 
The method can be successful, if a suitable cascade decay chain 
is identified in the data. An example of such a decay chain is presented in 
Fig.~\ref{fig:ABCD}, where we show the sequence of three two-body 
decays $D\to C+q$, $C\to B+\ell_n$ and $B\to A+\ell_f$.
\FIGURE[ht]{
{
\unitlength=1.5 pt
\SetScale{1.5}
\SetWidth{1.0}      
\normalsize    
{} \qquad\allowbreak
\begin{picture}(300,100)(0,50)
\SetColor{Gray}
\Line(100,80)(120,120)
\Line(150,80)(170,120)
\Line(200,80)(220,120)
\SetColor{Red}
\Text( 75, 86)[c]{\Red{$D$}}
\Text(125, 86)[c]{\Red{$C$}}
\Text(175, 86)[c]{\Red{$B$}}
\Text(225, 86)[c]{\Red{$A$}}
\SetWidth{1.2}      
\Line(50,80)(250,80)
\Text(100, 55)[c]{\Black{$x=\frac{m_C^2}{m_D^2}$}}
\Text(150, 55)[c]{\Black{$y=\frac{m_B^2}{m_C^2}$}}
\Text(200, 55)[c]{\Black{$z=\frac{m_A^2}{m_B^2}$}}
\Text(120,129)[c]{\Black{$q$}}
\Text(170,129)[c]{\Black{$\ell_n$}}
\Text(220,129)[c]{\Black{$\ell_f$}}
\SetColor{Blue}
\Vertex(100,80){2}
\Vertex(150,80){2}
\Vertex(200,80){2}
\scriptsize
\Text(100, 73)[c]{\Blue{$c_LP_L+c_RP_R$}}
\Text(150, 73)[c]{\Blue{$b_LP_L+b_RP_R$}}
\Text(200, 73)[c]{\Blue{$a_LP_L+a_RP_R$}}
\end{picture}
}
\caption{The typical cascade decay chain under consideration in this paper.
At each vertex we assume the most general coupling (see Sec.~\ref{sec:th} for
the exact definitions) and we quote our results in terms of the 
dimensionless mass ratios $x$, $y$ and $z$.}
\label{fig:ABCD} 
}
Here $D$, $C$, $B$ and $A$ are some heavy particles with masses
$m_D$, $m_C$, $m_B$ and $m_A$, correspondingly.
For simplicity, throughout this paper we shall assume that 
all heavy particles are on-shell, i.e. 
\begin{equation}
m_D > m_C > m_B > m_A\ .
\end{equation}
We shall take the visible decay products to be a quark jet $q$ and 
two leptons (either electron or muon), in that order\footnote{Note
that this choice is made only for concreteness of the discussion
and does not represent a fundamental limitation to our method.
All of our results below can be readily applied in the general case 
where the visible particles are any 3 SM fermions, not necessarily 
a quark and two leptons. The generalisation of the method to the 
case where the set of visible SM particles includes SM gauge bosons 
and/or a Higgs boson is straightforward and will be presented in a 
future publication \cite{BKMP}.}.
For discussion purposes, the leptons are often referred to
as ``near'' ($\ell_n$) and ``far'' ($\ell_f$), although 
this distinction is difficult to make in the actual data.
Our setup follows closely the conventions of 
Refs.~\cite{Miller:2005zp,Barr:2004ze,Smillie:2005ar,Athanasiou:2006ef,Athanasiou:2006hv}.
Accordingly, we shall also find it convenient to express our 
results in terms of the mass ratios
\begin{equation}
x\equiv \frac{m_C^2}{m_D^2}, \qquad
y\equiv \frac{m_B^2}{m_C^2}, \qquad
z\equiv \frac{m_A^2}{m_B^2}.
\label{eq:xyz}
\end{equation}

For a variety of reasons, the particular decay sequence 
exhibited in Fig.~\ref{fig:ABCD} has attracted a lot 
of interest in the past and has been extensively studied 
both in relation to an eventual discovery of new physics
as well as precision measurements of the new physics parameters. 
Rather early on, it was realized that this decay chain 
commonly occurs in the most popular models of low energy 
supersymmetry, such as minimal supergravity (MSUGRA), 
minimal gauge mediation \cite{Bagger:1996bt},
minimal anomaly mediation \cite{Gherghetta:1999sw,Feng:1999hg},
minimal gaugino mediation \cite{Schmaltz:2000gy}, etc.
More recently it was pointed out that the same chain may 
also occur in a non-supersymmetric context, e.g. Universal 
Extra Dimensions (UED) \cite{Cheng:2002iz,Cheng:2002ab}
and Little Higgs theories with $T$-parity \cite{Cheng:2005as}.
Therefore, even if the observable SM particles (the quark jet 
and the two leptons) can be uniquely identified, there may 
still be several competing BSM interpretations. Recently there
has been a lot of effort on developing various techniques for
discriminating among different model scenarios \cite{Barr:2004ze,Smillie:2005ar,Athanasiou:2006ef,Athanasiou:2006hv,Csaki:2007xm,Kong:2006pi,Battaglia:2005zf,Battaglia:2005ma,Datta:2005zs,Datta:2005vx,Barr:2005dz,Meade:2006dw,Alves:2006df,Wang:2006hk,SA:2006jm,Smillie:2006cd,Kilic:2007zk,Alves:2007xt,Datta:2007xy,Buckley:2007th,Buckley:2008pp,Kane:2008kw}.
The crux of the problem is the fact that the spin of the missing particle A is 
unknown, and this gives rise to several distinct possibilities.
Furthermore, the spin of particle A, even if it were known, 
still does not completely fix the spins of the preceding particles 
$B$, $C$ and $D$. Indeed, since the SM particles in Fig.~\ref{fig:ABCD} 
are all spin 1/2 fermions, the particles $A$, $B$, $C$ and $D$ must 
alternate between bosons and fermions, but the exact values of their
spins are a priori unknown. In the spirit of 
Refs.~\cite{Athanasiou:2006ef,Athanasiou:2006hv},
here we shall limit our discussion\footnote{Our method is 
nevertheless completely general and can be immediately generalised 
for higher spin particles as well. } only to particles of spin 1 or less,
namely we shall consider spin 0 scalars (S), spin 1/2 fermions (F) 
and spin 1 vector particles (V). Table \ref{table:spins} lists the
6 spin configurations for the decay chain of Fig.~\ref{fig:ABCD}, 
which were also considered in~\cite{Athanasiou:2006ef,Athanasiou:2006hv}.
\TABULAR[ht]{|c|c|c|c|c|c|c|}{
\hline 
$S$ & Spins       &  D      &    C    &   B     &  A      & Example  \\  \hline \hline
1   & \Red{SFSF}  & \Red{S}calar  & \Red{F}ermion & \Red{S}calar  & \Red{F}ermion & $\tilde q\to\tilde\chi^0_2\to\tilde\ell\to\tilde\chi^0_1$ \\  \hline
2   & \Red{FSFS}  & \Red{F}ermion & \Red{S}calar  & \Red{F}ermion & \Red{S}calar  & $q_1\to Z_H\to \ell_1\to \gamma_H$ \\  \hline
3   & \Red{FSFV}  & \Red{F}ermion & \Red{S}calar  & \Red{F}ermion & \Red{V}ector  & $q_1\to Z_H\to \ell_1\to \gamma_1$ \\  \hline
4   & \Red{FVFS}  & \Red{F}ermion & \Red{V}ector  & \Red{F}ermion & \Red{S}calar  & $q_1\to Z_1\to \ell_1\to \gamma_H$ \\  \hline
5   & \Red{FVFV}  & \Red{F}ermion & \Red{V}ector  & \Red{F}ermion & \Red{V}ector  & $q_1\to Z_1\to \ell_1\to \gamma_1$ \\  \hline
6   & \Red{SFVF}  & \Red{S}calar  & \Red{F}ermion & \Red{V}ector  & \Red{F}ermion & --- \\  \hline
}{\label{table:spins} Possible spin configurations of the heavy particles 
$D$, $C$, $B$ and $A$ in the decay chain of Fig.~\ref{fig:ABCD}.
The last column gives one typical SUSY or UED example. In the following 
we shall use the subscript $S$ to label these 6 possibilities.}
Five of these six possibilities can be readily accommodated in either 
supersymmetric or UED models. The last column of Table \ref{table:spins}
gives some typical examples involving the squarks $\tilde q$, 
sleptons $\tilde\ell$ and neutralinos $\tilde\chi_i^0$ in supersymmetry, 
the KK quarks $q_1$, KK leptons $\ell_1$ and KK gauge bosons $Z_1$ and 
$\gamma_1$ in 5D (or 6D) UED \cite{Appelquist:2000nn}, 
and the spinless gauge bosons $\gamma_H$ and $Z_H$ 
in 6D UED \cite{Dobrescu:2007xf}. The last case in 
Table \ref{table:spins} (SFVF) would 
require either a scalar leptoquark or a new gauge boson carrying lepton 
number. Nevertheless, we include it in our study for completeness and
also to connect to the results of \cite{Athanasiou:2006ef,Athanasiou:2006hv}.
We should emphasize from the start that we list the supersymmetry 
and UED examples in Table~\ref{table:spins} only as an illustration
and in what follows we shall never restrict ourselves to any particular model.
In particular, we shall not assume any features of the mass spectrum or the
couplings which might be expected in SUSY or UED. For example, we shall 
not assume a degenerate mass spectrum for the cases which might be 
expected in UED models, nor shall we assume any specific chirality structure 
of the couplings as predicted in supersymmetry or UED. We shall instead 
keep the spectrum completely arbitrary and also use the most general 
parametrization for the couplings of the heavy partners. Furthermore, 
we shall not make any assumptions about the nature of particle A --
it may or may not be the lightest heavy partner, and it may or may not be 
stable. While the dark matter problem mentioned at the beginning does
provide good theoretical motivation to look for missing energy signals,
particle A here does not at all have to be the dark matter WIMP, e.g. it may very well
decay to other heavy particle states, or even directly to SM particles. 
Consequently, the results presented in this paper will be completely 
general and can be applied to any model of new physics which 
exhibits a decay chain of the type shown in Fig.~\ref{fig:ABCD}.

The main goal of this paper is to assess the possibility of discriminating 
between the six different alternatives in Table \ref{table:spins}, using 
the experimentally observable invariant mass distributions of the visible
particles (the quark and the two leptons) in Fig.~\ref{fig:ABCD}. 
If such a discrimination could be made in a completely model-independent 
fashion, one could honestly claim a true measurement of the spins of the 
new particles. As a byproduct of our method, we shall also obtain an 
independent measurement of certain combinations of couplings and mixing 
angles of the heavy partners. The invariant mass distributions 
(of the quark and leptons) are convenient because they 
are Lorentz invariant quantities, and are certainly sensitive to the 
spins of the new particles. However, extracting spin, coupling and/or 
mixing angle information out of them is a highly nontrivial task 
and to the best of our knowledge has not been demonstrated up to now 
in a model-independent setup like ours. The main difficulties 
can be classified into two categories, experimental 
and theoretical, which we shall now discuss in some detail.

\subsection{Experimental challenges}
\label{sec:exp}

This class of problems is related to the ability of the experiment to 
uniquely identify the particles coming from the cascade of Fig.~\ref{fig:ABCD}.
\begin{itemize}
\item[{\bf E1}] {\em Jet combinatorics.} The events in which the cascade decay of 
Fig.~\ref{fig:ABCD} occurs, will also typically contain a number of additional 
jets. Some of those may come from initial state radiation, others may 
originate from the opposite cascade in the same event, and there may also 
be jets appearing from the decays of heavier particles into particle $D$. 
This poses a severe combinatorics problem: which one of the many jets in 
the event is the correct one to assign to the $D$ decay in Fig.~\ref{fig:ABCD}?
Some of the existing spin studies in the literature simply take for granted that 
the correct jet can be somehow identified, others select the jet by matching to
the true quark jet in the event generator output, which is of course unobservable.
The severity of the jet combinatorics problem is rather model dependent
and how well it can be dealt with in practice depends on the individual case at hand.
For example, if the mass splitting between $D$ and $C$ is relatively large, 
one might expect the jet from the $D$ decay to be among the hardest in the event, 
and this fact can be used to improve the purity of the sample. Fortunately,
there exists a method (the mixed event technique) which should, at least in 
principle, remove the effect from the wrong jet combinations \cite{Hinchliffe:1996iu}. 
More recently, the method has been successfully applied to measuring SUSY masses 
at the SPS1a study point \cite{Ozturk:2007ap}. A subtraction by a mixed event technique is 
particularly well suited for our purposes, since our method for spin measurements
only relies on the shapes of the global distributions, and we do not
need to guess the correct jet on an event by event basis.  
\item[{\bf E2}] {\em Lepton combinatorics.} There is an analogous combinatorics problem
related to the selection of the two leptons in the cascade of Fig.~\ref{fig:ABCD}.
First, in general, there may be additional isolated leptons in the event, 
so one might consider requiring two and only two leptons per event.
However, even then, it is not guaranteed that those two leptons are 
coming from the process in Fig.~\ref{fig:ABCD}: for example, each of the two 
leptons may come from a different cascade. Fortunately, there is again a universal
method (opposite lepton flavor subtraction) which solves both of these
lepton combinatorics problems \cite{Hinchliffe:1996iu}. One forms the 
linear combination of $\{e^+e^-\}+\{\mu^+\mu^-\}-\{e^+\mu^-\}-\{\mu^+e^-\}$,
in which the effects of the uncorrelated leptons in the signal (as well as all 
SM backgrounds involving top quarks, b-jets and $W$ bosons) 
cancel out\footnote{The method is not limited
to dilepton events and can also be applied to events with 3 or more leptons.
In that case one would use all possible dilepton combinations, but include a weight 
factor for their contribution to any given distribution, so that the total
weight of any given event, summed over all dilepton combinations, is 1.}.
In what follows we shall be assuming that the measured invariant mass 
distributions have already been properly subtracted to take care of the 
above mentioned jet and lepton combinatorial problems.  
\item[{\bf E3}] {\em Quark-antiquark jet ambiguity.} The cascade shown in Fig.~\ref{fig:ABCD}
consists of two separate processes. In the first one we produce a particle $D$, 
which decays to a quark jet and a particle $C$. In the conjugate process, the 
antiparticle of $D$ is produced and it decays to an antiquark jet and the 
antiparticle of $C$. Since the two types of jets appear identical in the 
detector\footnote{If $q$ is a heavy flavor, the distinction \emph{can} 
be made (statistically). To be conservative, we ignore this possibility 
in order to demonstrate that our method works even in the worst case scenario
of jet ambiguity.}, 
we cannot distinguish between these two cases, and the observable invariant 
mass distributions are the sum of the individual contributions from these 
two processes. This is a problem since, as we shall see, the sum tends 
to wash out to some extent the spin 
correlations which may have been originally present. In section~\ref{sec:generic}
we shall first present our formulas for the individual quark and antiquark 
jet distributions, but from section~\ref{sec:observable} onwards we shall 
always be adding up the quark and antiquark contributions together, and
we shall use the term ``jet'' to refer to either a quark or an antiquark.
For example, when we discuss a ``jet-lepton'' distribution $\{j\ell\}$ we shall
always imply that it was constructed by adding up the individual 
quark-lepton and antiquark-lepton distributions $\{q\ell\}+\{\bar{q}\ell\}$,
so that this quark-antiquark ambiguity does not represent a problem.
\item[{\bf E4}] {\em Near and far lepton ambiguity.} While the charge of the two leptons
can be measured very well, a priori one does not know which of them is 
the ``near'' lepton $\ell_n$ (i.e., coming from the decay of $C$) and 
which is the ``far'' lepton $\ell_f$ (i.e., coming from the decay of $B$).
Strictly speaking, once the mass spectrum of $A$, $B$, $C$ and $D$ is known,
one can select a subsample of the original events, in which
$\ell_n$ and $\ell_f$ can be uniquely identified. This can be done simply 
by ordering the two invariant masses $m_{j\ell^+}$ and $m_{j\ell^-}$
as $m_{j\ell}^{high}\equiv \max \{m_{j\ell^+},m_{j\ell^-}\}$
and $m_{j\ell}^{low}\equiv \min \{m_{j\ell^+},m_{j\ell^-}\}$,
and selecting only those events for which $m_{j\ell}^{high}$ happens 
to be above the observed kinematic endpoint of the $m_{j\ell}^{low}$ 
distribution. For that limited sample of events one can 
unambiguously identify $\ell_n$ and $\ell_f$. However, the price to pay
is that the statistics becomes very limited, especially if the
kinematic endpoints of the $m_{j\ell}^{high}$  and $m_{j\ell}^{low}$ 
distributions are close to each other. We therefore choose not to 
apply this trick, and instead we shall consider the combined
$m_{j\ell_n}$ and $m_{j\ell_f}$ distributions for each of the
two possible lepton charges. This allows us not only to avoid 
the near-far lepton ambiguity, but also to use the spin information 
contained in the $m_{j\ell_f}$ distribution. Previous studies 
on spin measurements have concentrated on the spin correlations 
between the jet and the near lepton, for which relatively simple 
and compact analytical expressions can be derived. The jet-far lepton contribution 
was regarded to a large extent as an annoying background which tends 
to wash out the jet-near lepton correlations. Our approach is 
very different: we actually treat both $m_{j\ell_n}$  and
$m_{j\ell_f}$ distributions on the same footing. Since we
have derived the most general expressions for both $m_{j\ell_n}$
and $m_{j\ell_f}$, in our method we are in effect able to fit 
{\em separately} to each one, and we do not even need to make the 
$\ell_n$-$\ell_f$ discrimination on an event by event basis.
In this sense our method is using all of the available information 
about spins which is present in the data.
\end{itemize}
Additionally, there are the usual complications on the experimental side, 
such as SM backgrounds, detector acceptance and resolution, triggering etc.
All of these factors should be taken into account when trying to decide 
{\em how well} our method will work in any particular case. But the main 
advantage of our method is that it is completely general, and can 
always be applied, even in the extremely complex environment
of a hadron collider experiment.

\subsection{Theoretical issues}
\label{sec:th}

Even if none of the experimental issues {\bf E1}-{\bf E4} discussed above
ever existed, e.g. we had a perfect detector, and we could 
somehow identify on an event by event basis with absolute certainty
which particular jet and two leptons came from the cascade in 
Fig.~\ref{fig:ABCD}, and furthermore, we could 
discriminate $q$ from $\bar{q}$ as well as $\ell_n$ from $\ell_f$;
even in that idealized case, there would still have been a long way to 
go towards a clean spin measurement, i.e. a discrimination between the
6 cases of Table~\ref{table:spins}. The problem is that the measured invariant 
mass distributions depend on all of the following 4 factors:
\begin{enumerate}
\item[{\bf T1}] {\em Mass spectrum}. It is well known that the shapes of the
observed invariant mass distributions in general depend on the heavy partner
spectrum. In fact this has been used in the past to make mass measurements of
the heavy partner masses, especially in the case when one of the heavy 
particles in the chain is off-shell \cite{Birkedal:2005cm,Gjelsten:2006tg}.
Mass measurements are therefore a useful (but not necessary -- see below) 
first step towards determining the spins. For simplicity, throughout
this paper we assume that all masses $m_A$, $m_B$, $m_C$ and $m_D$ have already been
determined from kinematic endpoints. This assumption is common with all previous 
spin studies. It appears rather feasible, since the mass measurements only 
require the extraction of the kinematic endpoints, which are sharp features
in the invariant mass distributions, and those are likely to be seen in the data
much earlier than the actual shape of the distributions. However, we should
emphasize that our assumption about the known mass spectrum was made only
for simplicity, and to keep the discussion focused on the more challenging
measurements like the spins, couplings and mixing angles.
Our method in fact does not require any prior knowledge of the mass spectrum.
When the mass spectrum is a priori unknown, the fits described in Sec.~\ref{sec:method}
would actually pick up the correct values of the masses, 
in addition to the spin and coupling measurements.
\item[{\bf T2}] {\em Particle-antiparticle ambiguity ($D/\bar{D}$).} 
This problem is related to the experimental issue {\bf E3}
from the previous subsection. Since we do not know if the jet was initiated 
by a quark or an antiquark, we also do not know whether the heavy particle
cascade was initiated by a particle $D$ or its antiparticle $\bar{D}$.
At a $p\bar{p}$ collider such as the Tevatron, the symmetry of the initial 
state implies that the fraction $f$ of $D$ particles produced in the data
should be equal to the fraction $\bar{f}$ of antiparticles $\bar{D}$.
Unfortunately, at a $pp$ collider like the LHC, the initial state is not 
symmetric, so one may expect an excess of particles over antiparticles:
$f>\bar{f}$, but the precise value of this excess $\Delta f \equiv f-\bar{f}$ is a priori unknown.
Therefore at the LHC $f$ is in principle an unknown parameter, which
significantly affects the observable $\{j\ell^+\}$ and $\{j\ell^-\}$ invariant mass 
distributions. Most previous studies of spin measurements have fixed 
$f$ to the value for the corresponding study point \cite{Barr:2004ze,Smillie:2005ar}. 
However, in the absence of an independent measurement of $f$, this is 
unjustified. The influence of $f$ on the spin extraction was
considered in \cite{Kong:2006pi,SA:2006jm}, where $f$ was left as a floating parameter
and consequently the extraction of the spins became much more difficult.
In what follows we shall follow a similar approach, namely, we shall not 
make any assumptions about the value of $f$ when we discuss measurements 
at the LHC and we shall instead treat $f$ as a free input parameter. 
Only in Sec.~\ref{sec:Tevatron}, where we apply our method to the Tevatron,
we shall take $f=\bar{f}$. Naturally, $\bar{f}$ is trivially related to $f$ as
\begin{equation}
f + \bar{f}=1\ .
\label{unit_f}
\end{equation}
\item[{\bf T3}] {\em Chirality of the fermion couplings.} 
Note that the three SM particles in Fig.~\ref{fig:ABCD} are all fermions, 
whose couplings to the heavy partners at each vertex are a priori unknown.
The observed invariant mass distributions depend on the 
chirality of those couplings, and this presents a formidable challenge 
in measuring the spins. The problem is that any given set of measured 
invariant mass distributions could in principle be explained by
one spin configuration with a certain choice of chiralities,
or a {\em different} spin configuration with a {\em different} 
choice of chiralities for the fermion couplings. 
To the best of our knowledge, none of the existing spin studies 
have accounted for this ambiguity in a consistent and fully model-independent 
way. Our main objective in this paper is to devise a method
for spin measurements which makes no assumptions about the  
chirality of the couplings at each vertex in Fig.~\ref{fig:ABCD}.
Correspondingly, we shall keep those couplings completely arbitrary,
and parameterize them in the most general way in terms of 
independent chirality coefficients at each vertex. 
For example, in the case of an interaction between 
a heavy spin 1/2 fermion $F$, a heavy scalar $\Phi$ and a SM fermion $f$
we take the interaction Lagrangian to be
\begin{equation}
\mathcal{L}(F,f,\Phi) = \bar{\Psi}_{F}(g_L P_{L}+g_R P_{R})\Psi_{f}\Phi + h.c.
\label{Lag_scalar}
\end{equation}
where $g_L$ and $g_R$ are arbitrary (and in general complex) coefficients. 
In general, there are three different sets of $\{g_L,g_R\}$, one 
at each vertex of Fig.~\ref{fig:ABCD}. We shall denote them as
$\{c_L,c_R\}$, $\{b_L,b_R\}$ and $\{a_L,a_R\}$, as shown in Fig.~\ref{fig:ABCD}.
Similarly, in case of an interaction between a heavy spin 1/2 fermion $F$, 
a heavy vector boson $A_\mu$ and a SM fermion $f$ we use the interaction Lagrangian
\begin{equation}
\mathcal{L}(F,f,A_\mu) = \bar{\Psi}_{F}\gamma^\mu (g_L P_{L}+g_R P_{R})\Psi_{f} A_\mu + h.c.
\label{Lag_vector}
\end{equation}
where just like before the coefficients $\{g_L,g_R\}=\{c_L,c_R\}$, 
$\{b_L,b_R\}$ or $\{a_L,a_R\}$, depending on the vertex.
In what follows we present our results in terms of these 
most general coefficients $\{c_L,c_R\}$, $\{b_L,b_R\}$ and $\{a_L,a_R\}$.
According to our convention, the couplings $\{c_L,c_R\}$
are always associated with the $D$-$C$-$q$ vertex, the 
couplings $\{b_L,b_R\}$ are always associated with the $C$-$B$-$\ell_n$ vertex,
and the couplings $\{a_L,a_R\}$ are always associated with the 
$B$-$A$-$\ell_f$ vertex.
We shall not be specifying explicitly whether a given pair such as $\{a_L,a_R\}$
parameterizes the interaction (\ref{Lag_scalar}) or the interaction 
(\ref{Lag_vector}), since that should be clear from the context.

We shall see below that the shapes of the invariant mass distributions
only depend on the {\em relative} chirality of each vertex, 
therefore it is convenient to unit normalize the couplings as
\begin{eqnarray}
|a_L|^2+|a_R|^2 &=& 1 \ ,  \label{unit_a} \\
|b_L|^2+|b_R|^2 &=& 1 \ ,  \label{unit_b} \\
|c_L|^2+|c_R|^2 &=& 1 \ ,  \label{unit_c}
\end{eqnarray}
In that case, the {\em relative} chirality at each vertex is parameterized
in terms of a single parameter, which can be taken as an angle:
\begin{equation}
\tan\varphi_a = \frac{|a_R|}{|a_L|} \ , \qquad 
\tan\varphi_b = \frac{|b_R|}{|b_L|} \ , \qquad
\tan\varphi_c = \frac{|c_R|}{|c_L|} \ .
\label{phi_angles}
\end{equation}
By convention, we shall take all three of these angles 
to be defined in the range $[0,\frac{\pi}{2}]$ 
(as opposed to $[\pi,\frac{3\pi}{2}]$).
The angles $\varphi_a$, $\varphi_b$ and $\varphi_c$
encode all of the relevant\footnote{At this point it may be useful to 
do a quick count of the relevant degrees of freedom.
For example, consider the $B$-$A$-$\ell_f$ vertex
parameterized by $\{a_L,a_R\}$.
Since $a_L\equiv |a_L|e^{\phi_L}$ and $a_R\equiv |a_R|e^{\phi_R}$ 
are in general complex parameters,
originally there are four degrees of freedom 
($|a_L|$, $|a_R|$, $\phi_L$ and $\phi_R$) parameterizing 
each of the SM fermion interactions (\ref{Lag_scalar},\ref{Lag_vector}).
One combination of $|a_L|$ and $|a_R|$  is eliminated through 
the normalisation condition (\ref{unit_a}), while
(\ref{phi_angles}) simply parameterizes the 
other combination of $|a_L|$ and $|a_R|$ in terms of $\varphi_a$.
The remaining two degrees of freedom, the phases $\phi_L$ and 
$\phi_R$, remain arbitrary and cannot be measured from the invariant 
mass distributions that we are considering here. Instead, 
they will have to be measured by some other means.} 
model dependence, e.g. the nature of the 
interaction and the mixing angles of the heavy partner
mass eigenstates. It is worth emphasizing that we consider
the couplings $g_L$ and $g_R$ in eqs.~(\ref{Lag_scalar}, \ref{Lag_vector}) 
to be the couplings in the mass eigenstate basis for the heavy
partners. Therefore, whenever there is mixing among the
heavy partner states, our couplings $g_L$ and $g_R$
are in general matrices which are related to the
couplings $g_L^{(0)}$ and $g_R^{(0)}$ in the interaction 
eigenstate basis through rotations by the corresponding 
mixing angles
\begin{equation}
g_{L,R} \equiv {\bf U_F}^\dagger\ g_{L,R}^{(0)}\ {\bf U_B}\, ,
\label{eqn:mixing}
\end{equation}
where the matrix ${\bf U_F}$ (${\bf U_B}$) diagonalises the
mass matrix of the corresponding heavy fermion (boson).
Due to this mixing, in general we do not expect our couplings 
$g_L$ and $g_R$ to be purely chiral, even in models where 
one starts with purely chiral couplings $g_L^{(0)}$ and $g_R^{(0)}$ 
in the interaction eigenstate basis. The effect of 
heavy fermion mixing ${\bf U_F}$ in a specific UED model 
was previously considered in \cite{Kilic:2007zk}, and here 
we generalise the discussion to the case of arbitrary 
heavy fermion mixing ${\bf U_F}$, arbitrary 
heavy boson mixing ${\bf U_B}$, and arbitrary 
couplings $g_L^{(0)}$ and $g_R^{(0)}$.
Clearly, there is an enormous number of model-dependent parameters 
contained in  ${\bf U_F}$, ${\bf U_B}$, $g_L^{(0)}$ and $g_R^{(0)}$,
and it will be rather hopeless to try to measure them all at once.
One of the main results of this paper will be to identify which 
particular combinations of these coupling and mixing angle 
parameters can be experimentally measured from the invariant 
mass distributions of the three SM fermions 
(in our case, $q$, $\ell_n$ and $\ell_f$),
and to propose the actual method for measuring them.
We shall find that there are three such combinations,
which we shall call $\alpha$, $\beta$ and $\gamma$ 
(for details, see Secs.~\ref{sec:method} and \ref{sec:couplings}).
Each one of them is potentially experimentally accessible, and 
represents some combination of couplings and mixing angles as
illustrated in eq.~(\ref{eqn:mixing}). It is in this sense that 
our method yields a measurement of the couplings and mixing angles
of the heavy partners, as advertised in the abstract.
\item[{\bf T4}] {\em Spins.} Finally, the invariant mass distributions also
contain information about the spins of the heavy particles along the decay chain.
For example, pure phase space predicts flat (in $m^2$) 
invariant mass distributions for SM particle pairs 
originating from adjacent vertices in the decay chain. 
Deviation from this pure phase space 
prediction implies some kind of spin correlations \cite{Barr:2004ze}, 
but what type? Conversely, observing distributions
which are consistent with the pure phase space prediction
does not necessarily mean that all particles involved in 
the decay are scalars -- spin correlations may have been
present for the individual subprocesses (to be defined below)
but may have been washed out when added up to form the
experimentally observable distributions. Below we shall 
encounter examples of both of these situations.
\end{enumerate}
The general approach in previous spin studies has been to
compare the data from a given study point within one specific model 
to the corresponding data obtained from another model 
alternative with different choice of spins for the heavy partners.
A common flaw in all such studies was that three of the four 
relevant factors, namely {\bf T1}, {\bf T2} and {\bf T3},
were fixed to be {\em identical} in the two models, so that
any remaining difference can be interpreted as a manifestation 
of spins (the factor {\bf T4} above). However, this is not the 
correct approach when it comes to actual pure measurements of spins
in a model-independent fashion. Since the chirality parameters 
$\varphi_a$, $\varphi_b$ and $\varphi_c$ and 
the particle-antiparticle ratio $f$ are not independently measured prior
to the attempted spin determination, they need not have the same 
values for each of the different spin configurations under study
(in our case, the 6 ones listed in Table~\ref{table:spins})
and should be allowed to float. Therefore, the proper question to ask
instead is: 
\begin{quote}
Given the data, which (and how many) spin configuration gives a good fit to it
{\em for some choice} of the chirality parameters $\varphi_a$, $\varphi_b$ 
and $\varphi_c$, and {\em for some choice} of the particle-antiparticle 
ratio $f$?
\end{quote}

The main result of this paper is that we provide the tools 
needed to address this question in a completely model-independent way, namely 
in order to determine whether a given spin configuration ``S'' 
is consistent with the data or not,
we do not need to specify the values of $f$ and $\bar{f}$, nor do we
need to specify the chirality of the couplings $\varphi_a$, $\varphi_b$ and $\varphi_c$. 
In other words, we have divided the question posed above into two parts:
for a given mass spectrum (i.e. factor {\bf T1} is known), 
\begin{itemize}
\item Q1: What is the spin, i.e. what is factor {\bf T4}?
\item Q2: What are the particle-antiparticle fractions 
$f$ and $\bar{f}$ (item {\bf T2} above) and what are the 
couplings and mixing angles (item {\bf T3} above)?
\end{itemize} 
Our method allows us to provide an independent answer to the spin 
question Q1 {\em regardless of the answer} to the follow up question Q2.
In this sense we are able to make a pure measurement of spin
in a model-independent way. Of course, as we shall see below, 
the {\em actual} answer to the question Q1 may not be unique,
and sometimes there are cases where more than one particular spin 
configuration may fit the data. In fact in Sec~\ref{sec:degenerate} 
we shall show that the model pairs \{FSFS, FSFV\}
as well as \{FVFS, FVFV\} are quite often indistinguishable.

Since we have decoupled the spin issue {\bf T4} from the $f$-$\bar{f}$
issue {\bf T2},
our method is not limited to $pp$ colliders such as LHC, and is equally 
applicable to the Tevatron. In contrast, the lepton charge asymmetry
proposed by Barr \cite{Barr:2004ze} is greatly affected by the value
of $f$, for example it 
is predicted to be identically zero at the Tevatron and has no 
discriminating power there with regards to spins. In this sense our 
method provides a pure measurement of the spins and the spins alone.
What is more, in the process of answering the spin question, we also get
a measurement of some combination of the couplings and $f$ and $\bar{f}$.
In this sense our method is also the first and most general attempt
to measure mixing angles of heavy partners (e.g. superpartners) 
at the LHC.

The paper is organised as follows. In Section \ref{sec:generic}
we describe the main idea of our method and derive the main building 
blocks for the spin measurement. In particular, we give exact analytical 
expressions for all relevant invariant mass distributions 
(including $\{q\ell_f^\pm\}$ and $\{\bar{q}\ell_f^\pm\}$) in the most general case 
of arbitrary couplings, arbitrary $f$ and $\bar{f}$, and arbitrary 
mass spectrum, for each of the six cases from Table~\ref{table:spins}.
Our results in Sec.~\ref{sec:generic} generalize those of 
Refs.~\cite{Smillie:2005ar,Athanasiou:2006ef,Athanasiou:2006hv,Miller:2005zp}.
In Sec.~\ref{sec:observable} we reorganise our results from Sec.~\ref{sec:generic} 
to form the {\em experimentally observable} invariant mass distributions
$\{j\ell^+\}$, $\{j\ell^-\}$ and $\{\ell^+\ell^-\}$.
We also derive the exact combinations of couplings and mixing 
angles which are being measured as a byproduct of the spin 
measurement\footnote{Readers who are only interested in the practical applications of our results,
and would prefer to skip these mathematical derivations, are invited to jump 
directly to Secs.~\ref{sec:method} and \ref{sec:numerics}, which are 
self-contained and can be read independently from the more technical
sections \ref{sec:generic} and \ref{sec:observable}.}. Section \ref{sec:method}
begins by summarising the key analytical results from the previous two sections, 
and outlines our method for spin and coupling measurements.
In Sec.~\ref{sec:degenerate} we prove analytically the degeneracy of
the \{FSFS, FSFV\} and \{FVFS, FVFV\} model pairs -- we derive the 
relation between the couplings and mixing angles within each pair of models 
which would result in {\em identical} observable invariant mass distributions
for those model pairs.
In Sec.~\ref{sec:Tevatron} we specify our results to the case of 
$p\bar{p}$ colliders such as the Tevatron and show that our spin 
analysis method can be just as successful there. Finally,
in Sec.~\ref{sec:numerics} we provide an illustration of
an actual idealised measurement, using a mass spectrum 
and couplings as for the SPS1a study point in supersymmetry. 
Assuming that the data comes from each 
one of the 6 models from Table~\ref{table:spins} in turn, we then
demonstrate how well the remaining 5 possibilities can be ruled in or 
out. This results in a total of 36 different case studies, the
results of which are presented and analysed in that section.
In Sec.~\ref{sec:conclusions} we summarize our main conclusions, 
and discuss the pros and cons of our method in comparison to
other proposals for spin measurements in the literature.

\section{General expressions for the invariant mass distributions}
\label{sec:generic}

\subsection{Preliminaries}

The basic idea behind our method is the following.
For any given spin configuration $S$, we write the invariant mass distribution 
of a pair of SM particles from Fig.~\ref{fig:ABCD} as
\begin{equation}
\left( \frac{dN}{d\hat{m}^2_{p}}\right)_S = \sum_{I=1}^{2}\sum_{J=1}^{2} K_{IJ}^{(p)}(f,\varphi_a,\varphi_b,\varphi_c) 
{\mathcal F}_{S;IJ}^{(p)}(\hat{m}_p^2;x,y,z)\ ,
\label{mastereq}
\end{equation}
where the index $p$ denotes one of the five possible SM particle pairs: 
$p=\{j\ell_n^-, j\ell_n^+, j\ell_f^-, j\ell_f^+, \ell^+\ell^-\}$;
$\hat{m}_{p}$ is the unit-normalised invariant mass
\begin{equation}
\hat{m}_p \equiv \frac{m_{p}}{m_{p}^{max}}\ ,\qquad 0\le \hat{m}_{p}\le 1\ , 
\label{def_mhat}
\end{equation}
i.e. the invariant mass $m_{p}$ scaled by the value of the corresponding 
kinematic endpoint $m_{p}^{max}$, which has already been measured from 
the corresponding $m_{p}$ distribution. 
The mass ratios $x$, $y$ and $z$ were already defined in (\ref{eq:xyz}), while
$\{IJ\}$ is a pair of indices denoting one out of four possible classes of 
subprocesses $P_{IJ}$ which will be discussed in detail below in Sec.~\ref{sec:classification}. 
The coefficients $K_{IJ}^{(p)}$ and the functions ${\mathcal F}_{S;IJ}^{(p)}$ 
will be explicitly defined later in Sec.~\ref{sec:imd}. 

The general expression (\ref{mastereq}) corroborates our discussion in Sec.~\ref{sec:th} --
we see that the invariant mass distributions indeed depend simultaneously on
all of the four factors ({\bf T1}-{\bf T4}) discussed earlier. 
However, notice that the coefficients $K_{IJ}$ in the expansion
(\ref{mastereq}) only depend on the particle/antiparticle fraction $f$ and
the chiralities $\varphi_a$, $\varphi_b$ and $\varphi_c$, i.e. factors 
{\bf T2} and {\bf T3}. On the other hand, 
the functions ${\mathcal F}_{S;IJ}^{(p)}(\hat{m}^2;x,y,z)$
only depend on the mass spectrum (factor {\bf T1}) and the spin 
(factor {\bf T4}). Once the spectrum is measured and the 
mass ratios $x$, $y$ and $z$ become known, the functions
${\mathcal F}_{S;IJ}^{(p)}$ only depend on $\hat{m}$
and provide a unique basis which can be fitted to the data 
for {\em each} of the measured distributions $\{p\}$. 
Since the functions ${\mathcal F}$ do not depend on the 
model dependent parameters $f$, $\varphi_a$, $\varphi_b$ and $\varphi_c$,
this fit can be done in a completely model-independent way, without
any prior knowledge about the nature of the particles $A$, $B$, $C$ and $D$,
the nature of their couplings, or the size of their mixing angles.
For each of the 6 possible spin configurations $S$, this fit may or 
may not yield a good match: then, those spin configurations 
which give a bad fit to the data will be ruled out. 
Conversely, the spin configurations which give a good fit 
will be ruled in, and furthermore, the values of the
fitted coefficients $K$ will represent {\em a measurement} 
of the couplings and mixing angles of the heavy partners.

\subsection{Classification of helicity combinations}
\label{sec:classification}

Table~\ref{table:processes} lists all possible helicity combinations (32 altogether)
contributing to the process of Fig.~\ref{fig:ABCD}. The 8 combinations shown 
in blue have been previously considered in \cite{Smillie:2005ar,Athanasiou:2006ef,Athanasiou:2006hv}.
The remaining 24 combinations shown in red are being considered here for the first time.
We find it convenient to classify all possibilities into four categories $P_{IJ}, I,J=1,2$,
where each category gives rise to the same functional dependence for the
three invariant mass distributions of interest: $\{j\ell^\pm_n\}$, $\{j\ell^\pm_f\}$ and 
$\{\ell^\pm_n\ell^\mp_f\}$. We name these four categories
as follows:
\begin{itemize}
\item Processes of type $P_{11}$. These include all cases where the {\em physical} 
helicities of the (anti)quark jet and near lepton are the same, while the {\em physical} 
helicities of the two leptons are opposite. The four processes of type 1 in the 
nomenclature of Refs.~\cite{Smillie:2005ar,Athanasiou:2006ef,Athanasiou:2006hv}
fall into this set. In addition in this group we find four new combinations
involving right-handed quarks.
\item Processes of type $P_{21}$. These include all cases where the {\em physical} 
helicities of the (anti)quark jet and near lepton as well as the {\em physical}
helicities of the two leptons are opposite. The four processes of type 2 in the 
nomenclature of Refs.~\cite{Smillie:2005ar,Athanasiou:2006ef,Athanasiou:2006hv}
fall into this set. Again, there are four new cases involving right-handed quarks.
Note that the processes of type $P_{21}$ are simply obtained from those of type $P_{11}$
by interchanging $q \leftrightarrow \bar{q}$ while keeping the chirality labels fixed.
\item Processes of type $P_{12}$. Here the {\em physical} helicities of the 
(anti)quark jet and near lepton as well as the {\em physical} helicities of the 
two leptons are the same. These processes are obtained from those of $P_{11}$ by
changing the chirality label of the far lepton: $L\leftrightarrow R$ for $\ell^\pm_f$.
\item Processes of type $P_{22}$. Here the {\em physical} helicities of the 
(anti)quark jet and near lepton are opposite, while the {\em physical} helicities of the 
two leptons are the same. These processes can be obtained from $P_{12}$ by interchanging
$q \leftrightarrow \bar{q}$, or alternatively, from $P_{21}$ by
changing the chirality label of the far lepton: $L\leftrightarrow R$ for $\ell^\pm_f$.
\end{itemize}
All processes falling into the last two categories are new, and more importantly, as we
shall see below, they give a qualitatively new functional dependence of the dilepton 
and $j\ell_f$ invariant mass distributions which was not exhibited in the previous studies 
\cite{Smillie:2005ar,Athanasiou:2006ef,Athanasiou:2006hv}. 

\TABULAR[ht]{||c|c||c|c||}{
\hline\hline
\multicolumn{2}{||c||}{Processes $P_{11}$}  & 
\multicolumn{2}{c||}{Processes $P_{12}$}       \\ \hline
\Blue{$\{      q_L,\ell^-_L,\ell^+_L\}$} & \Blue{$\{\bar{q}_L,\ell^+_L,\ell^-_L\}$}       &
\Red{$\{      q_L,\ell^-_L,\ell^+_R\}$}  &  \Red{$\{\bar{q}_L,\ell^+_L,\ell^-_R\}$} \\  
$     f |c_L|^2|b_L|^2|a_L|^2$ & 
$\bar{f}|c_L|^2|b_L|^2|a_L|^2$ & 
$     f |c_L|^2|b_L|^2|a_R|^2$ &
$\bar{f}|c_L|^2|b_L|^2|a_R|^2$  \\ \hline
\Blue{$\{\bar{q}_L,\ell^-_R,\ell^+_R\}$} & \Blue{$\{      q_L,\ell^+_R,\ell^-_R\}$}       &  
\Red{$\{\bar{q}_L,\ell^-_R,\ell^+_L\}$}  &  \Red{$\{      q_L,\ell^+_R,\ell^-_L\}$}  \\  
$\bar{f}|c_L|^2|b_R|^2|a_R|^2$ & 
$     f |c_L|^2|b_R|^2|a_R|^2$ & 
$\bar{f}|c_L|^2|b_R|^2|a_L|^2$ &
$     f |c_L|^2|b_R|^2|a_L|^2$ \\ \hline
\Red{$\{      q_R,\ell^-_R,\ell^+_R\}$} &  \Red{$\{\bar{q}_R,\ell^+_R,\ell^-_R\}$}        &
\Red{$\{      q_R,\ell^-_R,\ell^+_L\}$} &  \Red{$\{\bar{q}_R,\ell^+_R,\ell^-_L\}$} \\
$     f |c_R|^2|b_R|^2|a_R|^2$ &  
$\bar{f}|c_R|^2|b_R|^2|a_R|^2$ &  
$     f |c_R|^2|b_R|^2|a_L|^2$ &
$\bar{f}|c_R|^2|b_R|^2|a_L|^2$  \\  \hline
\Red{$\{\bar{q}_R,\ell^-_L,\ell^+_L\}$} &  \Red{$\{      q_R,\ell^+_L,\ell^-_L\}$}        &  
\Red{$\{\bar{q}_R,\ell^-_L,\ell^+_R\}$} &  \Red{$\{      q_R,\ell^+_L,\ell^-_R\}$} \\
$\bar{f}|c_R|^2|b_L|^2|a_L|^2$ &  
$     f |c_R|^2|b_L|^2|a_L|^2$ &  
$\bar{f}|c_R|^2|b_L|^2|a_R|^2$ &
$     f |c_R|^2|b_L|^2|a_R|^2$  \\  \hline\hline
\Blue{$\{\bar{q}_L,\ell^-_L,\ell^+_L\}$} & \Blue{$\{      q_L,\ell^+_L,\ell^-_L\}$}       &  
\Red{$\{\bar{q}_L,\ell^-_L,\ell^+_R\}$} &  \Red{$\{      q_L,\ell^+_L,\ell^-_R\}$} \\
$\bar{f}|c_L|^2|b_L|^2|a_L|^2$  & 
$     f |c_L|^2|b_L|^2|a_L|^2$  & 
$\bar{f}|c_L|^2|b_L|^2|a_R|^2$  &
$     f |c_L|^2|b_L|^2|a_R|^2$  \\  \hline
\Blue{$\{      q_L,\ell^-_R,\ell^+_R\}$} & \Blue{$\{\bar{q}_L,\ell^+_R,\ell^-_R\}$}     &  
\Red{$\{      q_L,\ell^-_R,\ell^+_L\}$} &  \Red{$\{\bar{q}_L,\ell^+_R,\ell^-_L\}$}   \\
$     f |c_L|^2|b_R|^2|a_R|^2$ & 
$\bar{f}|c_L|^2|b_R|^2|a_R|^2$ & 
$     f |c_L|^2|b_R|^2|a_L|^2$ &
$\bar{f}|c_L|^2|b_R|^2|a_L|^2$  \\  \hline
\Red{$\{\bar{q}_R,\ell^-_R,\ell^+_R\}$} &  \Red{$\{      q_R,\ell^+_R,\ell^-_R\}$}   &  
\Red{$\{\bar{q}_R,\ell^-_R,\ell^+_L\}$} &  \Red{$\{      q_R,\ell^+_R,\ell^-_L\}$} \\
$\bar{f}|c_R|^2|b_R|^2|a_R|^2$  &  
$     f |c_R|^2|b_R|^2|a_R|^2$  &  
$\bar{f}|c_R|^2|b_R|^2|a_L|^2$  &
$     f |c_R|^2|b_R|^2|a_L|^2$  \\  \hline
\Red{$\{      q_R,\ell^-_L,\ell^+_L\}$} &  \Red{$\{\bar{q}_R,\ell^+_L,\ell^-_L\}$}  &  
\Red{$\{      q_R,\ell^-_L,\ell^+_R\}$} &  \Red{$\{\bar{q}_R,\ell^+_L,\ell^-_R\}$} \\
$     f |c_R|^2|b_L|^2|a_L|^2$ &  
$\bar{f}|c_R|^2|b_L|^2|a_L|^2$ &  
$     f |c_R|^2|b_L|^2|a_R|^2$ & 
$\bar{f}|c_R|^2|b_L|^2|a_R|^2$  \\  \hline
\multicolumn{2}{||c||}{Processes $P_{21}$}     & 
\multicolumn{2}{c||}{Processes $P_{22}$}       \\ \hline\hline
}{\label{table:processes} Classification of all possible helicity combinations
contributing to the process of Fig.~\ref{fig:ABCD}. The combinations shown 
in blue have been previously considered in \cite{Smillie:2005ar,Athanasiou:2006ef,Athanasiou:2006hv}.
The combinations shown in red are being considered here for the first time.
Under each helicity combination, we also show the associated prefactor
contributing to $K_{IJ}^{(p)}$ in eq.~(\ref{mastereq}).
}

It is worth noting that in the case of a heavy fermion (F), 
there is a distinction between the Dirac and Majorana case.
For a Dirac fermion, half of the processes within each category $P_{IJ}$
of Table~\ref{table:processes} are absent, since the adjacent 
SM fermions must be a particle and an antiparticle. For a Majorana fermion, 
there is no such restriction, and all processes exhibited in
Table~\ref{table:processes} are in principle allowed.

\subsection{Invariant mass distributions}
\label{sec:imd}

In principle, there are 9 invariant mass distributions that we can form:
\begin{eqnarray}
\left( \frac{dN}{d\hat{m}^2_{q\ell_n^\pm}}\right)_S &=& \frac{1}{2} \sum_{I=1}^{2}\sum_{J=1}^{2} K_{IJ}^{(q\ell_n^\pm)}(f,\varphi_a,\varphi_b,\varphi_c) 
{\mathcal F}_{S;IJ}^{(j\ell_n)}(\hat{m}^2_{q\ell_n^\pm};x,y,z)\ ,
\label{qln}   \\
\left( \frac{dN}{d\hat{m}^2_{\bar{q}\ell_n^\pm}}\right)_S &=& \frac{1}{2} \sum_{I=1}^{2}\sum_{J=1}^{2} K_{IJ}^{(\bar{q}\ell_n^\pm)}(f,\varphi_a,\varphi_b,\varphi_c) 
{\mathcal F}_{S;IJ}^{(j\ell_n)}(\hat{m}^2_{\bar{q}\ell_n^\pm};x,y,z)\ ,
\label{qbln}    \\
\left( \frac{dN}{d\hat{m}^2_{q\ell_f^\pm}}\right)_S &=& \frac{1}{2} \sum_{I=1}^{2}\sum_{J=1}^{2} K_{IJ}^{(q\ell_f^\pm)}(f,\varphi_a,\varphi_b,\varphi_c) 
{\mathcal F}_{S;IJ}^{(j\ell_f)}(\hat{m}^2_{q\ell_f^\pm};x,y,z)\ ,
\label{qlf}   \\
\left( \frac{dN}{d\hat{m}^2_{\bar{q}\ell_f^\pm}}\right)_S &=& \frac{1}{2} \sum_{I=1}^{2}\sum_{J=1}^{2} K_{IJ}^{(\bar{q}\ell_f^\pm)}(f,\varphi_a,\varphi_b,\varphi_c) 
{\mathcal F}_{S;IJ}^{(j\ell_f)}(\hat{m}^2_{\bar{q}\ell_f^\pm};x,y,z)\ ,
\label{qblf}  \\
\left( \frac{dN}{d\hat{m}^2_{\ell\ell}}\right)_S &=& \frac{1}{2} \sum_{I=1}^{2}\sum_{J=1}^{2} K_{IJ}^{(\ell\ell)}(f,\varphi_a,\varphi_b,\varphi_c) 
{\mathcal F}_{S;IJ}^{(\ell\ell)}(\hat{m}^2_{\ell\ell};x,y,z)\ ,
\label{ll}
\end{eqnarray}
where the factor of $\frac{1}{2}$ on the right hand side 
was introduced for future convenience. Note that it is the 
same set of functions ${\mathcal F}_{S;IJ}^{(j\ell_n)}$ which enter both 
the $\{q\ell_n\}$ and $\{\bar{q}\ell_n\}$ distributions
\begin{equation}
{\mathcal F}_{S;IJ}^{(q      \ell_n)}(\hat{m}^2;x,y,z)
=
{\mathcal F}_{S;IJ}^{(\bar{q}\ell_n)}(\hat{m}^2;x,y,z)
\equiv
{\mathcal F}_{S;IJ}^{(j      \ell_n)}(\hat{m}^2;x,y,z)\ ,
\end{equation}
and similarly, it is the same set of functions ${\mathcal F}_{S;IJ}^{(j\ell_f)}$ 
which enter the $\{q\ell_f\}$ and $\{\bar{q}\ell_f\}$ distributions:
\begin{equation}
{\mathcal F}_{S;IJ}^{(q      \ell_f)}(\hat{m}^2;x,y,z)
=
{\mathcal F}_{S;IJ}^{(\bar{q}\ell_f)}(\hat{m}^2;x,y,z)
\equiv
{\mathcal F}_{S;IJ}^{(j      \ell_f)}(\hat{m}^2;x,y,z)\ .
\end{equation}

In the following two subsections we shall separately define and discuss the
functions ${\mathcal F}_{S;IJ}^{(p)}$ and the coefficients $K_{IJ}^{(p)}$ 
appearing in the general expressions (\ref{qln}-\ref{ll}).

\subsubsection{The functions ${\mathcal F}_{S;IJ}^{(p)}$}

Eqs.~(\ref{qln}-\ref{ll}) show that all invariant mass distributions can be written
in terms of three sets of basis functions: ${\mathcal F}_{S;IJ}^{(j\ell_n)}(\hat{m}^2;x,y,z)$,
${\mathcal F}_{S;IJ}^{(j\ell_f)}(\hat{m}^2;x,y,z)$ and 
${\mathcal F}_{S;IJ}^{(\ell\ell)}(\hat{m}^2;x,y,z)$.
We shall define the basis functions to be unit normalized:
\begin{eqnarray}
\int_0^\infty {\mathcal F}_{S;IJ}^{(j\ell_n)}(\hat{m}^2;x,y,z) \, d\hat{m}^2 = 1\ , \label{unit_qln} \\
\int_0^\infty {\mathcal F}_{S;IJ}^{(j\ell_f)}(\hat{m}^2;x,y,z) \, d\hat{m}^2 = 1\ , \label{unit_qlf} \\
\int_0^\infty {\mathcal F}_{S;IJ}^{(\ell\ell)}(\hat{m}^2;x,y,z)\, d\hat{m}^2 = 1\ . \label{unit_ll} 
\end{eqnarray}
With this normalisation, all basis functions 
${\mathcal F}_{S;IJ}^{(j\ell_n)}(\hat{m}^2;x,y,z)$,
${\mathcal F}_{S;IJ}^{(\ell\ell)}(\hat{m}^2;x,y,z)$ \allowbreak
and \allowbreak
${\mathcal F}_{S;IJ}^{(j\ell_f)}(\hat{m}^2;x,y,z)$
are defined in Appendix~\ref{app:FIJ}.

A few comments regarding the ${\mathcal F}_{S,IJ}^{(p)}$ functions
are in order. Recall that half of the processes belonging to 
category $P_{11}$ and $P_{21}$ (in the classification of Sec.~\ref{sec:classification})
have been previously considered in \cite{Smillie:2005ar,Athanasiou:2006ef,Athanasiou:2006hv},
so that the functions ${\mathcal F}_{S,11}^{(p)}$ and ${\mathcal F}_{S,21}^{(p)}$ 
in principle already appear there. We find agreement with 
\cite{Smillie:2005ar,Athanasiou:2006ef,Athanasiou:2006hv} for the case of 
${\mathcal F}_{S,11}^{(p)}$ and ${\mathcal F}_{S,21}^{(p)}$, 
and we supplement those results with the remaining two types of functions
${\mathcal F}_{S,12}^{(p)}$ and ${\mathcal F}_{S,22}^{(p)}$.
We shall now comment individually on each type $p$ of basis functions
${\mathcal F}_{S,IJ}^{(p)}$.

Table~\ref{table:Fqln} in Appendix~\ref{app:FIJ}
shows that the ${\mathcal F}_{S,IJ}^{(j\ell_n)}$
functions are pairwise equal:
\begin{eqnarray}
{\mathcal F}_{S,11}^{(j\ell_n)}(\hat{m}^2;x,y,z)&=&{\mathcal F}_{S,12}^{(j\ell_n)}(\hat{m}^2;x,y,z) \label{eq:Fqln11}\ , \\ [2mm]
{\mathcal F}_{S,21}^{(j\ell_n)}(\hat{m}^2;x,y,z)&=&{\mathcal F}_{S,22}^{(j\ell_n)}(\hat{m}^2;x,y,z) \label{eq:Fqln22}\ .
\end{eqnarray}
These relations are easy to understand: processes $P_{I2}$
differ from processes $P_{I1}$ only by the chirality label of 
the far lepton $\ell_f$. However, the $j\ell_n$ distribution does not know 
about the far lepton, therefore the ${\mathcal F}_{S;IJ}^{(j\ell_n)}$ function 
should be the same for both $J=1$ and $J=2$. Table~\ref{table:Fqln}
has essentially already appeared in \cite{Athanasiou:2006ef}
(see Tables 10 and 11) and we reproduce it here just for completeness.

On the other hand, 
Table~\ref{table:Fll} of Appendix~\ref{app:FIJ} contains some new results
for the ${\mathcal F}_{S,IJ}^{(\ell\ell)}$ functions.
In this case there are still only two independent functions, but the
functional relationship is different from (\ref{eq:Fqln11},\ref{eq:Fqln22}):
\begin{eqnarray}
{\mathcal F}_{S,11}^{(\ell\ell)}(\hat{m}^2;x,y,z)&=&{\mathcal F}_{S,21}^{(\ell\ell)}(\hat{m}^2;x,y,z) \label{eq:Fll11}\ , \\ [2mm]
{\mathcal F}_{S,12}^{(\ell\ell)}(\hat{m}^2;x,y,z)&=&{\mathcal F}_{S,22}^{(\ell\ell)}(\hat{m}^2;x,y,z) \label{eq:Fll22}\ .
\end{eqnarray}
Again, the reason behind these relations is easy to understand intuitively.
Processes $P_{1J}$ are related to processes $P_{2J}$ by simply interchanging 
$q\leftrightarrow \bar{q}$, which, of course, does not affect the two leptons
which are further down the cascade decay chain. Because of (\ref{eq:Fll11}), 
Refs.~\cite{Smillie:2005ar,Athanasiou:2006ef,Athanasiou:2006hv}
found identical results for ${\mathcal F}_{S,11}^{(\ell\ell)}$ and
${\mathcal F}_{S,21}^{(\ell\ell)}$
(corresponding to processes of type 1 and 2 in their notation), but missed
the functions ${\mathcal F}_{S,12}^{(\ell\ell)}$ and 
${\mathcal F}_{S,22}^{(\ell\ell)}$.
This was a direct consequence of the underlying model dependence, and
in particular factor {\bf T3}: the studies
\cite{Smillie:2005ar,Athanasiou:2006ef,Athanasiou:2006hv}
assumed very specific fixed values of the chirality coefficients
(namely, $c_L=1$, $c_R=0$, $b_L=0$, $b_R=1$, $a_L=0$, $a_R=1$ for the 
supersymmetry example and $c_L=1$, $c_R=0$, $b_L=1$, $b_R=0$, 
$a_L=1$, $a_R=0$ for the UED example)
and therefore their results, while correct, are only valid within
this limited model-dependent context. In contrast, deriving the complete set of functions
${\mathcal F}_{S,IJ}^{(\ell\ell)}$ for all possible sets of processes $P_{IJ}$ allows us to 
address the spin question Q1 raised in the Introduction 
in a completely model-independent fashion.

Similar remarks hold for the ${\mathcal F}_{S;IJ}^{(j\ell_f)}$
functions in Appendix~\ref{app:FIJ}. Here again 
the functions ${\mathcal F}_{S;11}^{(j\ell_f)}$ and ${\mathcal F}_{S;21}^{(j\ell_f)}$
agree\footnote{The only discrepancy we found was in the
constant coefficient in front of the $\ln y$ and $\ln \hat{m}^2$ 
terms in the ${\mathcal F}_{6;11}^{(j\ell_f)}$ function:
in eq.~(B.9) of Ref.~\cite{Athanasiou:2006ef} it is 
listed as $-(z+4y)$ while we find $-(1+4y)z$.
Since our results agree with the numerical results
of Figs.~5a and 5b in \cite{Athanasiou:2006ef}, we
believe that eq.~(B.9) in \cite{Athanasiou:2006ef} has a typo.} 
with the results of \cite{Athanasiou:2006ef}, while the
functions ${\mathcal F}_{S;12}^{(j\ell_f)}$ and ${\mathcal F}_{S;22}^{(j\ell_f)}$
are new. However, whether (and what type of) relations
exist between the four functions ${\mathcal F}_{S;IJ}^{(j\ell_f)}$
varies from case to case (i.e. the value of the spin configuration index $S$). 
In the three cases (SFSF, FSFS and FSFV) where there is an intermediate heavy 
scalar between the emitted jet and far lepton, the ${\mathcal F}_{S;IJ}^{(j\ell_f)}$
set is again reduced to only two independent functions, however, the 
exact functional relations are also $S$-dependent: for $S=1$ (SFSF) we find
\begin{eqnarray}
{\mathcal F}_{1,11}^{(j\ell_f)}(\hat{m}^2;x,y,z)&=&{\mathcal F}_{1,12}^{(j\ell_f)}(\hat{m}^2;x,y,z) \ , \\ [2mm]
{\mathcal F}_{1,21}^{(j\ell_f)}(\hat{m}^2;x,y,z)&=&{\mathcal F}_{1,22}^{(j\ell_f)}(\hat{m}^2;x,y,z) \ ,
\end{eqnarray}
while for $S=2$ (FSFS) and $S=3$ (FSFV) we find 
\begin{eqnarray}
{\mathcal F}_{S,11}^{(j\ell_f)}(\hat{m}^2;x,y,z)&=&{\mathcal F}_{S,21}^{(j\ell_f)}(\hat{m}^2;x,y,z) \quad {{\rm for}\ S=2,3} \ , \\ [2mm]
{\mathcal F}_{S,12}^{(j\ell_f)}(\hat{m}^2;x,y,z)&=&{\mathcal F}_{S,22}^{(j\ell_f)}(\hat{m}^2;x,y,z) \quad {{\rm for}\ S=2,3}\ .
\end{eqnarray}
In the remaining 3 cases $S=4,5,6$ (i.e. FVFS, FVFV and SFVF)
we find that all four functions ${\mathcal F}_{S;IJ}^{(j\ell_f)}$
are independent.

\subsubsection{The coefficients $K_{IJ}^{(p)}$}

Having defined the complete sets of functions ${\mathcal F}_{S;IJ}^{(p)}$
entering the general expressions (\ref{qln}-\ref{ll}), it now remains to define the
coefficients $K_{IJ}^{(p)}(f;\varphi_a,\varphi_b,\varphi_c)$ entering those formulas.
Notice that these coefficients do not carry a spin index $S$, i.e. 
they are independent of the assumed spin configuration.
Therefore we only need to define them for each fermion pair $p=\{q\ell_n^\pm,\bar{q}\ell_n^\pm,q\ell_f^\pm,\bar{q}\ell_f^\pm,\ell\ell\}$.

Using the factors from Table~\ref{table:processes}, for the coefficients belonging to processes $P_{11}$
we readily obtain
\begin{eqnarray}
K_{11}^{(     q \ell_n^-)}&=&K_{11}^{(     q \ell_f^+)}=     f |c_L|^2|b_L|^2|a_L|^2+     f |c_R|^2|b_R|^2|a_R|^2\, , \label{K11a}\\
K_{11}^{(\bar{q}\ell_n^-)}&=&K_{11}^{(\bar{q}\ell_f^+)}=\bar{f}|c_L|^2|b_R|^2|a_R|^2+\bar{f}|c_R|^2|b_L|^2|a_L|^2\, , \label{K11b}\\
K_{11}^{(     q \ell_n^+)}&=&K_{11}^{(     q \ell_f^-)}=     f |c_L|^2|b_R|^2|a_R|^2+     f |c_R|^2|b_L|^2|a_L|^2\, , \label{K11c}\\
K_{11}^{(\bar{q}\ell_n^+)}&=&K_{11}^{(\bar{q}\ell_f^-)}=\bar{f}|c_L|^2|b_L|^2|a_L|^2+\bar{f}|c_R|^2|b_R|^2|a_R|^2\, . \label{K11d}
\end{eqnarray}
The corresponding coefficients for processes $P_{12}$ can be now simply obtained from (\ref{K11a}-\ref{K11d})
by the substitution $a_L\leftrightarrow a_R$: 
\begin{eqnarray}
K_{12}^{(     q \ell_n^-)}&=&K_{12}^{(     q \ell_f^+)}=     f |c_L|^2|b_L|^2|a_R|^2+     f |c_R|^2|b_R|^2|a_L|^2\, , \label{K12a} \\
K_{12}^{(\bar{q}\ell_n^-)}&=&K_{12}^{(\bar{q}\ell_f^+)}=\bar{f}|c_L|^2|b_R|^2|a_L|^2+\bar{f}|c_R|^2|b_L|^2|a_R|^2\, , \label{K12b} \\
K_{12}^{(     q \ell_n^+)}&=&K_{12}^{(     q \ell_f^-)}=     f |c_L|^2|b_R|^2|a_L|^2+     f |c_R|^2|b_L|^2|a_R|^2\, , \label{K12c} \\
K_{12}^{(\bar{q}\ell_n^+)}&=&K_{12}^{(\bar{q}\ell_f^-)}=\bar{f}|c_L|^2|b_L|^2|a_R|^2+\bar{f}|c_R|^2|b_R|^2|a_L|^2\, . \label{K12d}
\end{eqnarray}
Next, replacing $f\leftrightarrow \bar{f}$ and $q\leftrightarrow \bar{q}$ 
in (\ref{K11a}-\ref{K11d}) gives the corresponding coefficients for processes $P_{21}$:
\begin{eqnarray}
K_{21}^{(\bar{q}\ell_n^-)}&=&K_{21}^{(\bar{q}\ell_f^+)}=\bar{f}|c_L|^2|b_L|^2|a_L|^2+\bar{f}|c_R|^2|b_R|^2|a_R|^2\, , \label{K21a} \\
K_{21}^{(     q \ell_n^-)}&=&K_{21}^{(     q \ell_f^+)}=     f |c_L|^2|b_R|^2|a_R|^2+     f |c_R|^2|b_L|^2|a_L|^2\, , \label{K21b} \\
K_{21}^{(\bar{q}\ell_n^+)}&=&K_{21}^{(\bar{q}\ell_f^-)}=\bar{f}|c_L|^2|b_R|^2|a_R|^2+\bar{f}|c_R|^2|b_L|^2|a_L|^2\, , \label{K21c} \\
K_{21}^{(     q \ell_n^+)}&=&K_{21}^{(     q \ell_f^-)}=     f |c_L|^2|b_L|^2|a_L|^2+     f |c_R|^2|b_R|^2|a_R|^2\, . \label{K21d}
\end{eqnarray}
Finally, replacing $a_L\leftrightarrow a_R$ in (\ref{K21a}-\ref{K21d}) yields the coefficients
for processes $P_{22}$:
\begin{eqnarray}
K_{22}^{(\bar{q}\ell_n^-)}&=&K_{22}^{(\bar{q}\ell_f^+)}=\bar{f}|c_L|^2|b_L|^2|a_R|^2+\bar{f}|c_R|^2|b_R|^2|a_L|^2\, , \label{K22a} \\
K_{22}^{(     q \ell_n^-)}&=&K_{22}^{(     q \ell_f^+)}=     f |c_L|^2|b_R|^2|a_L|^2+     f |c_R|^2|b_L|^2|a_R|^2\, , \label{K22b} \\
K_{22}^{(\bar{q}\ell_n^+)}&=&K_{22}^{(\bar{q}\ell_f^-)}=\bar{f}|c_L|^2|b_R|^2|a_L|^2+\bar{f}|c_R|^2|b_L|^2|a_R|^2\, , \label{K22c} \\
K_{22}^{(     q \ell_n^+)}&=&K_{22}^{(     q \ell_f^-)}=     f |c_L|^2|b_L|^2|a_R|^2+     f |c_R|^2|b_R|^2|a_L|^2\, . \label{K22d}
\end{eqnarray}

The coefficients $K_{IJ}^{(\ell\ell)}$ for the dilepton distributions can be expressed 
in various ways, for example in terms of the coefficients involving the near lepton $\ell_n$
\begin{equation}
K_{IJ}^{(\ell\ell)}=
 K_{IJ}^{(     q \ell_n^-)}
+K_{IJ}^{(\bar{q}\ell_n^-)}
+K_{IJ}^{(     q \ell_n^+)}
+K_{IJ}^{(\bar{q}\ell_n^+)} \ ;
\label{Kll_n}
\end{equation}
in terms of the coefficients involving the far lepton $\ell_f$:
\begin{equation}
K_{IJ}^{(\ell\ell)}=
 K_{IJ}^{(     q \ell_f^-)}
+K_{IJ}^{(\bar{q}\ell_f^-)}
+K_{IJ}^{(     q \ell_f^+)}
+K_{IJ}^{(\bar{q}\ell_f^+)} \ ;
\label{Kll_f}
\end{equation}
in terms of the coefficients involving the positively charged lepton $\ell^+$
\begin{equation}
K_{IJ}^{(\ell\ell)}=
 K_{IJ}^{(     q \ell_n^+)}
+K_{IJ}^{(\bar{q}\ell_n^+)}
+K_{IJ}^{(     q \ell_f^+)}
+K_{IJ}^{(\bar{q}\ell_f^+)} \ ;
\label{Kll_+}
\end{equation}
or finally, in terms of the coefficients involving the negatively charged lepton $\ell^-$:
\begin{equation}
K_{IJ}^{(\ell\ell)}=
 K_{IJ}^{(     q \ell_n^-)}
+K_{IJ}^{(\bar{q}\ell_n^-)}
+K_{IJ}^{(     q \ell_f^-)}
+K_{IJ}^{(\bar{q}\ell_f^-)} \ .
\label{Kll_-}
\end{equation}
All of the definitions (\ref{Kll_n}-\ref{Kll_-}) are equivalent because of the 
relations (\ref{K11a}-\ref{K22d}) existing between the various coefficients.
Notice the normalisation condition
\begin{equation}
\sum_{I=1}^{2}\sum_{J=1}^{2} K_{IJ}^{(\ell\ell)} = 2\ .
\end{equation}

With the definitions (\ref{K11a}-\ref{K22d}) and the conventions
(\ref{unit_qln}-\ref{unit_ll}) and (\ref{unit_a}-\ref{unit_c}), 
our distributions (\ref{qln}-\ref{ll}) are normalised as follows:
\begin{eqnarray}
\int_0^\infty \left( \frac{dN}{d\hat{m}^2_{q\ell_n^\pm}}\right)_S\, d\hat{m}^2_{q\ell_n^\pm}
&=& 
\int_0^\infty \left( \frac{dN}{d\hat{m}^2_{q\ell_f^\pm}}\right)_S\, d\hat{m}^2_{q\ell_f^\pm}
= 
\frac{f}{2} \ ,
\label{norm_ql}  \\ [2mm]
\int_0^\infty \left( \frac{dN}{d\hat{m}^2_{\bar{q}\ell_n^\pm}}\right)_S\, d\hat{m}^2_{\bar{q}\ell_n^\pm}
&=& 
\int_0^\infty \left( \frac{dN}{d\hat{m}^2_{\bar{q}\ell_f^\pm}}\right)_S\, d\hat{m}^2_{\bar{q}\ell_f^\pm}
= 
\frac{\bar{f}}{2} \ ,
\label{norm_qbl}  \\ [2mm]
\int_0^\infty \left( \frac{dN}{d\hat{m}^2_{\ell\ell}}\right)_S\, d\hat{m}^2_{\ell\ell}
&=& 
1\ .
\label{norm_ll}
\end{eqnarray}
It is now clear how the factor of $\frac{1}{2}$ in
eqs.~(\ref{qln}-\ref{ll}) is related to the normalisation:
the dilepton distribution (\ref{ll}), which is experimentally observable,
is unit normalised, as seen by eq.~(\ref{norm_ll}). On the other hand, 
eqs.~(\ref{norm_ql}) and (\ref{norm_qbl}) show that the individual 
$\{q\ell_n\}$, $\{\bar{q}\ell_n\}$, $\{q\ell_f\}$ and $\{\bar{q}\ell_f\}$
distributions are not unit normalised. However, this is not a problem, since
those distributions cannot be separately observed. In fact, as we shall 
see in the next section, the
normalisation (\ref{norm_ql},\ref{norm_qbl}) is precisely what is needed 
in order to unit normalise the {\em observable} invariant mass distributions
for $\{j\ell^+\}$ and $\{j\ell^-\}$. 

\section{Observable distributions in a $\{q,\ell^\pm,\ell^\mp\}$ chain}
\label{sec:observable}

\subsection{Invariant mass formulas in the $\{{\mathcal F}_{S;IJ}^{(p)} \}$ basis}
\label{sec:FIJbasis}

If we could identify the nature of the jet ($q$ versus $\bar{q}$)
on an event by event basis, we could use directly the distributions
(\ref{qln}-\ref{ll}) derived in the previous section. 
As mentioned in the Introduction, there may be cases where 
this is possible, e.g. if $q$ is a $b$-quark, or alternatively, 
if it is a lepton so that the decay chain of Fig.~\ref{fig:ABCD}
represents a {\em trilepton} signature. Here, however, we shall make
the conservative assumption, which also happens to be true in many models,
that $q$ is a light flavor quark, so that the experimental distinction between
a $q$ and $\bar{q}$ cannot be made. In that case, we have to add 
the corresponding distributions involving a $q$ and a $\bar{q}$:
\begin{eqnarray}
\left( \frac{dN}{d\hat{m}^2_{j\ell_n^\pm}}\right)_S 
&=&
 \left( \frac{dN}{d\hat{m}^2_{     q \ell_n^\pm}}\right)_S 
+\left( \frac{dN}{d\hat{m}^2_{\bar{q}\ell_n^\pm}}\right)_S \nonumber \\ [2mm]
&\equiv& \frac{1}{2}
\sum_{I=1}^{2}\sum_{J=1}^{2} K_{IJ}^{(j\ell_n^\pm)}(f,\varphi_a,\varphi_b,\varphi_c) 
\, {\mathcal F}_{S;IJ}^{(j\ell_n)}(\hat{m}^2_{j\ell_n^\pm};x,y,z)\ ,
\label{jln}   \\ [2mm]
\left( \frac{dN}{d\hat{m}^2_{j\ell_f^\pm}}\right)_S 
&=& 
 \left( \frac{dN}{d\hat{m}^2_{     q \ell_f^\pm}}\right)_S 
+\left( \frac{dN}{d\hat{m}^2_{\bar{q}\ell_f^\pm}}\right)_S \nonumber \\ [2mm]
&\equiv& \frac{1}{2}
\sum_{I=1}^{2}\sum_{J=1}^{2} K_{IJ}^{(j\ell_f^\pm)}(f,\varphi_a,\varphi_b,\varphi_c) 
\, {\mathcal F}_{S;IJ}^{(j\ell_f)}(\hat{m}^2_{j\ell_f^\pm};x,y,z)\ .
\label{jlf}   
\end{eqnarray}
Since the ${\mathcal F}_{S;IJ}^{(p)}$ functions do not depend on the
$q$-$\bar{q}$ ambiguity (factor {\bf E3}), the 
new set of coefficients $K_{IJ}^{(j\ell_n^\pm)}$
and $K_{IJ}^{(j\ell_f^\pm)}$ can be simply related to those
already introduced in the previous section:
\begin{eqnarray}
 K_{IJ}^{(     j \ell_n^\pm)}(f,\varphi_a,\varphi_b,\varphi_c) &=&
 K_{IJ}^{(     q \ell_n^\pm)}(f,\varphi_a,\varphi_b,\varphi_c)
+K_{IJ}^{(\bar{q}\ell_n^\pm)}(f,\varphi_a,\varphi_b,\varphi_c) \ , \label{Kjln} \\ [2mm]
 K_{IJ}^{(     j \ell_f^\pm)}(f,\varphi_a,\varphi_b,\varphi_c) &=&
 K_{IJ}^{(     q \ell_f^\pm)}(f,\varphi_a,\varphi_b,\varphi_c)
+K_{IJ}^{(\bar{q}\ell_f^\pm)}(f,\varphi_a,\varphi_b,\varphi_c) \ . \label{Kjlf} 
\end{eqnarray}
Substituting the definitions (\ref{K11a}-\ref{K22d}) into
(\ref{Kjln}) and (\ref{Kjlf}), we find that the $K_{IJ}^{(j\ell)}$ 
coefficients can be expressed in terms of the particle-antiparticle fraction $f$
and the relative chiralities $\varphi_a$, $\varphi_b$ and $\varphi_c$ as follows
\begin{eqnarray}
K_{11}^{(j\ell_n^-)}(f,\varphi_a,\varphi_b,\varphi_c) 
         &=& (     f |c_L|^2 + \bar{f}|c_R|^2)|b_L|^2|a_L|^2 + (\bar{f}|c_L|^2+     f |c_R|^2)|b_R|^2|a_R|^2 \ , \label{K11def}\\ [2mm]
K_{12}^{(j\ell_n^-)}(f,\varphi_a,\varphi_b,\varphi_c)
         &=& (     f |c_L|^2 + \bar{f}|c_R|^2)|b_L|^2|a_R|^2 + (\bar{f}|c_L|^2+     f |c_R|^2)|b_R|^2|a_L|^2 \ , \label{K12def}\\ [2mm]
K_{21}^{(j\ell_n^-)}(f,\varphi_a,\varphi_b,\varphi_c)
         &=& (\bar{f}|c_L|^2 +      f |c_R|^2)|b_L|^2|a_L|^2 + (     f |c_L|^2+\bar{f}|c_R|^2)|b_R|^2|a_R|^2 \ , \label{K21def}\\ [2mm]
K_{22}^{(j\ell_n^-)}(f,\varphi_a,\varphi_b,\varphi_c)
         &=& (\bar{f}|c_L|^2 +      f |c_R|^2)|b_L|^2|a_R|^2 + (     f |c_L|^2+\bar{f}|c_R|^2)|b_R|^2|a_L|^2 \ . \label{K22def}
\end{eqnarray}
The remaining $K_{IJ}^{(j\ell)}$ coefficients can be related to these as
\begin{eqnarray}
K_{11}^{(j\ell_n^-)} &=& K_{11}^{(j\ell_f^+)} = K_{21}^{(j\ell_n^+)} = K_{21}^{(j\ell_f^-)} \ , \label{K11rel} \\ [2mm]
K_{12}^{(j\ell_n^-)} &=& K_{12}^{(j\ell_f^+)} = K_{22}^{(j\ell_n^+)} = K_{22}^{(j\ell_f^-)} \ , \label{K12rel} \\ [2mm]
K_{21}^{(j\ell_n^-)} &=& K_{21}^{(j\ell_f^+)} = K_{11}^{(j\ell_n^+)} = K_{11}^{(j\ell_f^-)} \ , \label{K21rel} \\ [2mm]
K_{22}^{(j\ell_n^-)} &=& K_{22}^{(j\ell_f^+)} = K_{12}^{(j\ell_n^+)} = K_{12}^{(j\ell_f^-)} \ . \label{K22rel} 
\end{eqnarray}
It is important to notice that while the coefficients
$K_{IJ}^{(j\ell)}(f,\varphi_a,\varphi_b,\varphi_c)$ defined in 
(\ref{K11def}-\ref{K22rel}) depend on all four variables
$f$, $\varphi_a$, $\varphi_b$ and $\varphi_c$, the dependence on 
$f$ and $\varphi_c$ only appears through the combinations
$f|c_L|^2 + \bar{f}|c_R|^2 =f\cos^2\varphi_c+\bar{f}\sin^2\varphi_c$ 
and $\bar{f}|c_L|^2 + f |c_R|^2 = \bar{f} \cos^2\varphi_c+f\sin^2\varphi_c$.
We shall therefore find it convenient to introduce an alternative
chirality parameter $\tilde\varphi_c$ defined by the relations:
\begin{eqnarray}
\cos^2 \tilde{\varphi}_{c}=     f  \cos^2\varphi_c+\bar{f}\sin^2\varphi_c\ , \label{ct_def} \\[2mm]
\sin^2 \tilde{\varphi}_{c}=\bar{f} \cos^2\varphi_c+     f \sin^2\varphi_c\ , \label{st_def}
\end{eqnarray}
so that
\begin{equation}
\cos 2\tilde\varphi_c = (f-\bar{f})\, \cos 2\varphi_c\ .
\label{cos2ctilde}
\end{equation}
The relationship between the newly introduced parameter $\tilde\varphi_c$ 
and the original parameters $f$ and $\varphi_c$ is pictorially illustrated in
Fig.~\ref{fig:ctilde}.

\FIGURE[t]{
\epsfig{file=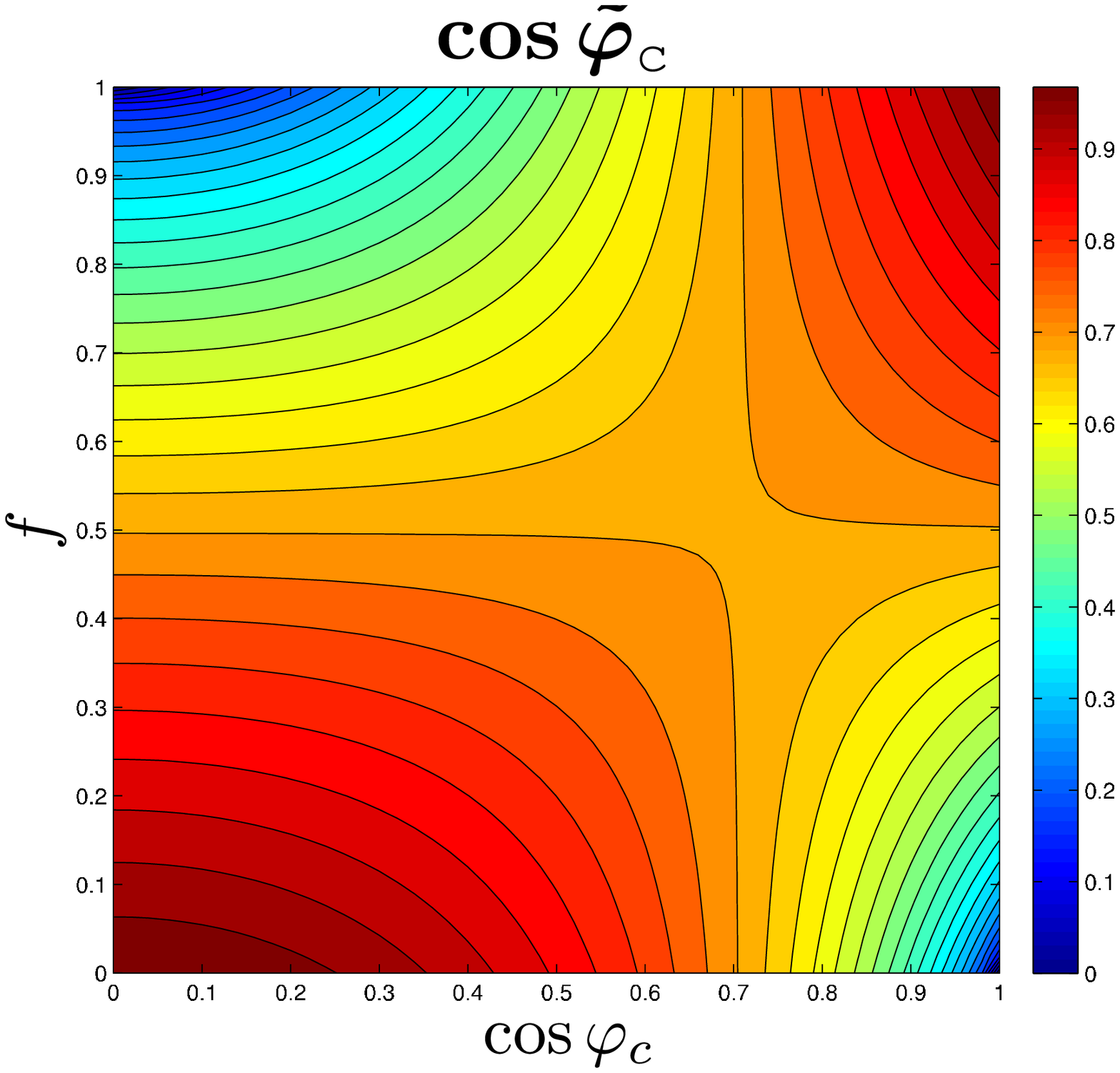,width=13cm}
\caption{\sl A contour plot of $\cos\tilde\varphi_c$ as a function of
$\cos\varphi_c$ and $f$.}
\label{fig:ctilde}}

In terms of the new parameter $\tilde\varphi_c$, the defining equations 
(\ref{K11def}-\ref{K22def}) for the $K_{IJ}^{(j\ell)}(f,\varphi_a,\varphi_b,\varphi_c)$ 
coefficients simply become
\begin{eqnarray}
K_{11}^{(j\ell_n^-)}(\varphi_a,\varphi_b,\tilde\varphi_c) 
         &=& \cos^2\tilde\varphi_c \cos^2\varphi_b \cos^2\varphi_a 
           + \sin^2\tilde\varphi_c \sin^2\varphi_b \sin^2\varphi_a  \ , \label{K11def2}\\ [2mm]
K_{12}^{(j\ell_n^-)}(\varphi_a,\varphi_b,\tilde\varphi_c)
         &=& \cos^2\tilde\varphi_c \cos^2\varphi_b \sin^2\varphi_a 
           + \sin^2\tilde\varphi_c \sin^2\varphi_b \cos^2\varphi_a  \ , \label{K12def2}\\ [2mm]
K_{21}^{(j\ell_n^-)}(\varphi_a,\varphi_b,\tilde\varphi_c)
         &=& \sin^2\tilde\varphi_c \cos^2\varphi_b \cos^2\varphi_a 
           + \cos^2\tilde\varphi_c \sin^2\varphi_b \sin^2\varphi_a  \ , \label{K21def2}\\ [2mm]
K_{22}^{(j\ell_n^-)}(\varphi_a,\varphi_b,\tilde\varphi_c)
         &=& \sin^2\tilde\varphi_c \cos^2\varphi_b \sin^2\varphi_a 
           + \cos^2\tilde\varphi_c \sin^2\varphi_b \cos^2\varphi_a  \ , \label{K22def2}
\end{eqnarray}
and the remaining relations (\ref{K11rel}-\ref{K22rel}) are unchanged.

Using the relations (\ref{K11def2}-\ref{K22def2}), and the normalisation conditions (\ref{unit_f})
and (\ref{unit_a}-\ref{unit_c}), it is easy to check that 
the $K_{IJ}^{(j\ell)}$ coefficients obey the following normalisation conditions
\begin{eqnarray}
\sum_{I=1}^2 \sum_{J=1}^2 K_{IJ}^{(j\ell_n^\pm)} &=& 1  \ ,  \label{unitKjln} \\ [2mm]
\sum_{I=1}^2 \sum_{J=1}^2 K_{IJ}^{(j\ell_f^\pm)} &=& 1  \ .  \label{unitKjlf}
\end{eqnarray}
Given the unit normalisation (\ref{unit_qln}-\ref{unit_ll})
of our basis functions ${\mathcal F}_{S;IJ}^{(p)}$,
eqs.~(\ref{unitKjln}) and (\ref{unitKjlf}) readily imply that 
the $\{j\ell_n^\pm\}$ and $\{j\ell_f^\pm\}$ distributions (\ref{jln}) and (\ref{jlf})
are automatically half-unit normalised\footnote{This can also be seen directly 
from the definitions (\ref{jln}) and (\ref{jlf}) of the $j\ell$ distributions 
and making use of eqs.~(\ref{norm_ql}), (\ref{norm_qbl}) and (\ref{unit_f}).}
\begin{eqnarray}
\int_0^\infty \left( \frac{dN}{d\hat{m}^2_{j\ell_n^\pm}}\right)_S\, d\hat{m}^2_{j\ell_n^\pm}
&=& \frac{1}{2} \ ,
\label{norm_jln}  \\ [2mm]
\int_0^\infty \left( \frac{dN}{d\hat{m}^2_{j\ell_f^\pm}}\right)_S\, d\hat{m}^2_{j\ell_f^\pm}
&=& \frac{1}{2} \ .
\label{norm_jlf} 
\end{eqnarray}

The last step in deriving the {\em experimentally observable}
invariant mass distributions is to recall that the near
and far lepton ($\ell_n$ and $\ell_f$) cannot be distinguished 
on an event by event basis, therefore we need to form the distributions
which are based on definite lepton charge:
\begin{eqnarray}
\left( \frac{dN}{dm^2_{j\ell^+}}\right)_S 
&\equiv&
 \left( \frac{dN}{dm^2_{j\ell_n^+}}\right)_S 
+\left( \frac{dN}{dm^2_{j\ell_f^+}}\right)_S \ ,
\label{defjl+}   \\ [2mm]
\left( \frac{dN}{dm^2_{j\ell^-}}\right)_S 
&\equiv&
 \left( \frac{dN}{dm^2_{j\ell_n^-}}\right)_S 
+\left( \frac{dN}{dm^2_{j\ell_f^-}}\right)_S \ .
\label{defjl-} 
\end{eqnarray}
When combining the jet-near lepton and the jet-far lepton distributions
in eqs.~(\ref{defjl+},\ref{defjl-}), 
one has to be careful since until now each individual distribution
was written in terms of its own 
unit-normalised invariant mass variable $\hat{m}_{j\ell_n}$
and $\hat{m}_{j\ell_f}$. In general, these two variables will be
different, since the kinematic endpoints 
$\hat{m}_{j\ell_n}^{max}$ and $\hat{m}_{j\ell_f}^{max}$,
to which they are normalised, will not coincide. 
Once this problem is identified, it can be handled in various ways, 
for example, by writing out the sums (\ref{defjl+},\ref{defjl-})  
in terms of the actual (i.e., not unit-normalised) invariant masses.
In this paper, we prefer to keep the $\hat{m}$ notation, 
and write all of our distributions in terms of unit-normalised 
invariant mass variables. To this end, we normalise any jet-lepton
invariant mass $m_{j\ell}$ to the endpoint
\begin{equation}
m_{j\ell}^{max}\equiv \max\{m_{j\ell_n}^{max},m_{j\ell_f}^{max}\}
\label{defmjlmax}
\end{equation}
of the {\em combined} jet-lepton distribution as follows:
\begin{equation}
\hat{m}_{j\ell^\pm}\equiv \frac{m_{j\ell^\pm}}{m_{j\ell}^{max}}\ .
\label{defmjlhat}
\end{equation}
Introducing the ratios
\begin{eqnarray}
r_n &\equiv& \frac{m_{j\ell}^{max}}{m_{j\ell_n}^{max}}\ , \label{rn}\\
r_f &\equiv& \frac{m_{j\ell}^{max}}{m_{j\ell_f}^{max}}\ , \label{rf} 
\end{eqnarray}
we can now write the combined jet-lepton distributions for each 
lepton charge in terms of the unit-normalised variable (\ref{defmjlhat}) as
\begin{eqnarray}
\left( \frac{dN}{d\hat{m}^2_{j\ell^+}}\right)_S 
&=&
\frac{1}{2} \sum_{I=1}^{2}\sum_{J=1}^{2} K_{IJ}^{(j\ell_n^+)}(\varphi_a,\varphi_b,\tilde\varphi_c) 
\, {r_n^2}\, {\mathcal F}_{S;IJ}^{(j\ell_n)}(r_n^2 \hat{m}_{j\ell^+}^2;x,y,z)
\nonumber \\ [2mm]
&+&
\frac{1}{2} \sum_{I=1}^{2}\sum_{J=1}^{2} K_{IJ}^{(j\ell_f^+)}(\varphi_a,\varphi_b,\tilde\varphi_c) 
\, {r_f^2}\, {\mathcal F}_{S;IJ}^{(j\ell_f)}(r_f^2 \hat{m}_{j\ell^+}^2;x,y,z)
\ ,
\label{jl+}   \\ [2mm]
\left( \frac{dN}{d\hat{m}^2_{j\ell^-}}\right)_S 
&=&
\frac{1}{2} \sum_{I=1}^{2}\sum_{J=1}^{2} K_{IJ}^{(j\ell_n^-)}(\varphi_a,\varphi_b,\tilde\varphi_c) 
\, {r_n^2}\, {\mathcal F}_{S;IJ}^{(j\ell_n)}(r_n^2 \hat{m}_{j\ell^-}^2;x,y,z)
\nonumber \\ [2mm]
&+&
\frac{1}{2} \sum_{I=1}^{2}\sum_{J=1}^{2} K_{IJ}^{(j\ell_f^-)}(\varphi_a,\varphi_b,\tilde\varphi_c) 
\, {r_f^2}\, {\mathcal F}_{S;IJ}^{(j\ell_f)}(r_f^2 \hat{m}_{j\ell^-}^2;x,y,z)
\ .
\label{jl-} 
\end{eqnarray}
Note that whenever the two endpoints $\hat{m}_{j\ell_n}^{max}$ and $\hat{m}_{j\ell_f}^{max}$
are different, one of the two ratios $r_n$ and $r_f$ is guaranteed to exceed 1, 
so that there will be a range of masses for which the corresponding argument
($r_n \hat{m}_{j\ell}$ or $r_f \hat{m}_{j\ell}$) in the ${\mathcal F}_{S;IJ}^{(j\ell)}$
functions would exceed 1 as well. This is why it was necessary to 
extend the range of definition of our 
${\mathcal F}_{S;IJ}^{(j\ell_n)}$ and ${\mathcal F}_{S;IJ}^{(j\ell_f)}$
functions in Appendix~\ref{app:FIJ} to be $0\le\hat{m}<\infty$,
although it seems trivial, since the functions vanish identically 
for $\hat{m}>1$.

As can be readily seen from eqs.~(\ref{norm_jln}) and (\ref{norm_jlf}),
both of these observable distributions are unit normalised
\begin{eqnarray}
\int_0^\infty \left( \frac{dN}{d \hat{m}^2_{j\ell^+}}\right)_S\, d \hat{m}^2_{j\ell^+}
&=& 1 \ ,
\label{norm_jl+}  \\ [2mm]
\int_0^\infty \left( \frac{dN}{d \hat{m}^2_{j\ell^-}}\right)_S\, d \hat{m}^2_{j\ell^-}
&=& 1 \ ,
\label{norm_jl-}  
\end{eqnarray}
just like the observable dilepton distribution (\ref{ll}) (see eq.~(\ref{norm_ll})).

This concludes the derivation of our first main result. 
It is worth recapitulating what we managed to 
achieve so far. We obtained exact analytical expressions for 
the three experimentally observable invariant mass distributions: 
dilepton (\ref{ll}), jet plus positive lepton (\ref{jl+})
and jet plus negative lepton (\ref{jl-}).
All three of our formulas are unit normalised and can be 
readily rescaled for the actual observed number of events
(which is the same for each of the three distributions).
Our formulas are written in terms of a set of known functions
${\mathcal F}_{S;IJ}^{(p)}$ which are explicitly defined 
in Appendix~\ref{app:FIJ}. The coefficients $K_{IJ}^{(p)}$ appearing
in our formulas are defined in eqs.~(\ref{K11def2}-\ref{K22def2}),
(\ref{K11rel}-\ref{K22rel}) and (\ref{Kll_n}-\ref{Kll_-}),
and depend on only three model-dependent parameters
$\varphi_a$, $\varphi_b$ and $\tilde\varphi_c$.
Those parameters are defined in eqs.~(\ref{phi_angles}) 
and (\ref{ct_def},\ref{st_def}), and are a priori unknown,
so that they must be measured from the data. 

The basic idea of our spin measurement method (whose main steps will
be presented in detail in the next section) will be to fit
our formulas to the shapes of the measured invariant 
mass distributions. Since there are 6 possible spin configurations, 
this fit will have to be repeated 6 times -- once for each value of $S$.
Since we have only three parametric degrees 
of freedom $\varphi_a$, $\varphi_b$ and $\tilde\varphi_c$,
with which we are trying to fit three whole {\em distributions},
one would expect that the fit will be successful only for the
correct spin configuration $S$ and for the remaining 5 
spin cases the fit will fail. Indeed we find that this expectation 
is generally correct, and in Sec.~\ref{sec:numerics}
we shall give explicit examples of how this procedure might 
work in practice. However, we also find that 
there are two pairs of ``twin'' spin scenarios,
discussed in Sec.~\ref{sec:degenerate}, which are often 
completely indistinguishable, even as a matter of principle.

\subsection{Invariant mass formulas in the $\{{\mathcal F}_{S;\alpha}^{(p)},
{\mathcal F}_{S;\beta}^{(p)}, {\mathcal F}_{S;\gamma}^{(p)}, {\mathcal F}_{S;\delta}^{(p)}\}$ basis}
\label{sec:Falphabasis}

While the fitting exercise just described 
can in principle be performed with our results 
written in terms of the ${\mathcal F}_{S;IJ}^{(p)}$ 
basis functions from Appendix \ref{app:FIJ}, 
we find that for the actual practical application
of our method, it is much more convenient to rewrite 
our results in a different functional basis.
We therefore introduce an alternative set of basis functions 
$\{{\mathcal F}_{S;\alpha}^{(p)}, {\mathcal F}_{S;\beta}^{(p)}, {\mathcal F}_{S;\gamma}^{(p)}, {\mathcal F}_{S;\delta}^{(p)}\}$
which are nothing but linear combinations of 
those appearing in our old set:
\begin{eqnarray}
\mathcal{F}_{S;\alpha}^{(p)}&=&\frac{1}{4}\left\{\mathcal{F}_{S;11}^{(p)}-\mathcal{F}_{S;12}^{(p)}+\mathcal{F}_{S;21}^{(p)}-\mathcal{F}_{S;22}^{(p)}\right\}\ , \label{Falpha} \\ [2mm]
\mathcal{F}_{S;\beta }^{(p)}&=&\frac{1}{4}\left\{\mathcal{F}_{S;11}^{(p)}+\mathcal{F}_{S;12}^{(p)}-\mathcal{F}_{S;21}^{(p)}-\mathcal{F}_{S;22}^{(p)}\right\}\ , \label{Fbeta}  \\ [2mm]
\mathcal{F}_{S;\gamma}^{(p)}&=&\frac{1}{4}\left\{\mathcal{F}_{S;11}^{(p)}-\mathcal{F}_{S;12}^{(p)}-\mathcal{F}_{S;21}^{(p)}+\mathcal{F}_{S;22}^{(p)}\right\}\ , \label{Fgamma} \\ [2mm]
\mathcal{F}_{S;\delta}^{(p)}&=&\frac{1}{4}\left\{\mathcal{F}_{S;11}^{(p)}+\mathcal{F}_{S;12}^{(p)}+\mathcal{F}_{S;21}^{(p)}+\mathcal{F}_{S;22}^{(p)}\right\}\ , \label{Fdelta} 
\end{eqnarray}
for any $p\in \{\ell\ell, j\ell_n, j\ell_f\}$.
Using the normalisation conditions (\ref{unit_qln}-\ref{unit_ll}),
it is easy to see that the newly defined functions
${\mathcal F}_{S;\alpha}^{(p)}$, ${\mathcal F}_{S;\beta}^{(p)}$ and ${\mathcal F}_{S;\gamma}^{(p)}$
are zero-normalised
\begin{eqnarray}
\int_0^\infty {\mathcal F}_{S;\alpha}^{(p)}(\hat{m}^2;x,y,z) \, d\hat{m}^2 = 0\ , \label{unit_Falpha} \\
\int_0^\infty {\mathcal F}_{S;\beta }^{(p)}(\hat{m}^2;x,y,z) \, d\hat{m}^2 = 0\ , \label{unit_Fbeta} \\
\int_0^\infty {\mathcal F}_{S;\gamma}^{(p)}(\hat{m}^2;x,y,z) \, d\hat{m}^2 = 0\ , \label{unit_Fgamma} 
\end{eqnarray}
while the function ${\mathcal F}_{S;\delta}^{(p)}$
is unit-normalised
\begin{eqnarray}
\int_0^\infty {\mathcal F}_{S;\delta}^{(p)}(\hat{m}^2;x,y,z) \, d\hat{m}^2 = 1\ . \label{unit_Fdelta} 
\end{eqnarray}
The explicit form of the new basis functions 
$\{{\mathcal F}_{S;\alpha}^{(p)}, {\mathcal F}_{S;\beta}^{(p)}, {\mathcal F}_{S;\gamma}^{(p)}, {\mathcal F}_{S;\delta}^{(p)}\}$
can be easily obtained by substituting the results from Appendix \ref{app:FIJ}
into the definitions~(\ref{Falpha}-\ref{Fdelta}). The result is given 
in Appendix~\ref{app:Falpha}. 

The advantage of the new set of basis functions becomes immediately apparent 
when we rewrite our results for the different invariant mass distributions:
\begin{eqnarray}
\left( \frac{dN}{d\hat{m}^2_{\ell\ell}}\right)_S &\equiv& L_S^{+-} = 
{\mathcal F}_{S;\delta}^{(\ell\ell)}(\hat{m}^2_{\ell\ell};x,y,z)
+ \alpha(\varphi_b,\varphi_a){\mathcal F}_{S;\alpha}^{(\ell\ell)}(\hat{m}^2_{\ell\ell};x,y,z) \ ,
\label{dN_dll} \\
\left( \frac{dN}{d\hat{m}^2_{j\ell_n^\pm}}\right)_S &=& 
\frac{1}{2}\biggl\{ {\mathcal F}_{S;\delta}^{(j\ell_n)}(\hat{m}^2_{j\ell_n};x,y,z)
\mp \beta(\tilde\varphi_c,\varphi_b){\mathcal F}_{S;\beta}^{(j\ell_n)}(\hat{m}^2_{j\ell_n};x,y,z) \biggr\} \ ,
\label{dN_djln} \\
\left( \frac{dN}{d\hat{m}^2_{j\ell_f^\pm}}\right)_S &=& 
\frac{1}{2}\biggl\{ {\mathcal F}_{S;\delta}^{(j\ell_f)}(\hat{m}^2_{j\ell_f};x,y,z)
+   \alpha(\varphi_b     ,\varphi_a){\mathcal F}_{S;\alpha}^{(j\ell_f)}(\hat{m}^2_{j\ell_f};x,y,z) \nonumber \\
&&\pm \beta(\tilde\varphi_c,\varphi_b){\mathcal F}_{S;\beta}^{(j\ell_f)}(\hat{m}^2_{j\ell_f};x,y,z) 
\pm \gamma(\varphi_a,\tilde\varphi_c){\mathcal F}_{S;\gamma}^{(j\ell_f)}(\hat{m}^2_{j\ell_f};x,y,z)\biggr\} \ ,
\label{dN_djlf} 
\end{eqnarray}
where $\alpha$, $\beta$ and $\gamma$ are constant coefficients related to the chirality parameters
(\ref{phi_angles}) as follows
\begin{eqnarray}
\alpha(\varphi_b,\varphi_a) &\equiv& \cos2\varphi_b\cos2\varphi_a\ ,
\label{def_alpha} \\
\beta( \tilde\varphi_c,\varphi_b) &\equiv& \cos2\tilde\varphi_c\cos2\varphi_b\ ,
\label{def_beta} \\
\gamma(\varphi_a,\tilde\varphi_c) &\equiv& \cos2\varphi_a\cos2\tilde\varphi_c\ .
\label{def_gamma} 
\end{eqnarray}
Each one of the $\alpha$, $\beta$ and $\gamma$ parameters can take values in the interval $[-1,1]$.
However, $\alpha$, $\beta$ and $\gamma$ are not completely unrelated.
Given their definitions (\ref{def_alpha}-\ref{def_gamma}), it is easy to see that
they must satisfy certain relations among themselves, and those are listed in Appendix~\ref{app:params}.

Using the normalisation conditions (\ref{unit_Falpha}-\ref{unit_Fdelta}),
one can easily show that all distributions (\ref{dN_dll}-\ref{dN_djlf})
are properly normalised as in eqs.~(\ref{norm_ll}, \ref{norm_jln}, \ref{norm_jlf}).
Eqs.~(\ref{dN_dll}-\ref{def_gamma}) represent our main theoretical result.
In the remainder of this section we shall discuss and interpret those equations.
In the subsequent sections we shall illustrate how Eqs.~(\ref{dN_dll}-\ref{def_gamma})
can be used for measurements of the spins, couplings and mixing angles.

There are several desirable features of the 
$\{{\mathcal F}_{S;\alpha}^{(p)}, {\mathcal F}_{S;\beta}^{(p)}, {\mathcal F}_{S;\gamma}^{(p)}, {\mathcal F}_{S;\delta}^{(p)}\}$
basis used to write down eqs.~(\ref{dN_dll}-\ref{dN_djlf}).
First, consider the ${\mathcal F}_{S;\delta}^{(p)}$ terms 
which appear without any parametric coefficients. 
In most cases for $S$ and $p$, the function ${\mathcal F}_{S;\delta}^{(p)}$
simply gives the invariant mass distribution as predicted
by pure phase space, i.e. where any spin correlations are 
ignored. This is true whenever there are only scalars and/or fermions
among the intermediate particles appearing between the SM fermion pair 
whose invariant mass is being calculated. However, 
if a heavy {\em vector boson} appears among the intermediate heavy particles, 
the ${\mathcal F}_{S;\delta}^{(p)}$ function always deviates from pure phase space. 
In fact this deviation cannot be compensated by a judicious choice of
the $\alpha$, $\beta$ and $\gamma$ parameters. Therefore, one of our
general conclusions will be that a heavy vector boson always leads
to deviations from pure phase space and conversely, whenever 
a pure phase space distribution is observed, a heavy vector boson 
can be ruled out.

Another nice feature of eqs.~(\ref{dN_dll}-\ref{dN_djlf})
is that the three parametric degrees of freedom
are now explicit in terms of the coefficients 
$\alpha$, $\beta$ and $\gamma$. Even more importantly,
it is immediately apparent which particular combination of 
the model-dependent parameters $\varphi_a$, 
$\varphi_b$ and $\tilde\varphi_c$
(i.e. which combination of couplings and mixing angles)
can be measured from any given distribution.
For example, the observable dilepton invariant mass distribution 
$L_S^{+-}$ given in eq.~(\ref{dN_dll})
only depends on $\alpha$, but does not depend on $\beta$ and $\gamma$.
Since the dilepton distribution is experimentally observable, this would 
allow a direct measurement of the $\alpha$ parameter from the
dilepton data alone, by fitting to the shape predicted
by (\ref{dN_dll}). Note that $\alpha(\varphi_b,\varphi_a)$
depends only on the chirality parameters $\varphi_b$ and $\varphi_a$ 
entering the corresponding vertices for the near ($\ell_n$) and
far ($\ell_f$) leptons. The fact that $\alpha$ (and as a consequence, 
the dilepton invariant mass shape (\ref{dN_dll})) 
does not depend on the chirality parameter
$\tilde\varphi_c$ associated with the quark vertex, 
should be intuitively obvious -- the two leptons 
are not affected by the preceding events higher up in the 
cascade decay chain (see Fig.~\ref{fig:ABCD}). The resulting measurement of $\alpha$ 
can be immediately interpreted in terms of the underlying
chirality parameters $\varphi_a$ and $\varphi_b$, as
illustrated in Fig.~\ref{fig:alpha}, leading to
one constraint among $\varphi_a$ and $\varphi_b$.
Clearly, the $\alpha(\varphi_b,\varphi_a)$ measurement alone is not
sufficient to pin down the precise values of $\varphi_a$ and $\varphi_b$.
However, once it is supplemented with the additional measurements of
$\beta(\tilde\varphi_c,\varphi_b)$ and $\gamma(\varphi_a,\tilde\varphi_c)$ 
as explained below, in principle all three parameters 
$\varphi_a$, $\varphi_b$ and $\tilde\varphi_c$ will be 
completely determined.
\FIGURE[t]{
\epsfig{file=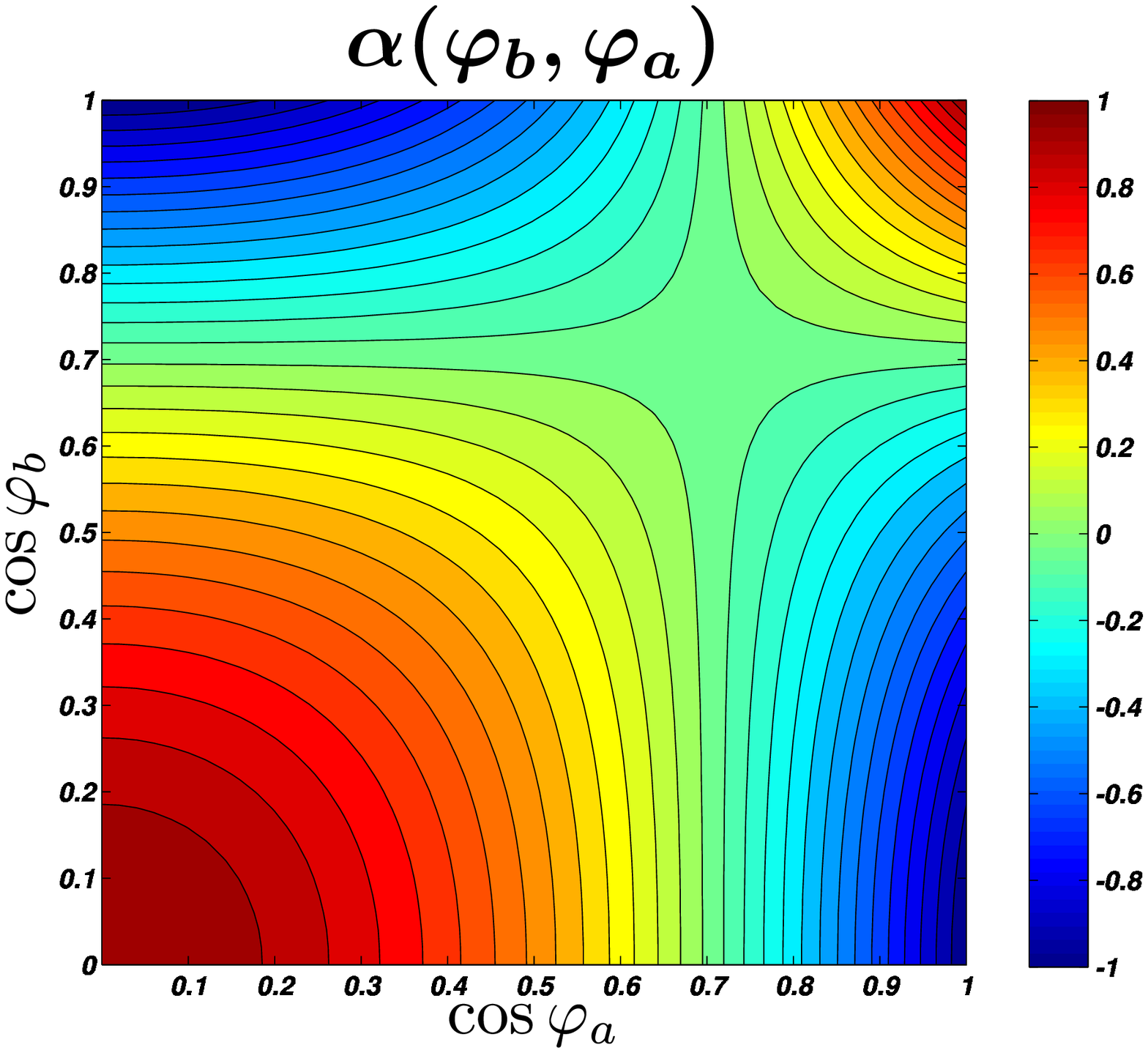,width=9cm}
\caption{\sl The parameter $\alpha(\varphi_b,\varphi_a)$ defined in (\ref{def_alpha}) 
as a function of $\cos\varphi_a$ and $\cos\varphi_b$.}
\label{fig:alpha}}

Similarly, we can see that the jet-near lepton invariant mass 
distribution (\ref{dN_djln}) only depends on the parameter $\beta$, and 
does not contain the parameters $\alpha$ or $\gamma$. Again notice from Fig.~\ref{fig:ABCD}
that $\beta( \tilde\varphi_c,\varphi_b)$ in turn depends only on the chirality parameters 
$\tilde\varphi_c$ and $\varphi_b$ associated with the corresponding vertices
for the quark ($q$) and the near lepton ($\ell_n$). This is also intuitively 
clear -- the jet and near lepton should not be affected by what happens 
later down the decay chain. 
A measurement of $\beta$ therefore can be immediately interpreted 
in terms of the underlying chirality parameters $\tilde\varphi_c$ 
and $\varphi_b$, and the relationship is exactly the same as the
one exhibited in Fig.~\ref{fig:alpha} 
between $\alpha(\varphi_b,\varphi_a)$ and its arguments.

However, as we already explained in Sec.~\ref{sec:FIJbasis},
the $\{j\ell_n\}$ invariant mass distribution (\ref{dN_djln})
is not separately observable, and instead has to be combined with the
$\{j\ell_f\}$ distribution given in (\ref{dN_djlf}) to form the
experimentally observable $\{j\ell^+\}$ and $\{j\ell^-\}$ distributions.
We see from eq.~(\ref{dN_djlf}) that the $\{j\ell_f\}$ distribution
depends on all three parameters $\alpha$, $\beta$ and $\gamma$,
which is again easy to understand intuitively -- the intermediate
lepton $\ell_n$ does affect its neighbors on both sides ($q$ and $\ell_f$).
Given the expressions (\ref{dN_djln}) and (\ref{dN_djlf}), we can immediately 
combine them using the same procedure as in eqs.~(\ref{jl+}) and (\ref{jl-}):
\begin{eqnarray}
\left( \frac{dN}{d\hat{m}^2_{j\ell^\pm}}\right)_S 
&=&
\frac{1}{2} \biggl\{
 {r_n^2}\, {\mathcal F}_{S;\delta}^{(j\ell_n)}(r_n^2 \hat{m}_{j\ell^\pm}^2;x,y,z)
+{r_f^2}\, {\mathcal F}_{S;\delta}^{(j\ell_f)}(r_f^2 \hat{m}_{j\ell^\pm}^2;x,y,z) \nonumber \\
&& \quad +\, \alpha\, {r_f^2}\, {\mathcal F}_{S;\alpha}^{(j\ell_f)}(r_f^2 \hat{m}_{j\ell^\pm}^2;x,y,z)
\pm \gamma\, {r_f^2}\, {\mathcal F}_{S;\gamma}^{(j\ell_f)}(r_f^2 \hat{m}_{j\ell^\pm}^2;x,y,z) \nonumber \\
&& \quad \pm\, \beta\, {r_f^2}\, {\mathcal F}_{S;\beta }^{(j\ell_f)}(r_f^2 \hat{m}_{j\ell^\pm}^2;x,y,z)
\mp \beta\,  {r_n^2}\, {\mathcal F}_{S;\beta }^{(j\ell_n)}(r_n^2 \hat{m}_{j\ell^\pm}^2;x,y,z) \biggr\} \ .
\label{dN_djl} 
\end{eqnarray}
Notice that the same $\beta$ and $\gamma$ terms in (\ref{dN_djl}) appear with 
opposite signs in the $\{j\ell^+\}$ and the $\{j\ell^-\}$ distribution.
This suggests that instead of the two individual distributions
(\ref{dN_djl}) we should be considering their sum 
\begin{eqnarray}
&&S_S^{+-}(\hat{m}_{j\ell}^2;x,y,z,\alpha) \equiv \left( \frac{dN}{d\hat{m}^2_{j\ell^+}}\right)_S
                      + \left( \frac{dN}{d\hat{m}^2_{j\ell^-}}\right)_S  \label{defS+-} \\[2mm]
&=&
    {r_n^2}\, {\mathcal F}_{S;\delta}^{(j\ell_n)}(r_n^2 \hat{m}_{j\ell}^2;x,y,z)
+\, {r_f^2}\, {\mathcal F}_{S;\delta}^{(j\ell_f)}(r_f^2 \hat{m}_{j\ell}^2;x,y,z)
+ \alpha\, {r_f^2}\, {\mathcal F}_{S;\alpha}^{(j\ell_f)}(r_f^2 \hat{m}_{j\ell}^2;x,y,z)
\nonumber
\end{eqnarray}
and their difference
\begin{eqnarray}
&& D_S^{+-}(\hat{m}_{j\ell}^2;x,y,z,\beta,\gamma) \equiv \left( \frac{dN}{d\hat{m}^2_{j\ell^+}}\right)_S
                      - \left( \frac{dN}{d\hat{m}^2_{j\ell^-}}\right)_S  \label{defD+-} \\[2mm]
&=&
 \gamma\, {r_f^2}\, {\mathcal F}_{S;\gamma}^{(j\ell_f)}(r_f^2 \hat{m}_{j\ell}^2;x,y,z)
+\beta \, {r_f^2}\, {\mathcal F}_{S;\beta }^{(j\ell_f)}(r_f^2 \hat{m}_{j\ell}^2;x,y,z)
-\beta \, {r_n^2}\, {\mathcal F}_{S;\beta }^{(j\ell_n)}(r_n^2 \hat{m}_{j\ell}^2;x,y,z)\ .
\nonumber
\end{eqnarray}
The normalisation conditions for the newly defined quantities
$S_S^{+-}$ and $D_S^{+-}$ are
\begin{eqnarray}
\int_0^\infty S_S^{+-}(\hat{m}_{j\ell}^2;x,y,z,\alpha)\, d \hat{m}^2_{j\ell} &=& 2 \ ,
\label{norm_S+-}  \\ [2mm]
\int_0^\infty D_S^{+-}(\hat{m}_{j\ell}^2;x,y,z,\beta,\gamma)\, d \hat{m}^2_{j\ell} &=& 0 \ .
\label{norm_D+-}  
\end{eqnarray}
Eq.~(\ref{defS+-}) reveals one of our most important results -- that the
sum of the two jet-lepton distributions depends on a single model-dependent parameter, 
and more importantly, this is the same parameter ($\alpha$) which also 
determines the dilepton invariant mass distribution. Therefore, once
$\alpha$ is measured from the relatively clean dilepton data, the 
experimentally observable $S_S^{+-}$ distribution is completely specified!
This is a very important result, and as we shall see later in 
our examples, the dilepton ($L^{+-}$) and $S_S^{+-}$ distributions by themselves
can often discriminate among the various spin alternatives.

Of course, the $D_S^{+-}$ distribution is also observable, and 
it can be used as an additional cross-check of the results obtained 
with the two $\alpha$-dependent distributions. 
The importance of the $D_S^{+-}$ distribution is that it can provide
a measurement of the other two model-dependent parameters $\beta$ and $\gamma$.
Note, however, that the $\gamma$ parameter can be measured only if
$S=4,5,6$, since for the remaining three cases we have
$${\mathcal F}_{S;\gamma}^{(j\ell_n)}={\mathcal F}_{S;\gamma}^{(j\ell_f)}=0
\qquad {\rm for\ }S=1,2,3,
$$
and $D_S^{+-}$ becomes $\gamma$-independent.
Similarly, the parameter $\beta$ can only be determined for
$S=1,4,5,6$ since for the remaining two cases $S=2,3$ 
$${\mathcal F}_{S;\beta}^{(j\ell_n)}={\mathcal F}_{S;\beta}^{(j\ell_f)}=0
\qquad {\rm for\ }S=2,3,
$$
and $D_S^{+-}$ becomes $\beta$-independent as well.

Now we are in a position to contrast our approach to 
previous spin discrimination studies based on the
lepton charge asymmetry \cite{Barr:2004ze}.
The latter is simply the ratio
\begin{equation}
A_S^{+-}(\hat{m}_{j\ell}^2;x,y,z,\alpha,\beta,\gamma) \equiv 
\frac{D_S^{+-}(\hat{m}_{j\ell}^2;x,y,z,\beta,\gamma)}{S_S^{+-}(\hat{m}_{j\ell}^2;x,y,z,\alpha)}\ .
\label{A+-}
\end{equation}
We can immediately see that, in general, $A_S^{+-}$
is a much more model-dependent quantity than either $S_S^{+-}$ or $D_S^{+-}$.
Indeed, as we just discussed, $S_S^{+-}$ depends on a single 
model-dependent parameter ($\alpha$),
$D_S^{+-}$ depends on two other model-dependent parameters ($\beta$ and $\gamma$), 
while, as evidenced by eq.~(\ref{A+-}),
$A_S^{+-}$ depends on all three of these ($\alpha$, $\beta$ {\em and} $\gamma$).
Second, the lepton charge asymmetry is not normalised to any particular 
constant numerical value, unlike the $S_S^{+-}$ and $D_S^{+-}$ distributions 
(see eqs.~(\ref{norm_S+-},\ref{norm_D+-})). But most importantly, 
$A_S^{+-}$ is a single distribution, derived from $S_S^{+-}$ and $D_S^{+-}$, 
therefore it is bound to contain less information
than the two separate distributions $S_S^{+-}$ and $D_S^{+-}$. 
Our explicit examples in Sec.~\ref{sec:numerics} will show that,
as might be expected, the useful information contained 
in $A_S^{+-}$ is approximately the same as the information contained in
$D_S^{+-}$. Therefore, by considering in addition the $S_S^{+-}$ distribution, 
as we are suggesting here, one is recovering the information which was lost 
when forming the ratio (\ref{A+-}). This information gain is most striking 
for the case of a $p\bar{p}$ collider like the Tevatron, as discussed 
in detail below in Sec.~\ref{sec:Tevatron}.

\section{The method}
\label{sec:method}

The starting point in our analysis is the set of analytical formulas 
(\ref{dN_dll}, \ref{defS+-}, \ref{defD+-}) derived in the previous section
for the three experimentally observable invariant mass distributions: dilepton $L_S^{+-}$,
and sum ($S_S^{+-}$) and difference ($D_S^{+-}$) of the $\{j\ell^+\}$ and the $\{j\ell^-\}$
distributions:
\begin{eqnarray}
&& L_S^{+-}(\hat{m}_{\ell\ell}^2;x,y,z,\alpha) \equiv \left( \frac{dN}{d\hat{m}^2_{\ell\ell}}\right)_S =
{\mathcal F}_{S;\delta}^{(\ell\ell)}(\hat{m}^2_{\ell\ell};x,y,z)
+ \alpha\, {\mathcal F}_{S;\alpha}^{(\ell\ell)}(\hat{m}^2_{\ell\ell};x,y,z) \ ,
\label{L+-}  \\[2mm]
&&S_S^{+-}(\hat{m}_{j\ell}^2;x,y,z,\alpha)
\equiv \left( \frac{dN}{d\hat{m}^2_{j\ell^+}}\right)_S
                      + \left( \frac{dN}{d\hat{m}^2_{j\ell^-}}\right)_S  \label{S+-} \\[2mm]
&=&
    {r_n^2}\, {\mathcal F}_{S;\delta}^{(j\ell_n)}(r_n^2 \hat{m}_{j\ell}^2;x,y,z)
+\, {r_f^2}\, {\mathcal F}_{S;\delta}^{(j\ell_f)}(r_f^2 \hat{m}_{j\ell}^2;x,y,z)
+ \alpha\, {r_f^2}\, {\mathcal F}_{S;\alpha}^{(j\ell_f)}(r_f^2 \hat{m}_{j\ell}^2;x,y,z)\ ,
\nonumber   \\ [2mm]
&& D_S^{+-}(\hat{m}_{j\ell}^2;x,y,z,\beta,\gamma) \equiv \left( \frac{dN}{d\hat{m}^2_{j\ell^+}}\right)_S
                  - \left( \frac{dN}{d\hat{m}^2_{j\ell^-}}\right)_S  \label{D+-} \\[2mm]
&=&
 \gamma\, {r_f^2}\, {\mathcal F}_{S;\gamma}^{(j\ell_f)}(r_f^2 \hat{m}_{j\ell}^2;x,y,z)
+\beta \, {r_f^2}\, {\mathcal F}_{S;\beta }^{(j\ell_f)}(r_f^2 \hat{m}_{j\ell}^2;x,y,z)
-\beta \, {r_n^2}\, {\mathcal F}_{S;\beta }^{(j\ell_n)}(r_n^2 \hat{m}_{j\ell}^2;x,y,z)\ .
\nonumber
\end{eqnarray}
The functions ${\mathcal F}_{S;\alpha}^{(p)}$, ${\mathcal F}_{S;\beta }^{(p)}$,
${\mathcal F}_{S;\gamma}^{(p)}$ and
${\mathcal F}_{S;\delta}^{(p)}$ are given in Appendix~\ref{app:Falpha},
while the constant model-dependent parameters
$\alpha$, $\beta$ and $\gamma$ were defined in 
eqs.~(\ref{def_alpha}-\ref{def_gamma}):
\begin{eqnarray}
\alpha(\varphi_b,\varphi_a) &=& \cos2\varphi_b\cos2\varphi_a\ ,
\label{def_alpha2} \\
\beta( \tilde\varphi_c,\varphi_b) &=& 
\cos2\tilde\varphi_c\cos2\varphi_b = (f-\bar{f}) \cos2\varphi_c \cos2\varphi_b \ ,
\label{def_beta2} \\
\gamma(\varphi_a,\tilde\varphi_c) &=& 
\cos2\varphi_a\cos2\tilde\varphi_c = (f-\bar{f}) \cos2\varphi_a \cos2\varphi_c\ ,
\label{def_gamma2} 
\end{eqnarray}
where in the last two equations we have used the relation (\ref{cos2ctilde}).
The angles $\varphi_a$, $\varphi_b$ and $\varphi_c$ were defined in eq.~(\ref{phi_angles})
and parameterise the relative chirality of the corresponding 
interaction vertex in Fig.~\ref{fig:ABCD}, while the particle-antiparticle
fractions $f$ and $\bar{f}$ were introduced in Sec.~\ref{sec:th}
and satisfy eq.~(\ref{unit_f}). 
Given the data for the three distributions (\ref{L+-}-\ref{D+-}), one then tries to
fit for the unknown model-dependent coefficients $\alpha$, $\beta$ and $\gamma$,
considering each of the six different spin possibilities $S$ one at a time.
The result will be 6 different sets of ``best fit'' values for these coefficients,
$\{\alpha_S, \beta_S, \gamma_S\}, S=\{1,...,6\}$, {\em and} an accompanying measure 
for the goodness of fit in each case. 
The fits can be done simultaneously for all three parameters, or alternatively, 
one can first determine $\alpha$ from the relatively cleaner $L_S^{+-}$
sample, and subsequently use this fitted value of $\alpha_S$ 
in eqs.~(\ref{S+-},\ref{D+-}). The goodness of fit for each $S$ will indicate 
whether this particular spin configuration is consistent with the data or not, 
and, given the expected experimental statistical and systematic errors,
one can also readily assign confidence level probabilities to those statements.
As we have been emphasizing throughout, this procedure is completely model-independent,
and in fact produces {\em an independent} measurement of the model-dependent parameters
$\alpha$, $\beta$ and $\gamma$, which can then be translated into
a measurement of the underlying theoretical model parameters 
$\varphi_a$, $\varphi_b$, $\varphi_c$ and $f$. For example,
when all three parameters $\alpha$, $\beta$ and $\gamma$ are measured and 
found to be non-zero, one can invert eqs.~(\ref{def_alpha2}-\ref{def_gamma2})
and solve for $\varphi_a$, $\varphi_b$ and $\varphi_c$
up to a two-fold ambiguity:
\begin{eqnarray}
\cos 2\varphi_a &=& \pm \frac{1}{\beta }\, \sqrt{ \alpha\beta\gamma }\, , \label{phia_sol} \\
\cos 2\varphi_b &=& \pm \frac{1}{\gamma}\, \sqrt{ \alpha\beta\gamma }\, , \label{phib_sol} \\
\cos 2\varphi_c &=& \pm \frac{1}{f-\bar{f}}\,\frac{1}{\alpha} \, \sqrt{\alpha\beta\gamma}\, , \label{phic_sol} 
\end{eqnarray}
where in all three equations one should take either the ``$+$''
or the ``$-$'' sign on the right-hand side.
The origin of this two-fold ambiguity is easy to understand. 
Observe that the defining equations (\ref{def_alpha2}-\ref{def_gamma2})
for $\alpha$, $\beta$ and $\gamma$ are invariant under
the simultaneous transformations
\begin{equation}
\varphi_a \to \frac{\pi}{2} - \varphi_a\, ,   \quad
\varphi_b \to \frac{\pi}{2} - \varphi_b\, ,   \quad
\varphi_c \to \frac{\pi}{2} - \varphi_c\, ,  \label{trans_abc}
\end{equation}
whose effect is precisely to flip the signs in the right-hand sides of 
eqs.~(\ref{phia_sol}-\ref{phic_sol}). Given the defining relation
(\ref{phi_angles}), the transformations (\ref{trans_abc}) 
are equivalent to the chirality exchange
\begin{equation}
|a_L| \leftrightarrow |a_R|\, , \quad
|b_L| \leftrightarrow |b_R|\, , \quad
|c_L| \leftrightarrow |c_R|\, . \label{LtoR}
\end{equation}
The physical meaning of eq.~(\ref{LtoR}) is clear -- we can only measure the 
chirality of the three different vertices in Fig.~\ref{fig:ABCD}
only {\em relative} to each other.
When choosing the plus signs in eqs.~(\ref{phia_sol}-\ref{phic_sol}),
we get a solution for the couplings with one particular set of chiralities, 
while choosing the minus sign in eqs.~(\ref{phia_sol}-\ref{phic_sol})
yields a solution where the couplings have just the opposite chiralities.
Since there is nothing to provide a reference point for the chiralities, 
it is impossible to remove this $L\leftrightarrow R$ ambiguity 
without making some model assumptions, 
or without considering additional independent measurements. Using the solutions
(\ref{phia_sol}-\ref{phic_sol}) and the definitions (\ref{phi_angles})
we can write down the general solution for the couplings 
in terms of the measured parameters $\alpha$, $\beta$ and $\gamma$, as
\begin{eqnarray}
|a_L| &=& \frac{1}{\sqrt{2}} \left( 1 \pm \frac{1}{\beta }\, \sqrt{ \alpha\beta\gamma } \right)^{\frac{1}{2}}, \label{aL}\\
|a_R| &=& \frac{1}{\sqrt{2}} \left( 1 \mp \frac{1}{\beta }\, \sqrt{ \alpha\beta\gamma } \right)^{\frac{1}{2}}, \label{aR}\\
|b_L| &=& \frac{1}{\sqrt{2}} \left( 1 \pm \frac{1}{\gamma}\, \sqrt{ \alpha\beta\gamma } \right)^{\frac{1}{2}}, \label{bL}\\
|b_R| &=& \frac{1}{\sqrt{2}} \left( 1 \mp \frac{1}{\gamma}\, \sqrt{ \alpha\beta\gamma } \right)^{\frac{1}{2}}, \label{bR}\\
|c_L| &=& \frac{1}{\sqrt{2}} \left( 1 \pm \frac{1}{f-\bar{f}}\, \frac{1}{\alpha}\, \sqrt{ \alpha\beta\gamma } \right)^{\frac{1}{2}}, \label{cL}\\
|c_R| &=& \frac{1}{\sqrt{2}} \left( 1 \mp \frac{1}{f-\bar{f}}\, \frac{1}{\alpha}\, \sqrt{ \alpha\beta\gamma } \right)^{\frac{1}{2}}, \label{cR}
\end{eqnarray}
where the appearance of the $\pm$ sign is due to the two-fold ambiguity 
just discussed. Here the two solutions are obtained by choosing the upper 
or lower sign in each equation, correspondingly. It is worth making a few 
comments regarding eqs.~(\ref{aL}-\ref{cR}), which represent our second main result.

Note that while in general $\alpha$, $\beta$ and $\gamma$ can have either sign, 
eqs.~(\ref{def_alpha}-\ref{def_gamma}) imply that the product $\alpha\beta\gamma$
is always non-negative. Furthermore, from eqs.~(\ref{def_alpha}-\ref{def_gamma})
it also follows that $|\alpha\beta | \le |\gamma|$,
$|\beta \gamma| \le |\alpha|$ and $|\gamma\alpha| \le |\beta |$.
Therefore all square roots in eqs.~(\ref{aL}-\ref{cR}) are well behaved and
never yield any imaginary solutions. It is interesting to note the 
dependence on the particle-antiparticle fraction $f$ discussed in
Sec.~\ref{sec:th}. We see that for any given measurement of 
$\alpha$, $\beta$ and $\gamma$, the effective couplings 
$|a_L|$, $|a_R|$, $|b_L|$ and $|b_R|$ associated with the 
particle A and particle B vertices of Fig.~\ref{fig:ABCD} can be 
uniquely determined, up to the two-fold $L\leftrightarrow R$ ambiguity (\ref{LtoR}).
In other words, the particle-antiparticle ambiguity {\bf T2} 
discussed in the Introduction only affects the determination of the
$|c_L|$ and $|c_R|$ couplings, as seen from eqs.~(\ref{cL}-\ref{cR}).
The values of the couplings $|c_L|$ and $|c_R|$ are not 
uniquely determined, and instead are parameterised as a function of $f$.
Although we do not know the exact value of $f$, consistency of 
eqs.~(\ref{cL}-\ref{cR}) restricts the allowed values of $f$ to be in the range
\begin{equation}
0\le f \le \frac{1}{2}\left( 1-\sqrt{\frac{\beta\gamma}{\alpha}} \right) \quad {\rm or} \quad
\frac{1}{2}\left( 1+\sqrt{\frac{\beta\gamma}{\alpha}} \right)  \le f \le 1\ .
\label{frange}
\end{equation}
The fact that the allowed range for $f$ splits into two 
separate intervals could already be seen in Fig.~\ref{fig:ctilde}:
notice that there are two disjoint branches in the $(\cos\varphi_c,f)$ plane
which are consistent with a given fixed value of $\tilde\varphi_c$,
i.e. with a given set of measured $\alpha$, $\beta$ and $\gamma$.
At a $pp$ collider like the LHC, in general we expect $f>\frac{1}{2}$,
so we should select the higher $f$ range in eq.~(\ref{frange}), while 
the lower $f$ range in eq.~(\ref{frange}) would be relevant for
a hypothetical $\bar{p}\bar{p}$ collider (``anti-LHC''):
\begin{eqnarray}
{\rm      LHC}\,(pp)             &:&\quad \frac{1}{2}\left( 1+\sqrt{\frac{\beta\gamma}{\alpha}} \right)  \le f \le 1\ , \label{frangeLHC}\\
{\rm anti-LHC}\,(\bar{p}\bar{p})&:&\quad 0\le f \le \frac{1}{2}\left( 1-\sqrt{\frac{\beta\gamma}{\alpha}} \right)\ . \label{frangeantiLHC}
\end{eqnarray}
While eq.~(\ref{frangeLHC}) is not a real measurement of the value of $f$
at the LHC, it nevertheless contains very important information.
For example, if the measured values of $\alpha$, $\beta$ and $\gamma$ 
happen to be such that $|\beta\gamma| \approx |\alpha|$, then $f$ becomes
very severely constrained, and the restriction (\ref{frangeLHC}) 
by itself is sufficient to yield a measurement of the value of $f$: $f\approx 1$.

In the following Section \ref{sec:numerics} we shall give 
numerous examples of how our method might work in practice.
But before we conclude this section we shall anticipate some 
general results which can be gleaned from our analytical formulas
(\ref{L+-}-\ref{D+-}). In particular, in Sec.~\ref{sec:degenerate} 
we shall show that the two pairs of spin configurations
FSFS and FSFV, as well as FVFS and FVFV, very often give identical results
for the invariant mass distributions, and cannot be differentiated without 
additional model assumptions. Then in Sec.~\ref{sec:Tevatron}
we shall show that our method is also applicable at the Tevatron, 
where in contrast the lepton charge asymmetry $A_S^{+-}$ is identically zero 
for all spin configurations $S$ and thus contains no useful information.

\subsection{The twin spin scenarios FSFS/FSFV and FVFS/FVFV}
\label{sec:degenerate}

Consulting the definitions of the functions in Appendix~\ref{app:Falpha},
one can see that 
\begin{eqnarray}
{\mathcal F}_{3;\alpha}^{(p)} &=& {\mathcal F}_{2;\alpha}^{(p)}\ \frac{1-2z}{1+2z}\ , \\[2mm]
{\mathcal F}_{3;\beta }^{(p)} &=& {\mathcal F}_{2;\beta }^{(p)} = 0 \ , \\[2mm]
{\mathcal F}_{3;\gamma}^{(p)} &=& {\mathcal F}_{2;\gamma}^{(p)} = 0 \ , \\[2mm]
{\mathcal F}_{3;\delta}^{(p)} &=& {\mathcal F}_{2;\delta}^{(p)} 
\end{eqnarray}
for any $p\in \{\ell\ell, j\ell_n, j\ell_f\}$. Therefore the relation
\begin{equation}
\alpha_2 = \alpha_3 \ \frac{1-2z}{1+2z}
\label{alpha2_dup}
\end{equation}
is sufficient to guarantee that {\em all} invariant mass distributions 
(\ref{L+-}-\ref{D+-}) are exactly the same in the case of $S=2$ (FSFS) and $S=3$ (FSFV):
\begin{eqnarray}
    L_2^{+-}\left(\hat{m}_{\ell\ell}^2;x,y,z,\alpha_3\frac{1-2z}{1+2z}\right) 
&=& L_3^{+-}\left(\hat{m}_{\ell\ell}^2;x,y,z,\alpha_3                 \right)\ ,\label{L23dup} \\[2mm]
    S_2^{+-}\left(\hat{m}_{j\ell}^2;x,y,z,\alpha_3\frac{1-2z}{1+2z}\right) 
&=& S_3^{+-}\left(\hat{m}_{j\ell}^2;x,y,z,\alpha_3                 \right)\ , \label{S23dup}\\[2mm]
    D_2^{+-}\left(\hat{m}_{j\ell}^2;x,y,z,\beta_2,\gamma_2\right) 
&=& D_3^{+-}\left(\hat{m}_{j\ell}^2;x,y,z,\beta_3,\gamma_3\right)\ . \label{D23dup}
\end{eqnarray}
Note that this exact duplication occurs {\em irrespective of} the values of the other 
two model-dependent parameters $\beta$ and $\gamma$. In other words, 
relations (\ref{L23dup}-\ref{D23dup}) hold identically for {\em any} values of 
the five parameters $\alpha_3$, $\beta_3$, $\gamma_3$, $\beta_2$ and $\gamma_2$.
As long as eq.~(\ref{alpha2_dup}) is true, the FSFS and FSFV models will  
yield identical invariant mass distributions for $L^{+-}$, $S^{+-}$ and $D^{+-}$.
This observation has very important implications for the eventual outcome of the 
spin measurement, if the data happens to come from one of those models, since
the exact duplication (\ref{L23dup}-\ref{D23dup}) then threatens to jeopardize 
our ability to discriminate among them. 
However, as we shall now see, whether discrimination is possible or not, 
depends on the actual values of $\alpha$ and $z$. 
Recall that the $\alpha$ parameter is defined in the range $[-1,1]$, 
while $z$ is defined in $(0,1)$, and therefore so is the ratio $|\frac{1-2z}{1+2z}|$.
Then, for any given value of $\alpha_3\in [-1,1]$, 
$\alpha_2$ as given by (\ref{alpha2_dup}) falls into its definition window, 
and an exact duplication takes place. However, the reverse is not true:
not every value of $\alpha_2$ would lead to a valid solution for $\alpha_3$
according to eq.~(\ref{alpha2_dup}), since for large enough values
of $|\alpha_2|$, the value of $|\alpha_3|$ would exceed 1, which is not allowed.

Our conclusion therefore is that the issue of confusing the two models
FSFS and FSFV depends on whether the data comes from FSFV and 
we are trying to fit it with FSFS, or whether the data comes from 
FSFS and we are trying to fit it with FSFV. In the former case
the two models will always be confused with each other, while
in the latter case, the confusion arises only if $\alpha_2$ happens to satisfy
\begin{equation}
\left|\alpha_2\right| \le \left| \frac{1-2z}{1+2z}\right| \ .
\label{alpha2_cond}
\end{equation}

A close inspection of Appendix~\ref{app:Falpha} also reveals a similar problem 
with the FVFS and FVFV spin configurations ($S=4$ and $S=5$). In this case, 
we notice the following relations
\begin{eqnarray}
{\mathcal F}_{5;\alpha}^{(p)} &=& {\mathcal F}_{4;\alpha}^{(p)}\ \frac{1-2z}{1+2z}\ , \\[2mm]
{\mathcal F}_{5;\beta }^{(p)} &=& {\mathcal F}_{4;\beta }^{(p)} \ , \\[2mm]
{\mathcal F}_{5;\gamma}^{(p)} &=& {\mathcal F}_{4;\gamma}^{(p)}\ \frac{1-2z}{1+2z}\ , \\[2mm]
{\mathcal F}_{5;\delta}^{(p)} &=& {\mathcal F}_{4;\delta}^{(p)} 
\end{eqnarray}
for any $p\in \{\ell\ell, j\ell_n, j\ell_f\}$. Therefore, the relations
\begin{eqnarray}
\alpha_4 &=& \alpha_5 \ \frac{1-2z}{1+2z} \ , \label{alpha4_dup} \\[2mm]
\beta_4  &=& \beta_5                      \ , \label{beta4_dup}  \\[2mm]
\gamma_4 &=& \gamma_5 \ \frac{1-2z}{1+2z}     \label{gamma4_dup}
\end{eqnarray}
would once again guarantee that {\em all} invariant mass distributions 
(\ref{L+-}-\ref{D+-}) are exactly the same in these two cases:
\begin{eqnarray}
    L_4^{+-}\left(\hat{m}_{\ell\ell}^2;x,y,z,\alpha_5\frac{1-2z}{1+2z}\right) 
&=& L_5^{+-}\left(\hat{m}_{\ell\ell}^2;x,y,z,\alpha_5                 \right)\ ,\label{L45dup} \\[2mm]
    S_4^{+-}\left(\hat{m}_{j\ell}^2;x,y,z,\alpha_5\frac{1-2z}{1+2z}\right) 
&=& S_5^{+-}\left(\hat{m}_{j\ell}^2;x,y,z,\alpha_5                 \right)\ , \label{S45dup}\\[2mm]
    D_4^{+-}\left(\hat{m}_{j\ell}^2;x,y,z,\beta_5,\gamma_5\frac{1-2z}{1+2z}\right) 
&=& D_5^{+-}\left(\hat{m}_{j\ell}^2;x,y,z,\beta_5,\gamma_5\right)\ . \label{D45dup}
\end{eqnarray}
Following the same logic as before, we conclude that 
whenever the data comes from FVFV, the model will always be
confused with FVFS. However, if the data comes from 
FVFS, the confusion arises only if $\alpha_4$ and $\gamma_4$ happen to satisfy
\begin{eqnarray}
\left| \alpha_4 \right| &\le& \left| \frac{1-2z}{1+2z} \right| \ , \label{alpha4_cond} \\[2mm]
\left| \gamma_4 \right| &\le& \left| \frac{1-2z}{1+2z} \right| \ . \label{gamma4_cond}
\end{eqnarray}
In addition to these two equations, the values of $\alpha_4$, $\beta_4$
and $\gamma_4$ must also satisfy the domain constraints 
(\ref{eqn:constraints1}-\ref{eqn:constraints4})
from Appendix~\ref{app:params}.

\subsection{Spin determination at the Tevatron}
\label{sec:Tevatron}

At a $p\bar{p}$ collider such as the Tevatron, the symmetry of the initial state implies
\begin{equation}
f = \bar{f} = \frac{1}{2}\ .
\label{fTevatron}
\end{equation}
On the surface, it may appear that this constraint eliminates only one out of the four
model-dependent degrees of freedom ($f$, $\varphi_a$, $\varphi_b$ and $\varphi_c$)
that we originally started with. However, as can be deduced from eqs.~(\ref{ct_def},\ref{st_def})
and also seen from Fig.~\ref{fig:ctilde}, the constraint (\ref{fTevatron})
in fact completely fixes the $\tilde\varphi_c$ parameter
\begin{equation}
\tilde\varphi_c = \frac{\pi}{4}
\end{equation}
and as a result both $\beta$ and $\gamma$ vanish identically:
\begin{equation}
\beta = \gamma = 0\ .
\end{equation}
In that case from eq.~(\ref{D+-}) we have
\begin{equation}
D_S^{+-} \equiv 0
\end{equation}
and a similar result holds for the lepton charge asymmetry (\ref{A+-}) 
\begin{equation}
A_S^{+-} \equiv 0\ .
\end{equation}
We see that at the Tevatron we do not learn anything from either $D_S^{+-}$
or from the lepton charge asymmetry $A_S^{+-}$.
However, our results for $L_S^{+-}$ and $S_S^{+-}$ still hold, and 
contain non-trivial spin information, so that the spin analysis 
following our method can still be performed.
In fact, our method can already be tested in the top quark 
semileptonic and dilepton samples at the Tevatron by looking at the
invariant mass distribution of the $b$-jet and the lepton \cite{KM}.
Indeed, our decay chain from Fig.~\ref{fig:ABCD} can be applied to
top quark decays, for example by identifying $C=t$, $B=W^+$ and $A=\nu_\ell$, 
and reinterpreting $\ell_n$ as the $b$-jet and $\ell_f$ as the
lepton coming from $W$ decay. In that case, the $m_{b\ell}$
distribution should be described by our formula (\ref{dN_dll}) 
for $L^{+-}_6$.
Alternatively, one can identify the particles in Fig.~\ref{fig:ABCD}
as $D=t$, $C=W^+$, $B=\nu_\ell$, 
$q=b$ and $\ell_n=\ell$. In this case, the $m_{b\ell}$ distribution
will be described by our formula (\ref{dN_djln}) applied for $S=4$ or $S=5$. 
In any case, one should observe the characteristic $\hat{m}^4$ term
in the invariant mass distribution (see the definition of 
${\mathcal F}_{6;\delta}^{(\ell\ell)}$ in Table~\ref{table:Fll2}
or the definition of ${\mathcal F}_{4;\delta}^{(j\ell)}$ and 
${\mathcal F}_{5;\delta}^{(j\ell)}$ in Table~\ref{table:Fqln2}), 
which would signal that the $W$ is spin 1 and therefore the top quark 
and the neutrino are both spin 1/2.

\section{Determination of spins and couplings: examples}
\label{sec:numerics}

In this section we shall give an explicit
demonstration how to apply our method in practice at the LHC. 
We shall work out in detail
6 different examples, namely, we shall assume in turn that
the observed data is coming from each one of the six spin configurations
from Table~\ref{table:spins}. Then we shall ask the question whether
this data is consistent with one of the remaining 5 alternatives.

Since we do not yet have real data available, 
we will have to use simulated data. We shall therefore have to pick some
values for the mass spectrum, couplings and particle-antiparticle fraction, 
namely we shall have to fix the values of $x$, $y$, $z$, $\varphi_a$,
$\varphi_b$, $\varphi_c$, and $f$. In order to allow comparisons to 
previous studies in the literature, we shall use the parameters 
of the SPS1a study point in supersymmetry. However, as advertised, 
we shall still perform the spin measurements in a model-independent way, 
i.e. as soon as we simulate our ``data'', we shall immediately ``forget''
how it was generated, and shall treat it as coming from a ``black box''
such as the actual collider experiment. 

For the SPS1a mass spectrum we take the values used in
Refs.~\cite{Smillie:2005ar,Athanasiou:2006ef} 
\begin{equation}
m_A= 96\ {\rm GeV}, \quad
m_B=143\ {\rm GeV}, \quad
m_C=177\ {\rm GeV}, \quad
m_D=537\ {\rm GeV}\ ,
\label{SPSmasses}
\end{equation}
which translate into
\begin{equation}
x=0.109, \quad
y=0.653, \quad
z=0.451\ .
\label{SPSxyz}
\end{equation}
SPS1a is characterised by the following approximate values for the coupling constants
\begin{equation}
a_L = 0,   \quad a_R = 1, \quad
b_L = 0,   \quad b_R = 1, \quad
c_L = 1,   \quad c_R = 0, 
\label{SPSabc}
\end{equation}
and particle-antiparticle fractions $f$ and $\bar{f}$ at the LHC
\begin{equation}
  f = 0.7,   \quad \bar{f} = 0.3 \ .
\label{SPSf}
\end{equation}
The spectrum (\ref{SPSmasses}) results in the following
kinematic endpoints\footnote{The kinematic endpoint 
$m_{j\ell\ell}^{max}$ is only needed for the extraction of the mass spectrum, 
while the actual $\{j\ell^+\ell^-\}$ distribution is not needed for our study.}
\begin{eqnarray}
m_{\ell\ell}^{max}   &=& m_D \sqrt{x(1-y)(1-z)} =  77.31\ {\rm GeV}  \label{mll_max} \, , \\
m_{j\ell_n}^{max}    &=& m_D \sqrt{ (1-x)(1-y)} = 298.77\ {\rm GeV}  \label{mjln_max} \, ,  \\
m_{j\ell_f}^{max}    &=& m_D \sqrt{ (1-x)(1-z)} = 375.76\ {\rm GeV}  \label{mjlf_max} \, , \\
m_{j\ell\ell}^{max} &=& m_D \sqrt{ (1-x)(1-y z)}= 425.94\ {\rm GeV}  \label{mjll_max} \, .
\end{eqnarray}
Since we assume that the spectrum has been measured, the values of these 
endpoints are also known in advance of the spin measurement. 
We are therefore still allowed to write the measured invariant mass 
distributions (\ref{L+-}-\ref{D+-})
in terms of the dimensionless invariant masses (\ref{def_mhat}).

Substituting the SPS1a parameter choice (\ref{SPSabc}) and (\ref{SPSf}) into the definitions
(\ref{def_alpha})-(\ref{def_gamma}) yields the following values of our model-dependent
parameters $\alpha$, $\beta$ and $\gamma$
\begin{equation}
\alpha=1 , \quad \beta=-0.4, \quad \gamma=-0.4 \, .
\label{SPSabg}
\end{equation}
Note that $\alpha=1$ necessarily implies $\beta = \gamma$,
in accordance with eqs.~(\ref{def_alpha})-(\ref{def_gamma}).

Eq.~(\ref{SPSabg}) defines the input values 
of the model-dependent parameters used in our study.
We should reiterate that there is nothing special about 
the SPS1a parameter choice, and we could have used any 
other study point instead. 

Using our method, we shall now perform 6 different exercises
of spin determination. For each exercise, we shall take the 
input ``data'' to be given in turn by one of the six models
from Table~\ref{table:spins}. We shall then try to fit the ``data''
to each of the remaining 5 spin configurations, using our general
analytical expressions (\ref{L+-}-\ref{D+-}) with {\em floating},
a priori unknown, parameters $\alpha$, $\beta$ and $\gamma$.
Although the fit can be done simultaneously for all three parameters
$\alpha$, $\beta$ and $\gamma$, we shall perform it sequentially, 
using the fact that the $L_S^{+-}$ and $S_S^{+-}$ distributions
depend only on $\alpha$ and not on $\beta$ and $\gamma$. 
Therefore, we shall start with the cleaner $L_S^{+-}$ sample
and first determine the value of $\alpha$, which we shall then 
use to compare the thus predicted $S_S^{+-}$ distribution to the
``data''. Quite often, it will be already at this stage that
one could rule out all but the correct spin configuration.
We shall encounter such examples below as well. Sometimes, however,
there may still be several alternatives left, in which case we need 
to also consider the $D^{+-}$ distribution, where we fit for the 
values of the coefficients $\beta$ and $\gamma$. Details of our
fitting procedure and examples of some fits are presented in 
Appendix~\ref{app:params}. Our results are summarised in 
Figs.~\ref{fig:ll}, \ref{fig:jl_sum} and \ref{fig:jl_diff},
which show our results for the $L_S^{+-}$, $S_S^{+-}$ and
$D_S^{+-}$ distributions, correspondingly. In each of
Figs.~\ref{fig:ll}, \ref{fig:jl_sum} and \ref{fig:jl_diff}
the solid (magenta) lines in each panel represent the input 
invariant mass distribution ($L_S^{+-}$, $S_S^{+-}$ or
$D_S^{+-}$, as appropriate) from our simulated ``data'', for each of the 6
spin configurations: a) SFSF; b) FSFS; c) FSFV; d) FVFS; e) FVFV; f) SFVF.
The other (dotted or dashed) lines are our best fits to this data, 
for each of the remaining 5 spin configurations from Table \ref{table:spins}.
The color code is the following. If the trial model 
fits the input data perfectly, we use a dashed (green) line. If
the fit fails to match the input data, we use (color-coded)
dotted lines. The best fit values of $\alpha$, $\beta$ and $\gamma$ 
for each case are also shown, except for those cases (labelled by ``NA'')
where they are left undetermined by the fit.
Dotted lines of the same color imply that they are identical to each other,
yet different from the input ``data''.

\FIGURE[p!]{
\epsfig{file=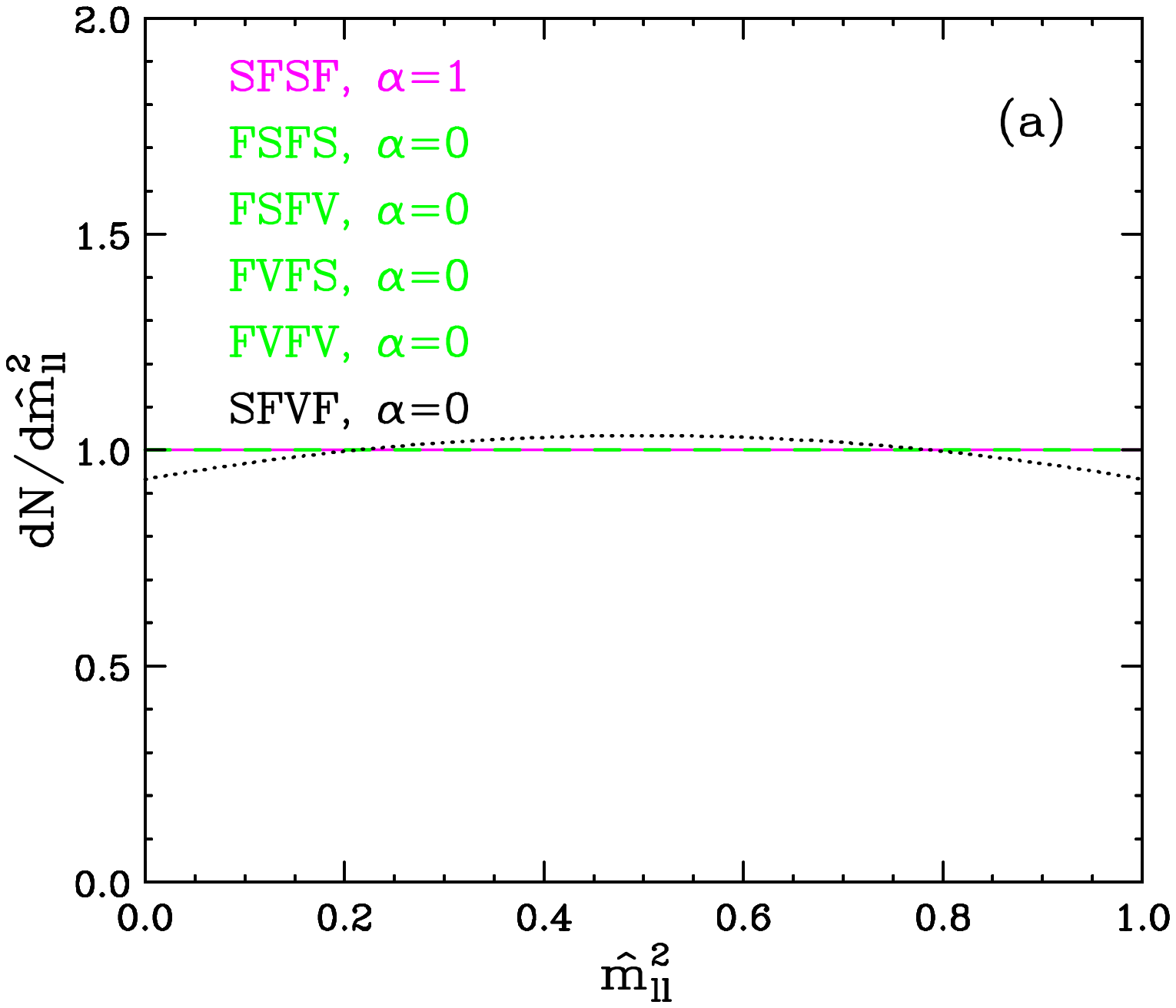,width=6.7cm}
\epsfig{file=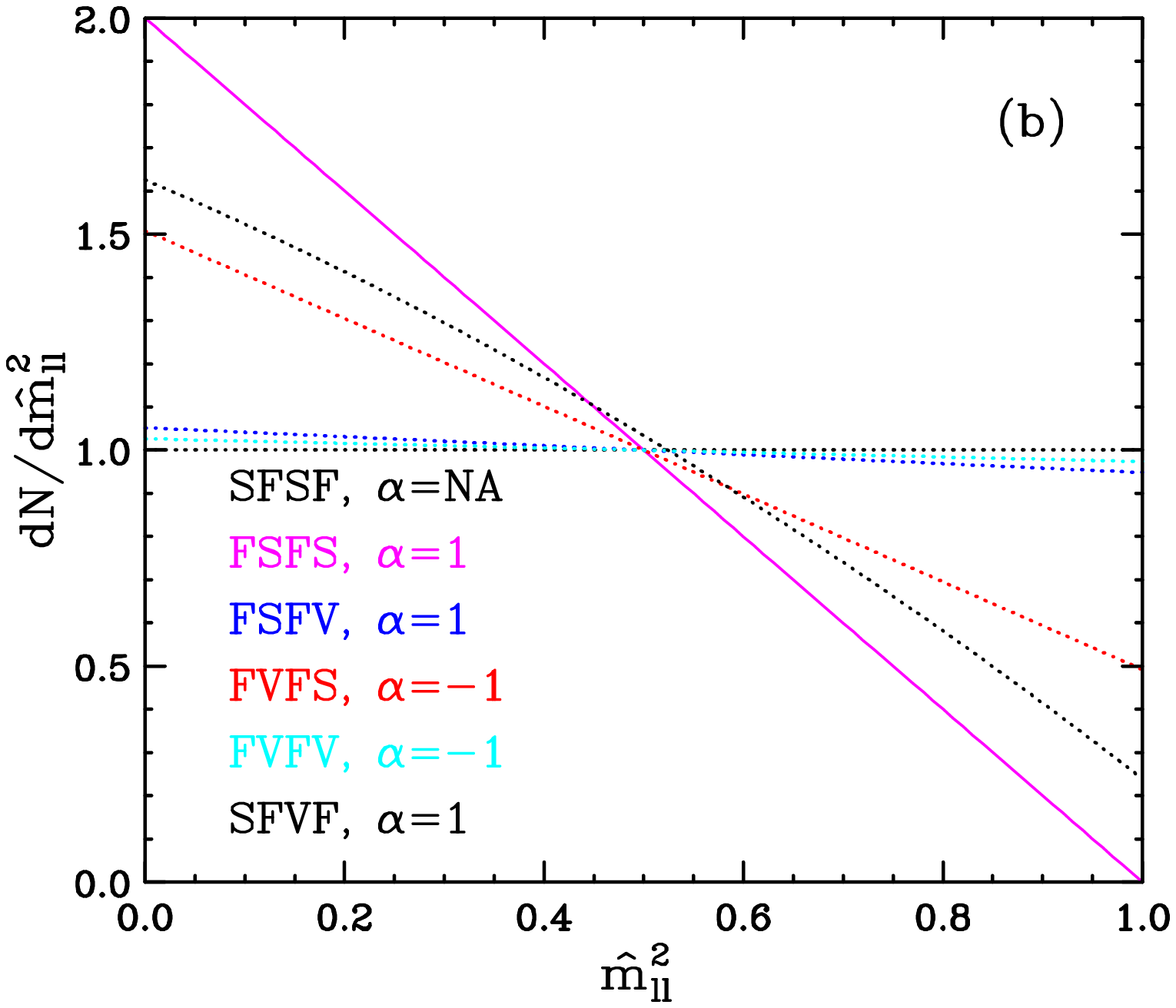,width=6.7cm}
\epsfig{file=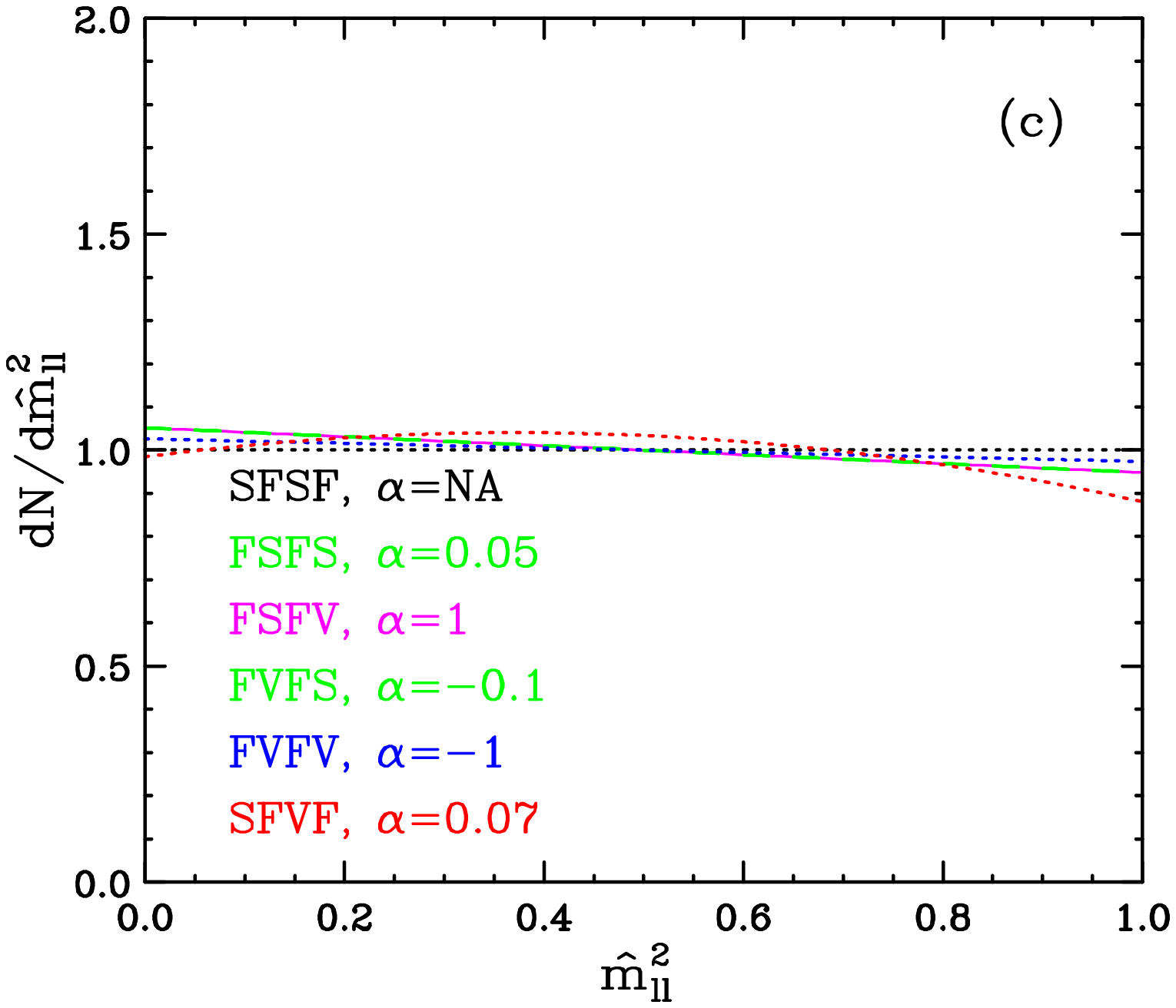,width=6.7cm}
\epsfig{file=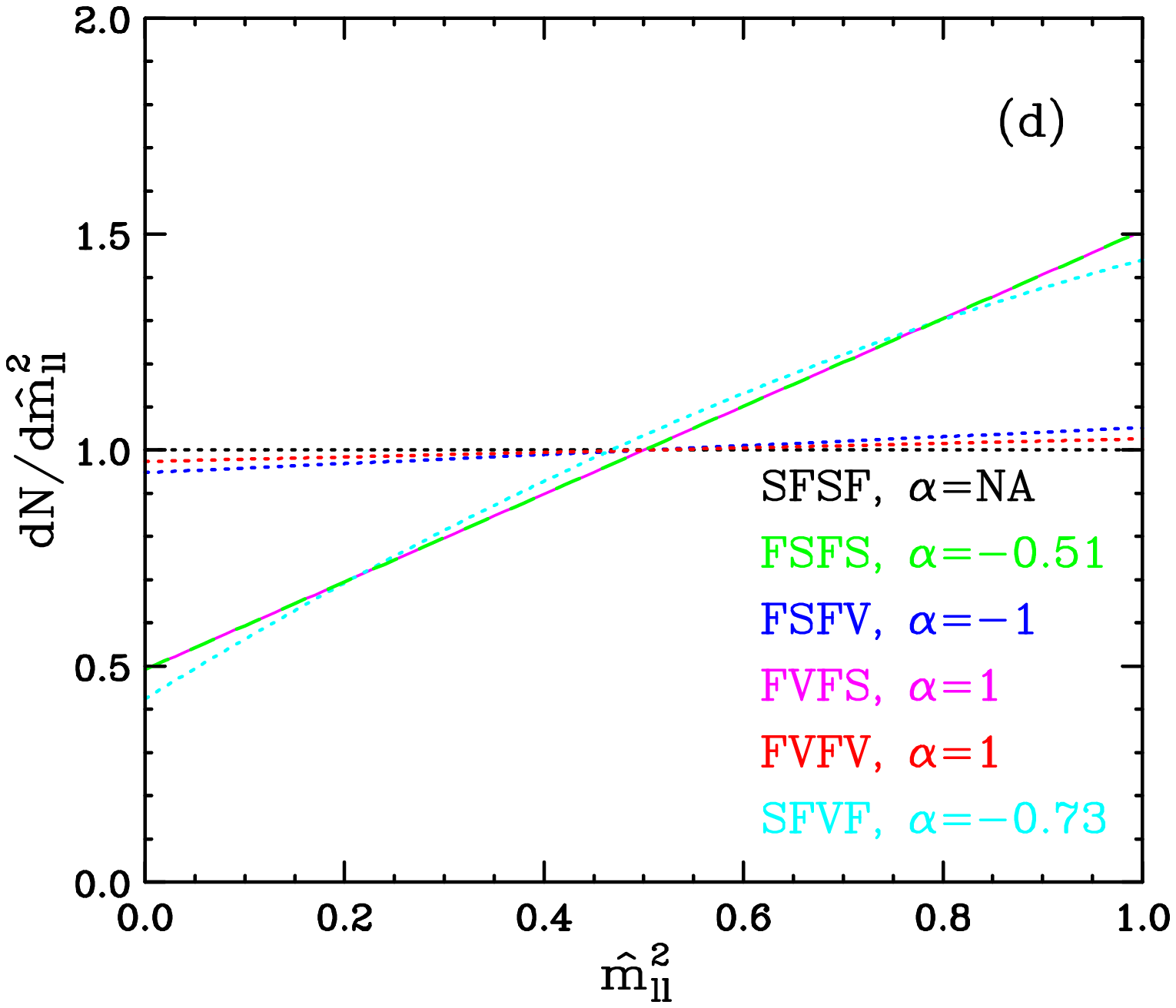,width=6.7cm}
\epsfig{file=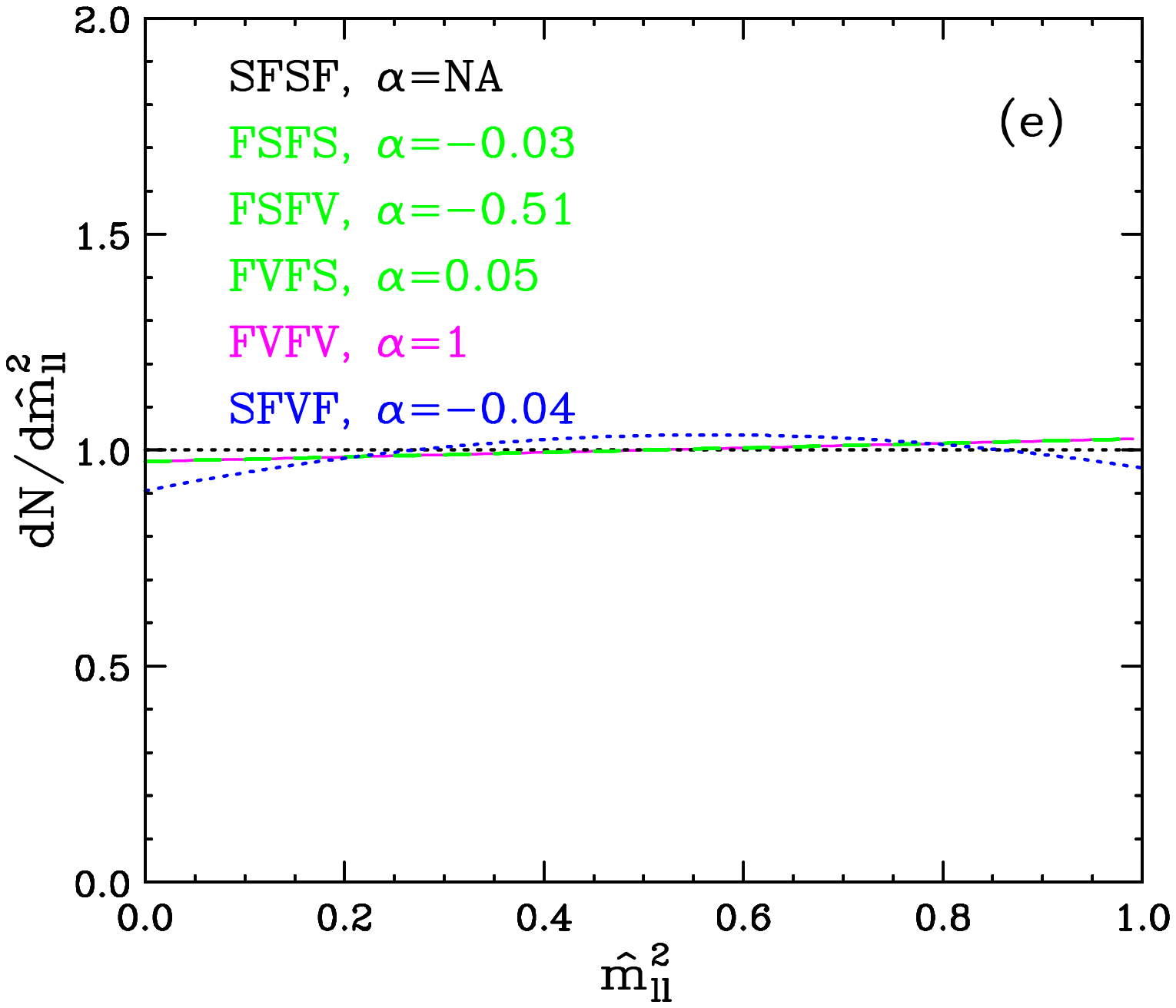,width=6.7cm}
\epsfig{file=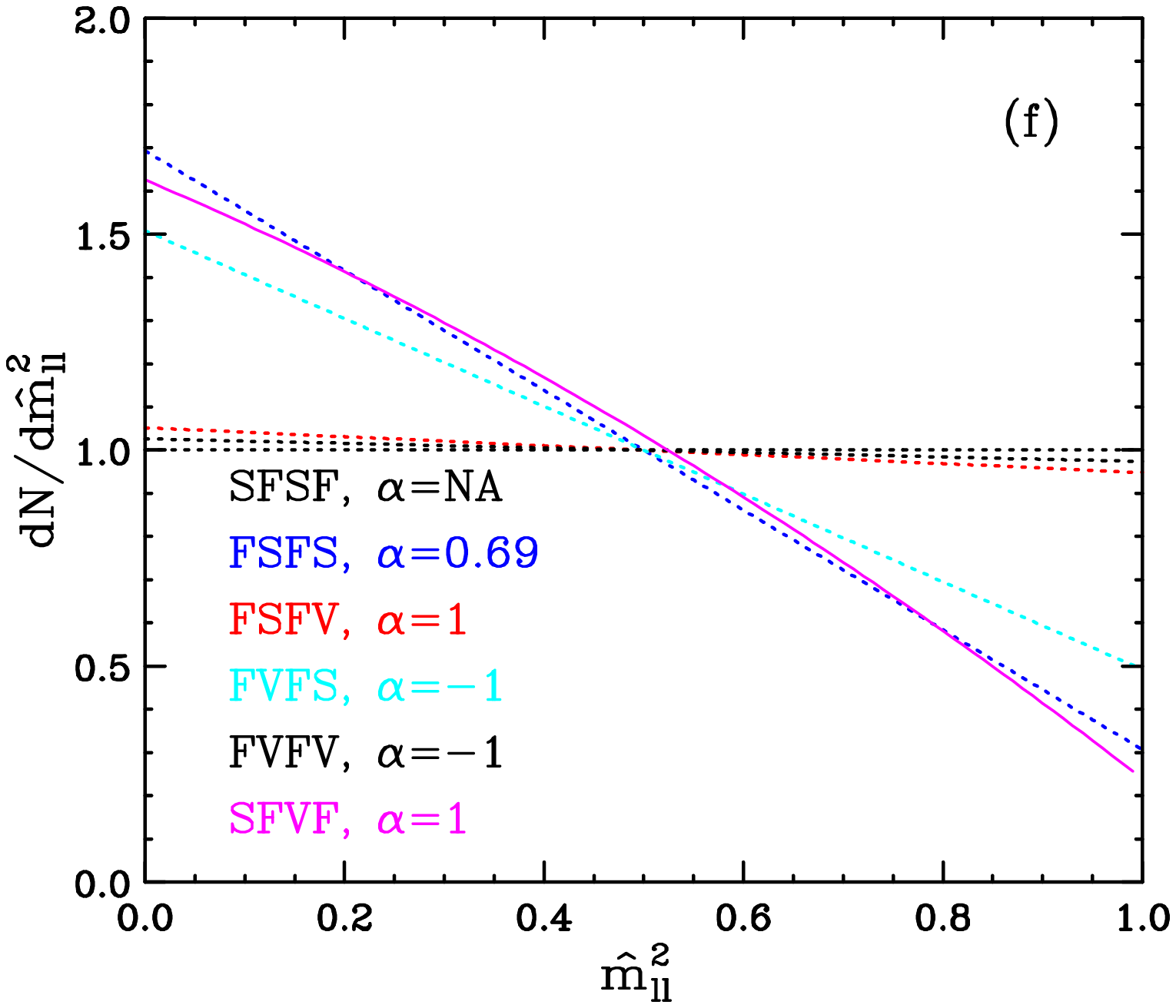,width=6.7cm}
\caption{\sl Dilepton invariant mass distributions ($L_S^{+-}$). 
The solid (magenta) line in each plot represents the input 
dilepton distribution from our simulated ``data'', for each of the 6
spin configurations: a) SFSF; b) FSFS; c) FSFV; d) FVFS; e) FVFV; f) SFVF.
The other (dotted or dashed) lines are our best fits to this data, 
for each of the remaining 5 spin configurations from Table \ref{table:spins}.
The color code is the following. If the trial model 
fits the input data perfectly, we use a dashed (green) line. If
the fit fails to match the input data, we use (color-coded)
dotted lines. The best fit value of $\alpha$ for each case is also shown,
except for cases where it is left undetermined (NA).
\label{fig:ll}}
}
\FIGURE[p!]{
\epsfig{file=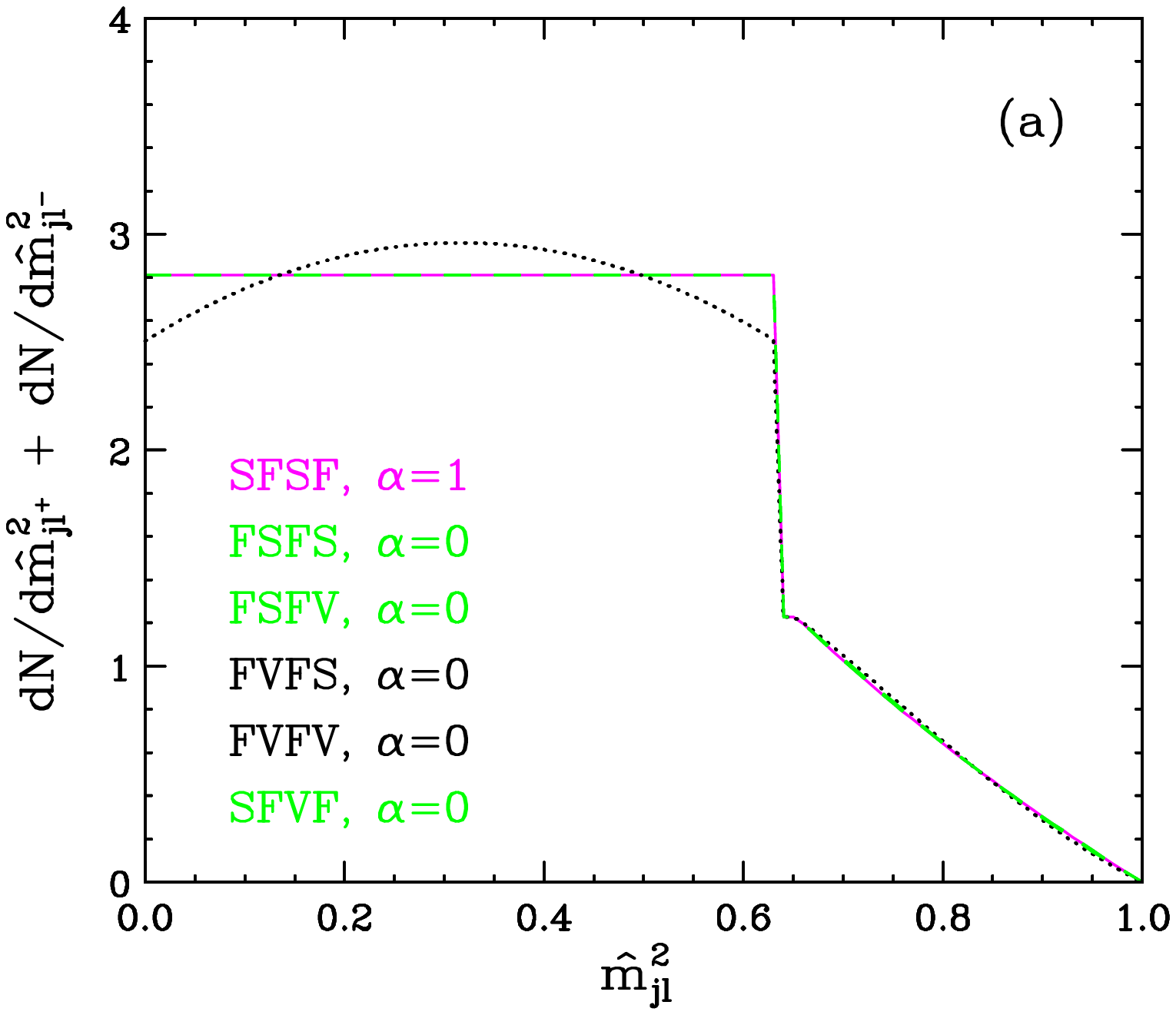,width=7.0cm}
\epsfig{file=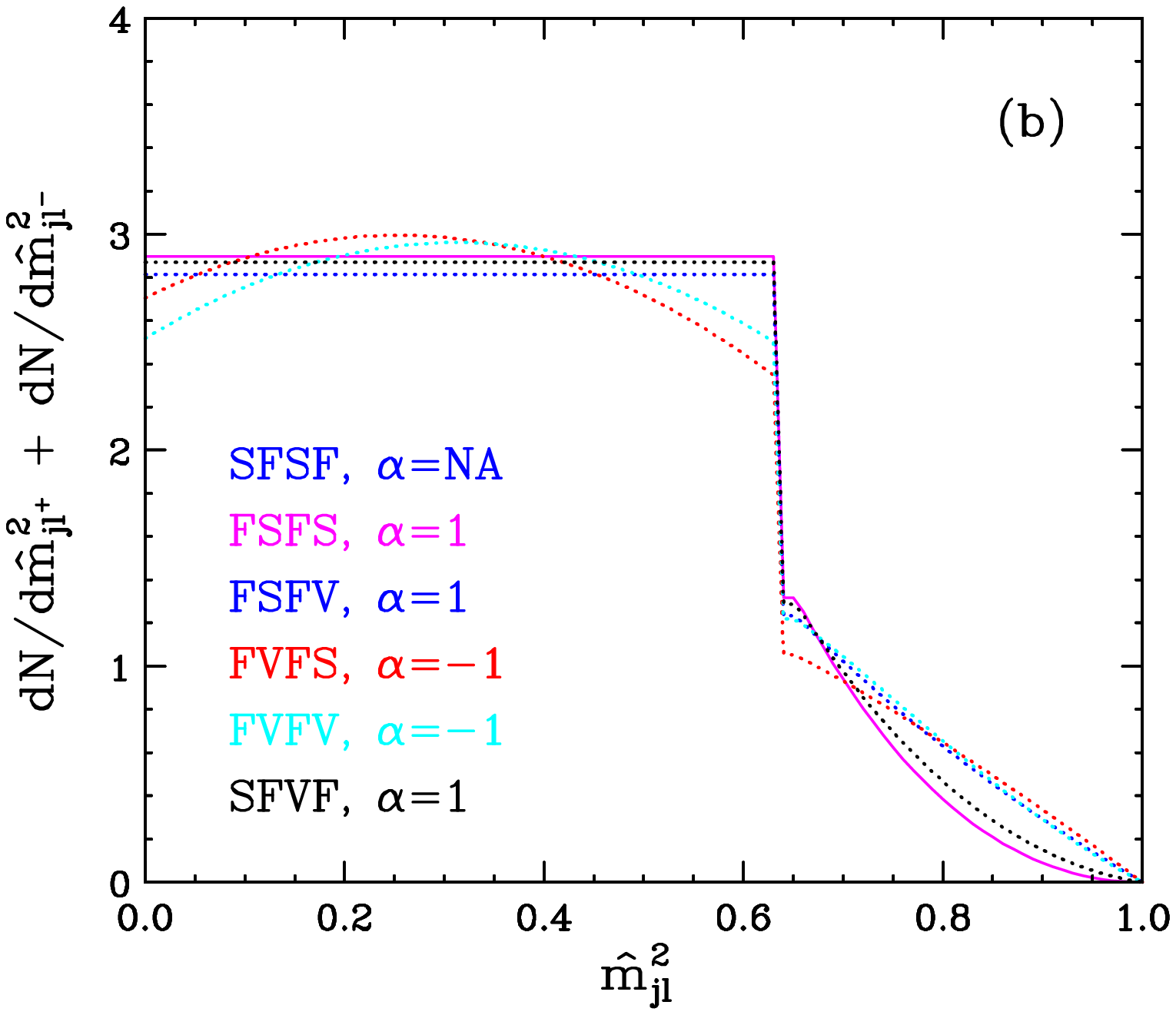,width=7.0cm}\\
\epsfig{file=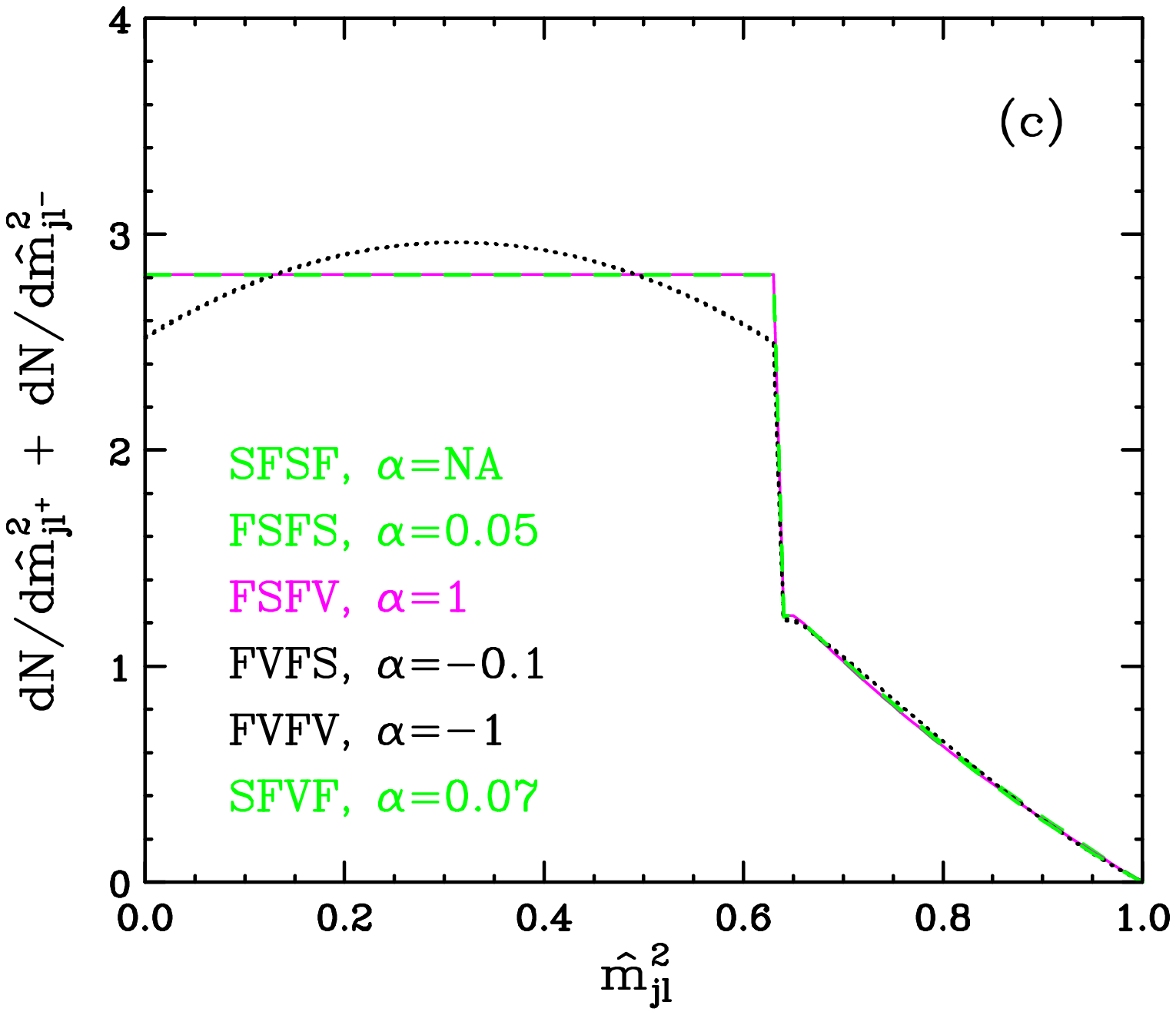,width=7.0cm}
\epsfig{file=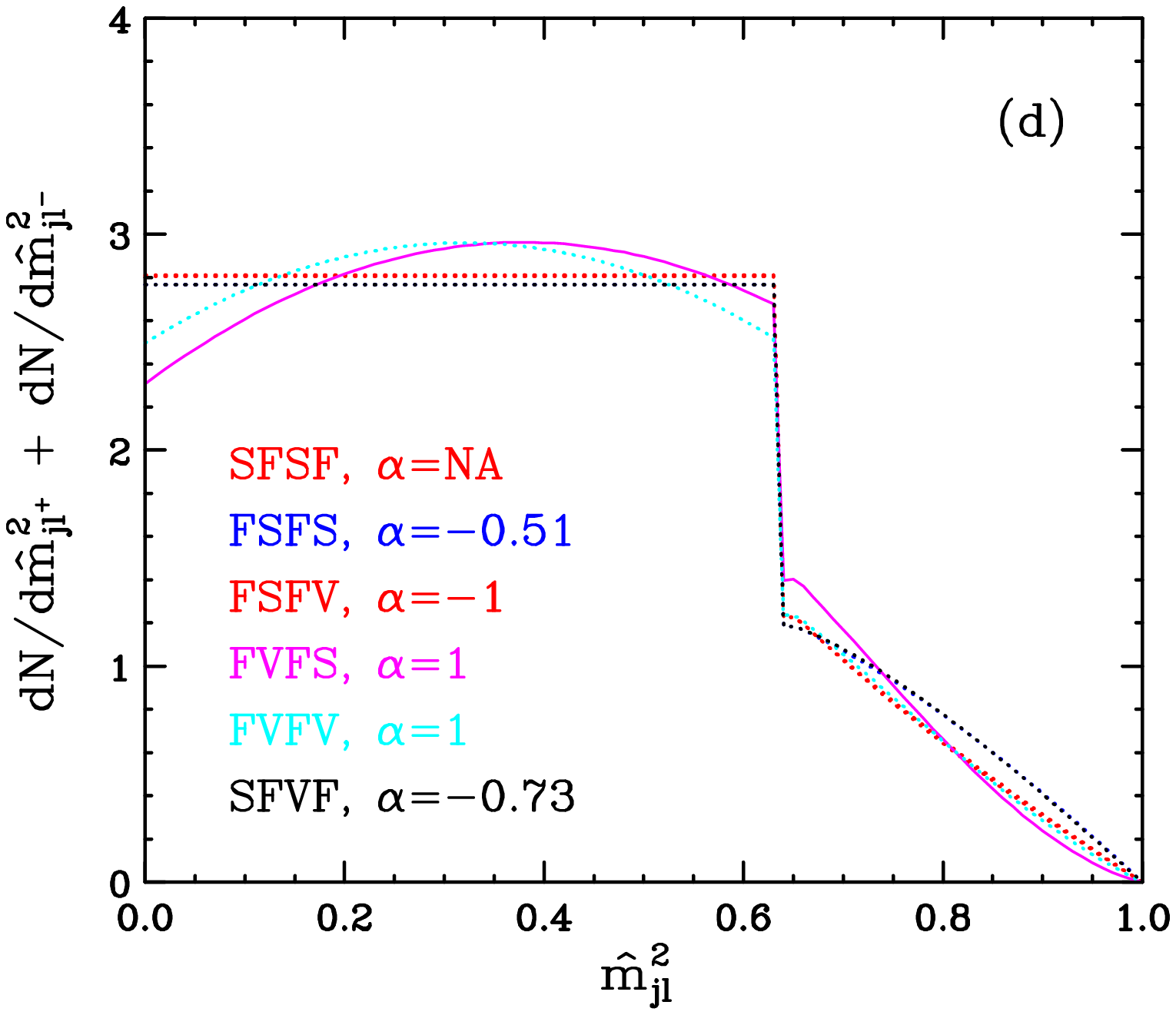,width=7.0cm}\\
\epsfig{file=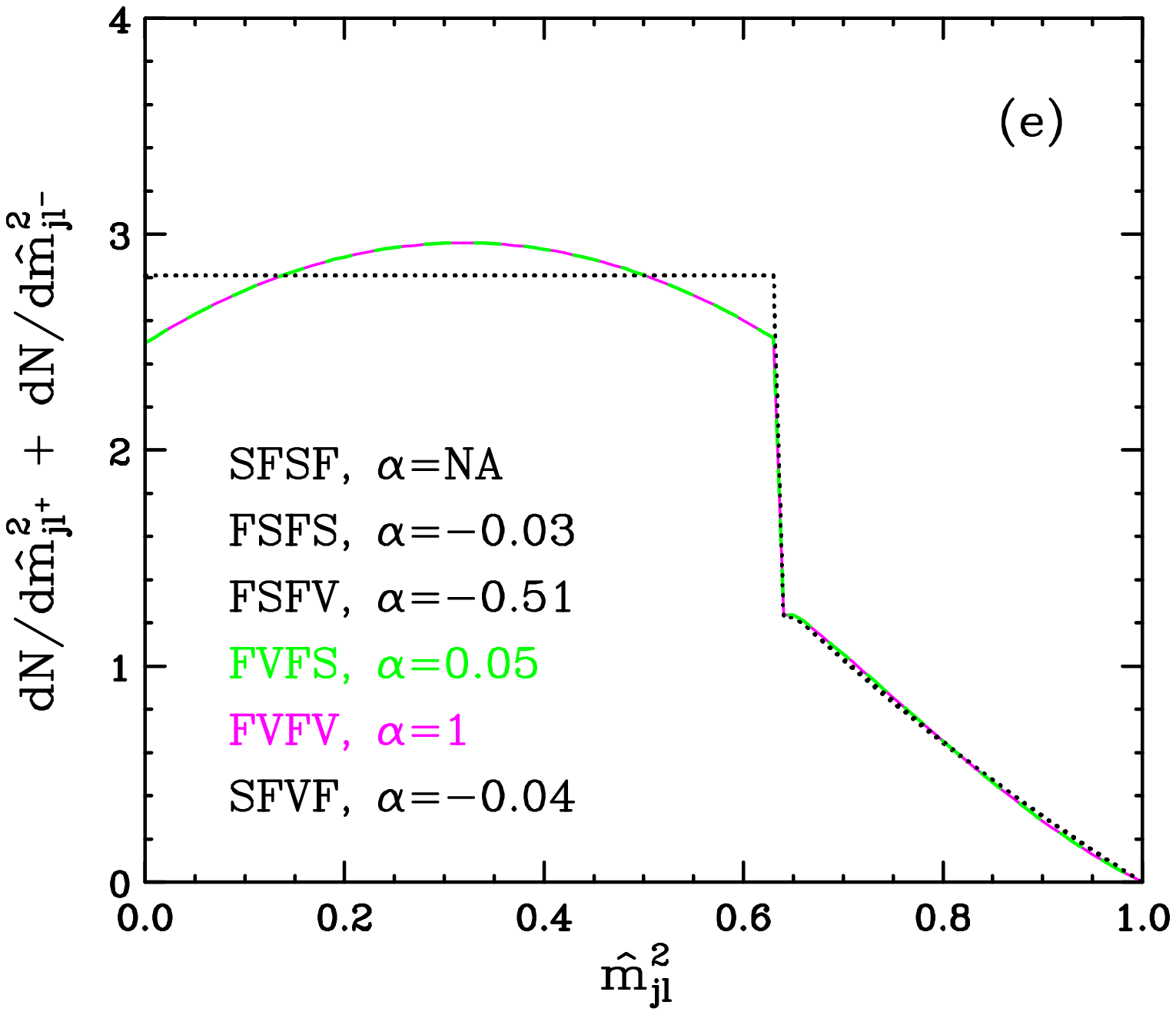,width=7.0cm}
\epsfig{file=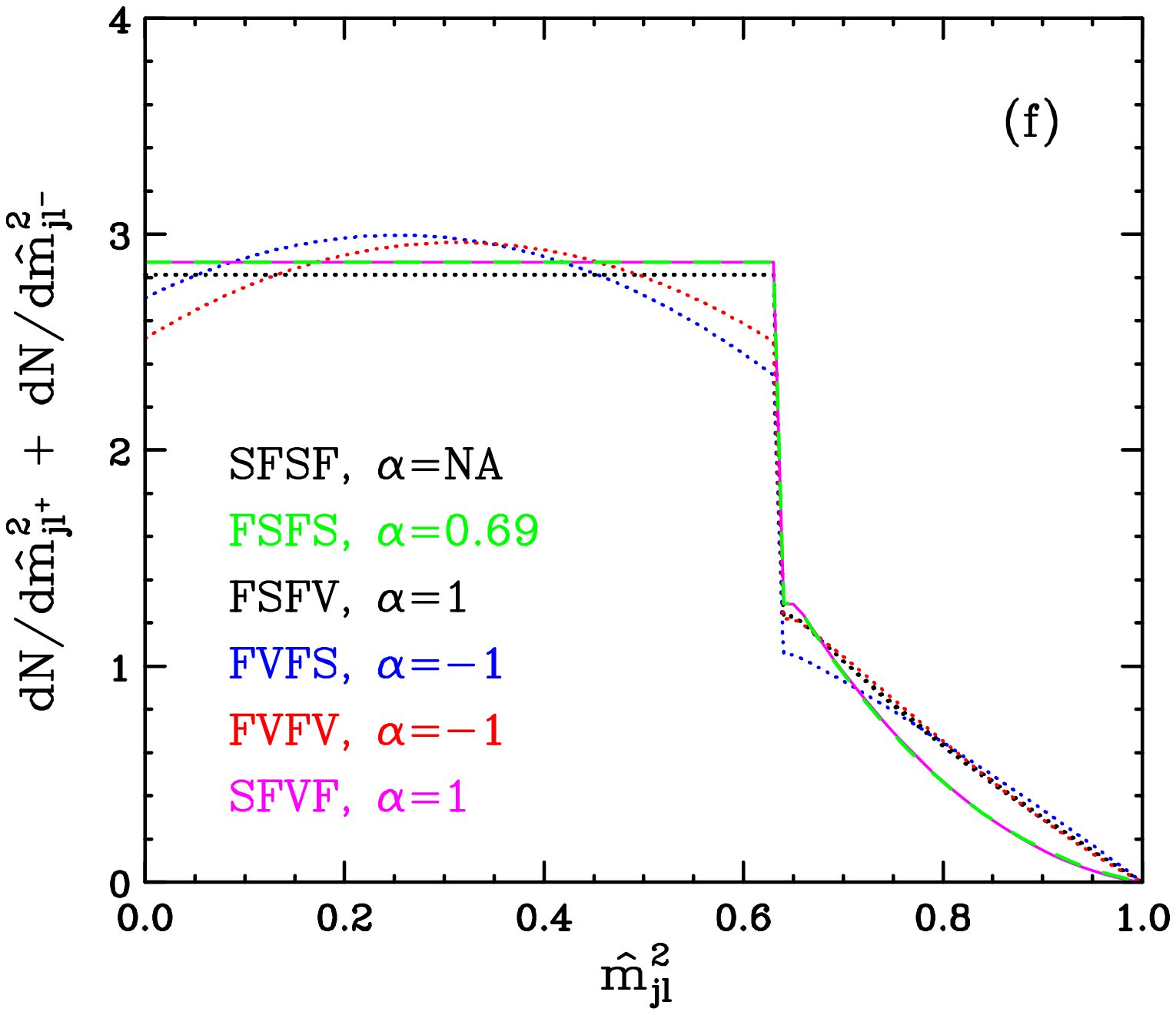,width=7.0cm}
\caption{\sl The same as in Fig.~\ref{fig:ll} but for $S^{+-}$ instead of $L^{+-}$. }
\label{fig:jl_sum}}
\FIGURE[p!]{
\epsfig{file=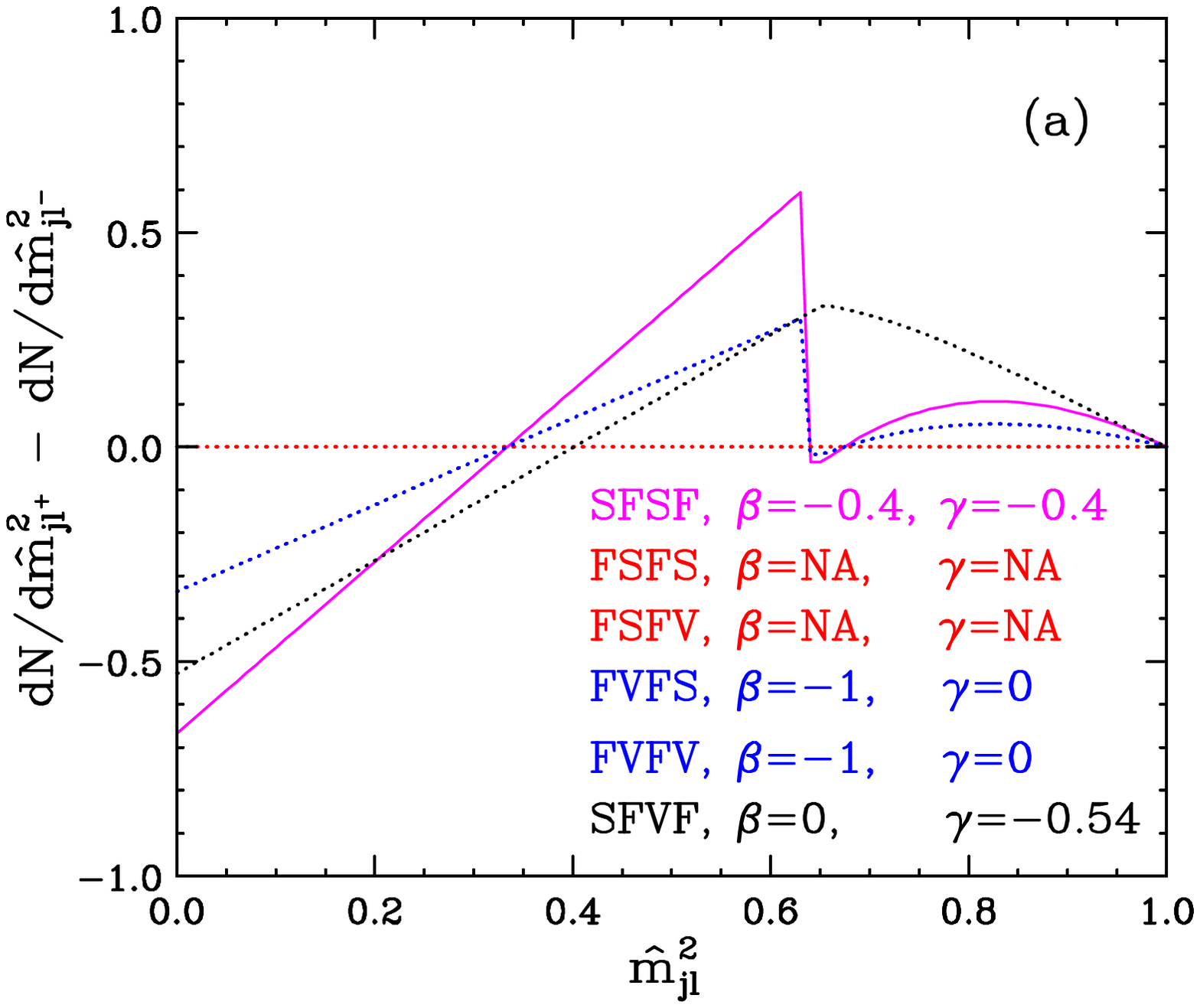,width=7.0cm}
\epsfig{file=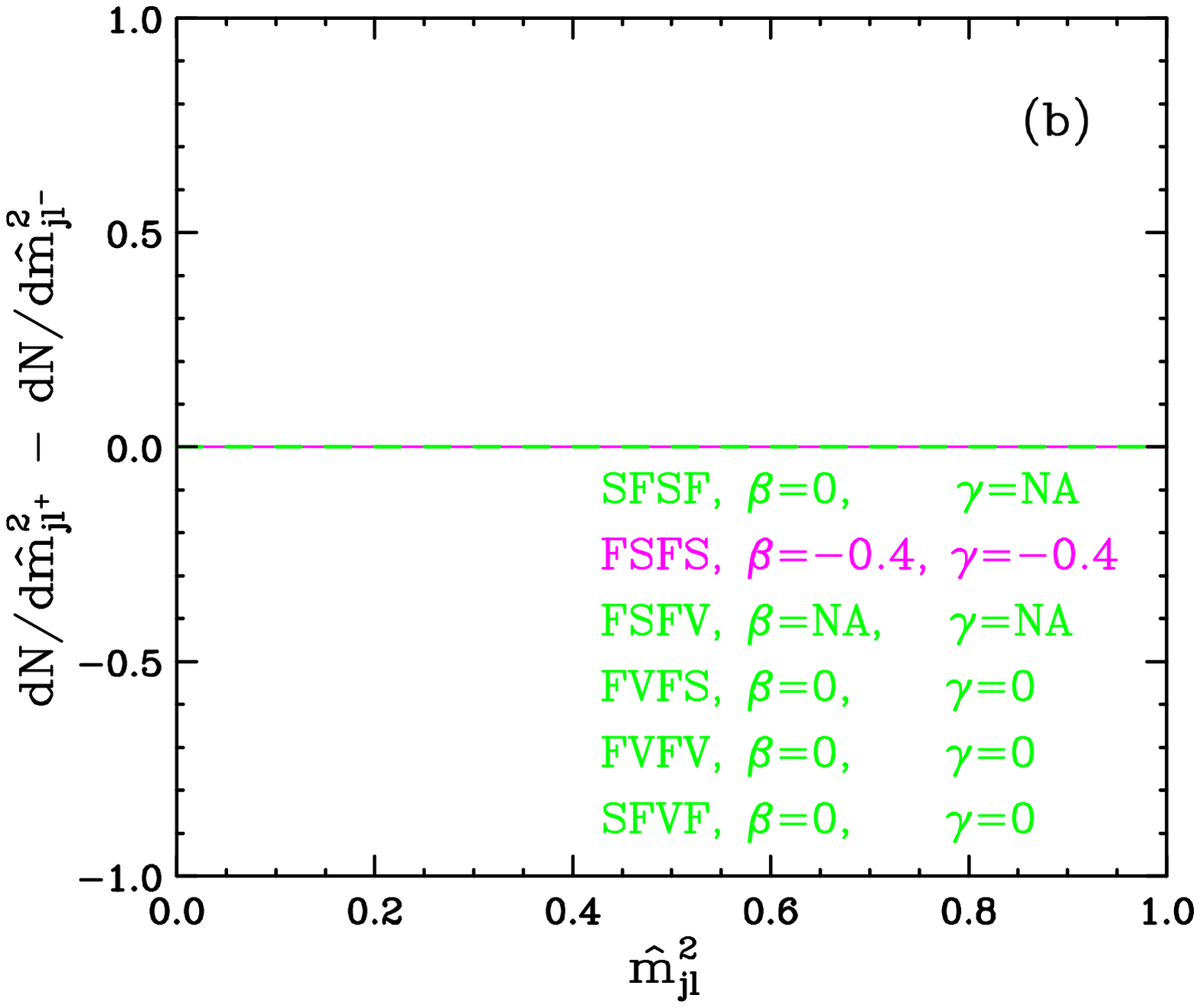,width=7.0cm}
\epsfig{file=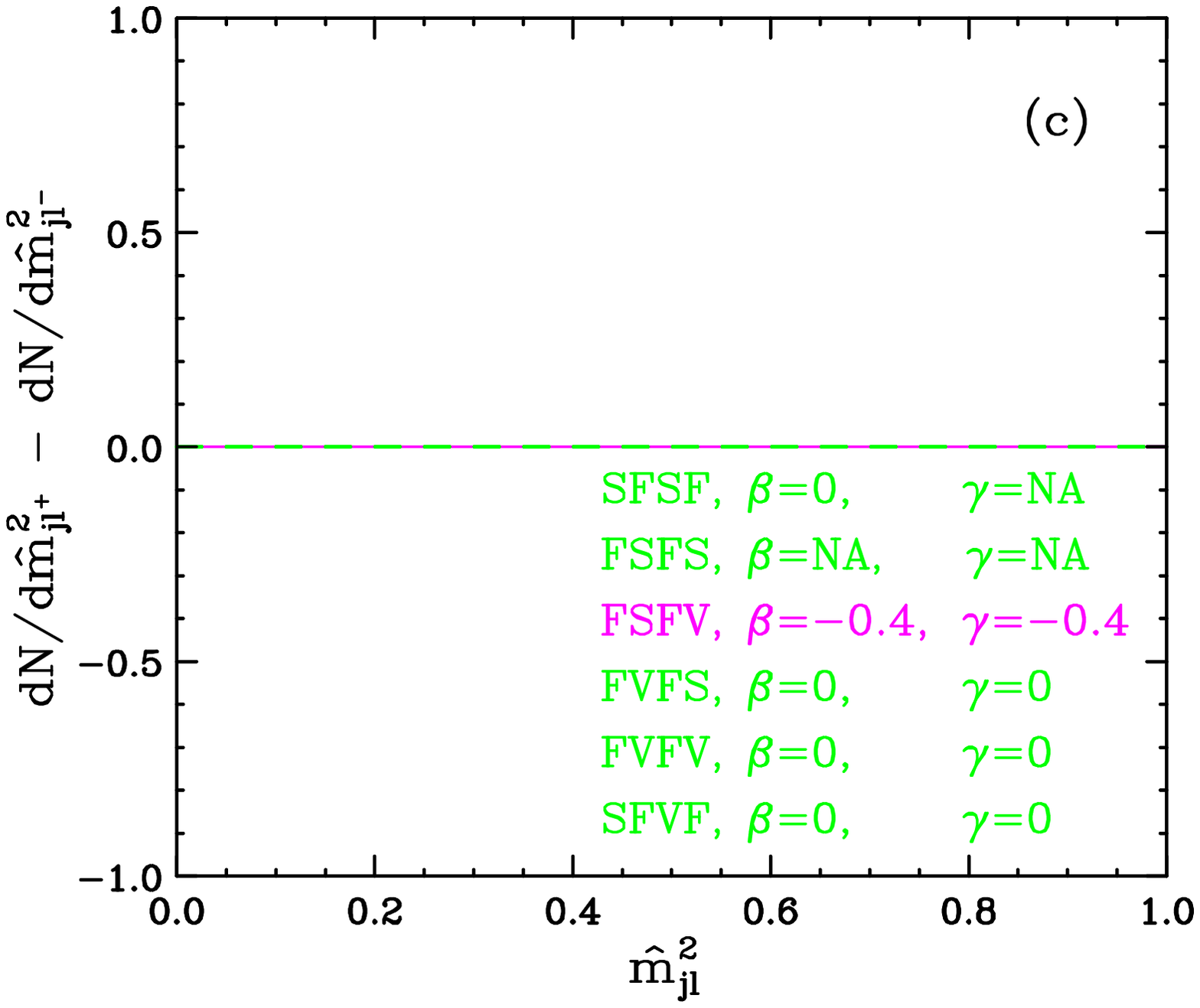,width=7.0cm}
\epsfig{file=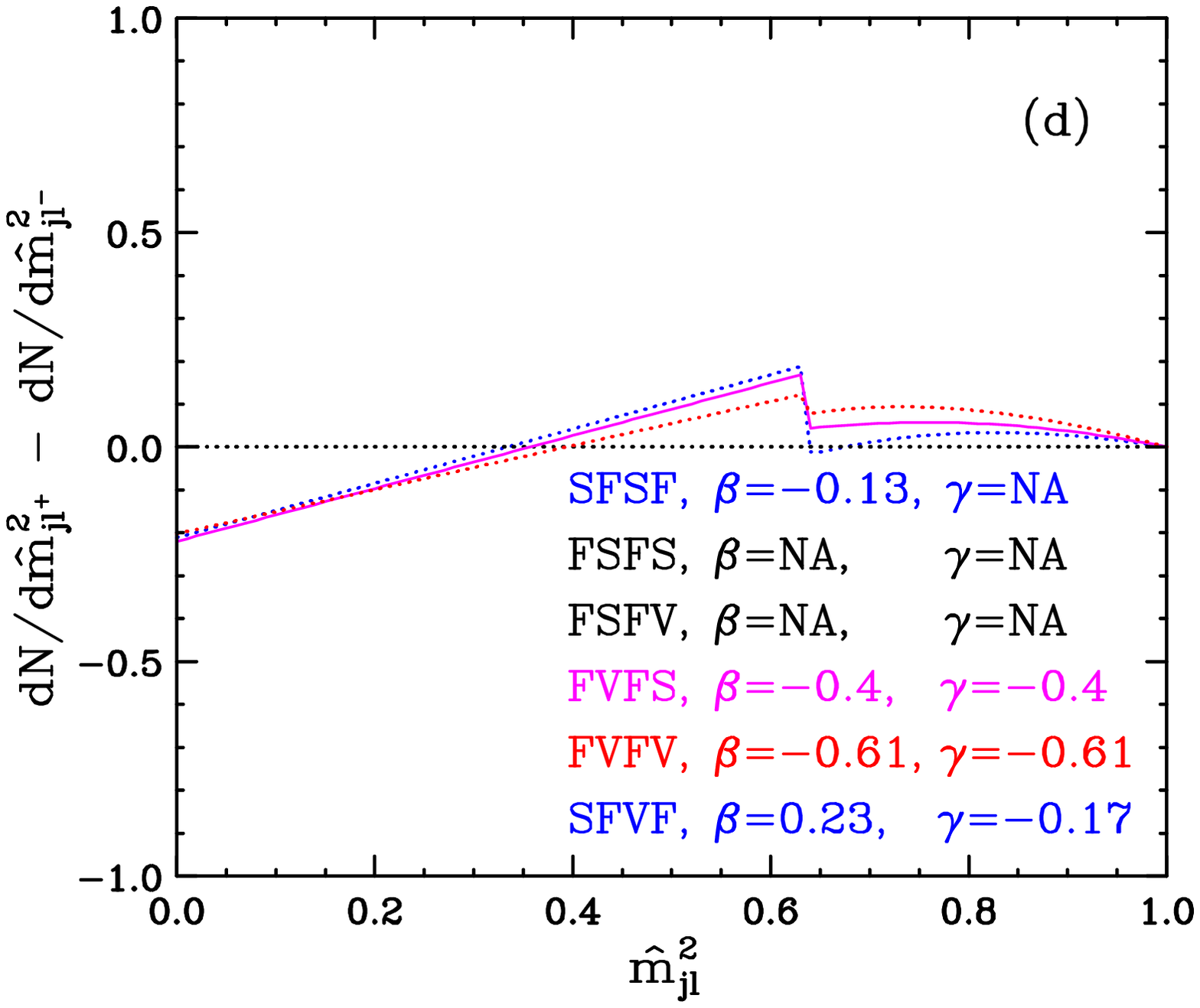,width=7.0cm}
\epsfig{file=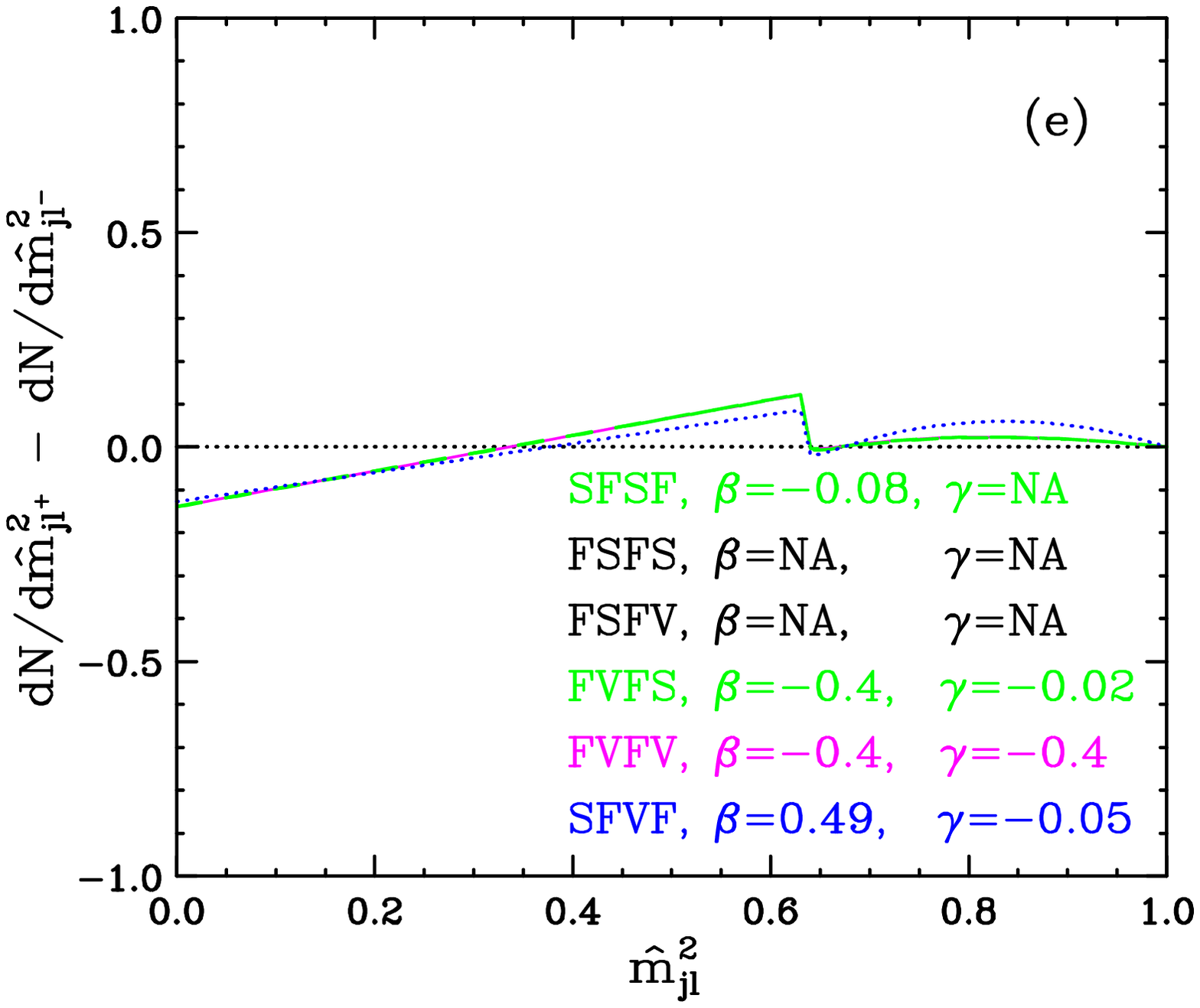,width=7.0cm}
\epsfig{file=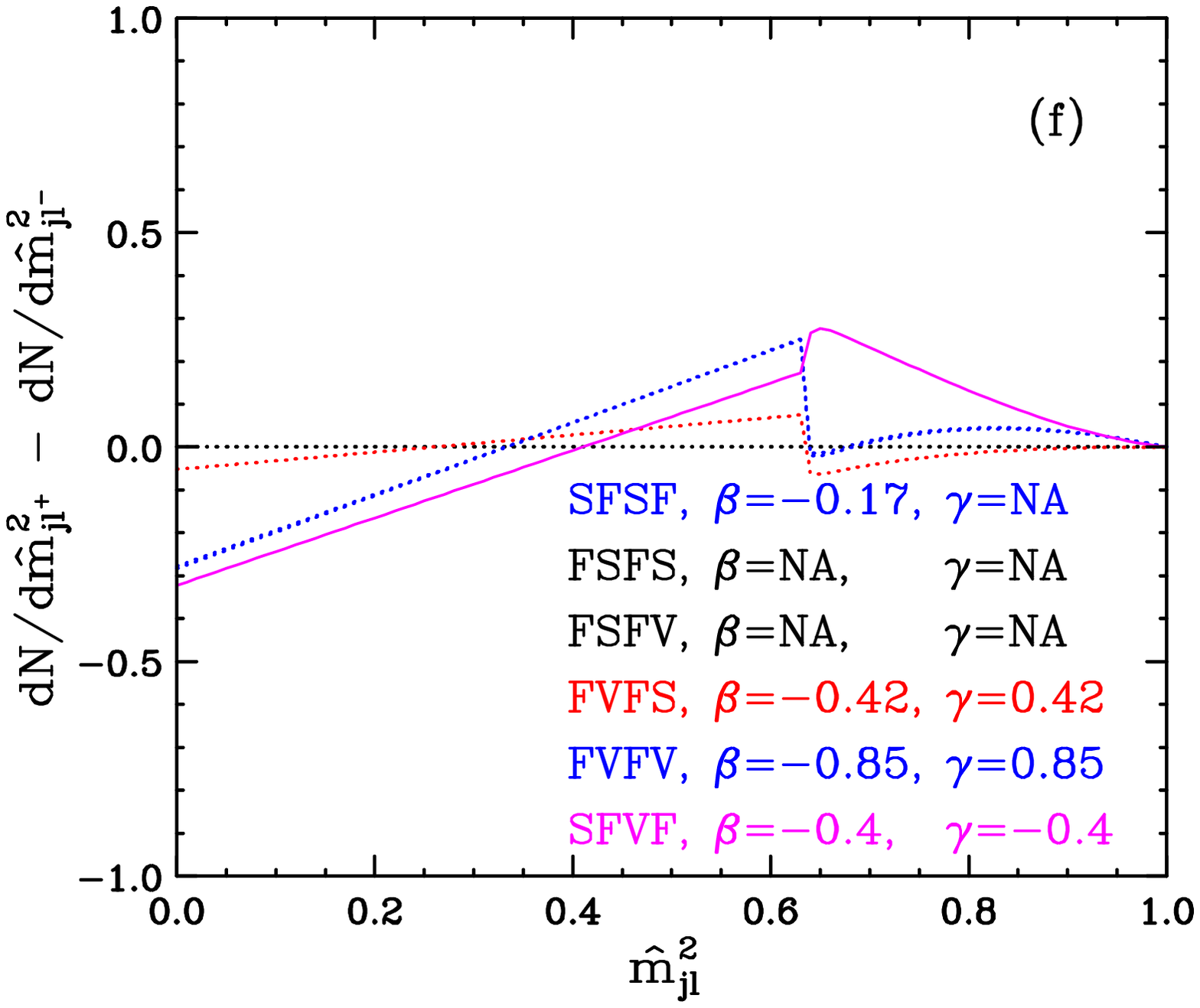,width=7.0cm}
\caption{\sl The same as in Fig.~\ref{fig:ll} but for $D^{+-}$ instead of $L^{+-}$.}
\label{fig:jl_diff}
}

\subsection{SFSF example ($S=1$)}

For the SPS1a parameters (\ref{SPSxyz}-\ref{SPSf}) (or alternatively,
(\ref{SPSabg})), eqs.~(\ref{L+-}-\ref{D+-})
predict the following observable invariant mass distributions
for the SFSF model:
%
%
\begin{eqnarray}
L_1^{+-} &=& 1 \, ,  \\
S_1^{+-} &=&
\left \{\begin{array}{lcccr}
 2.810                                 \hspace{4.5cm}   &&&  ~~~~~~~~~~
\hat{m}^2_{j\ell} \le 0.632   \\[2mm]
 1.228                                    &&& 0.632 \le
\hat{m}^2_{j\ell} \le 0.653   \\[2mm]
-2.880\,\log \hat{m}_{j\ell}^2           &&& 0.653 \le 
\hat{m}^2_{j\ell} \, , ~~~~~~~~~~~ 
\end{array}\right.   \\[3mm]
D_1^{+-} &=&
\left \{\begin{array}{lcccr}
-0.668 + 2.002\,\hat{m}_{j\ell}^2                                 &&&
~~~~~~~~~~\hat{m}^2_{j\ell} \le 0.632   \\[2mm]
-0.035                                                            &&&
0.632 \le \hat{m}^2_{j\ell} \le 0.653   \\[2mm]
6.633 - 6.633\,\hat{m}_{j\ell}^2 + 5.481\,\log \hat{m}_{j\ell}^2  &&&
0.653 \le \hat{m}^2_{j\ell} \, .~~~~~~~~~~~  
\end{array}\right.
\end{eqnarray}
These distributions are shown with solid (magenta) lines in Figs.~\ref{fig:ll}(a), 
\ref{fig:jl_sum}(a) and \ref{fig:jl_diff}(a), respectively.
Following our procedure described above, we first try to fit the
dilepton data in Fig.~\ref{fig:ll}(a). Due to the presence of an
intermediate scalar particle B, the $L^{+-}$ distribution
for the SFSF chain (S=1), is completely flat. However, that
does not necessarily mean that the spin of particle B is 
determined to be zero. In fact, as seen from Fig.~\ref{fig:ll}(a),
all other spin configurations except for $S=6$ (SFVF) can also 
fit this flat distribution, simply by choosing a vanishing $\alpha$ parameter.
Even the case of $S=6$ (SFVF), whose ``best fit'' prediction is
different from the input data, may still be difficult to 
discriminate in practice, once we factor in the finite statistics, 
detector resolution and combinatorial backgrounds.
The bad news, therefore, is that we cannot immediately 
determine the spins from the $L^{+-}$ distribution alone, 
but the good news is that, as anticipated, we got a measurement 
of the $\alpha$ parameter, which represents some combination of 
heavy particle couplings and mixing angles.

At this point it is worth comparing our Fig.~\ref{fig:ll}(a) to
Fig.~2a in Ref.~\cite{Athanasiou:2006ef}, where a very similar 
exercise was performed\footnote{Fig.~2a of Ref.~\cite{Athanasiou:2006ef}
is simply the collection of all six solid (magenta) lines
in our Fig.~\ref{fig:ll}(a)-(f), i.e. our input ``data'' for 
the six different spin configurations, using the same 
fixed SPS1a values (\ref{SPSxyz}-\ref{SPSf}) for the 
model dependent parameters.}. The two results are quite different,
for example we find that 4 out of the 5 ``wrong'' models
can perfectly fit the dilepton ``data'', while in 
Ref.~\cite{Athanasiou:2006ef} all 6 models give distinct dilepton 
shapes. Of course, neither of the two results is wrong,
and the difference simply arises due to our different philosophy. 
In Ref.~\cite{Athanasiou:2006ef} the parameters $\alpha$, 
$\beta$ and $\gamma$ (in our notation)
were all kept fixed to the SPS1a values (\ref{SPSabg}),
while here we are allowing them to float, since they would not have 
been measured in advance independently. As a result, we tend to get
much more similar distributions, indicating that once we factor in 
the experimental realism, the actual spin measurements might be even 
more challenging than previously anticipated.

Having extracted all the relevant information out of the 
$L^{+-}$ distribution, we now move on to studying the
$S^{+-}$ distribution. As we already explained in 
Sec.~\ref{sec:Falphabasis}, the advantage of considering
$S^{+-}$ as opposed to each one of the individual distributions
$\{j\ell^+\}$ and $\{j\ell^-\}$ is that $S^{+-}$ only depends 
on exactly the same parameter $\alpha$ as the dilepton distribution $L^{+-}$
(see eq.~\eqref{S+-}). Since we have just measured 
$\alpha$ by fitting to the $L^{+-}$ data, at this stage there 
are no free parameters left in the $S^{+-}$ distribution, and it is
uniquely predicted for each of the 5 ``wrong'' spin scenarios.
In Fig.~\ref{fig:jl_sum}(a) we plot the resulting predictions for 
the six spin models, using in each case the corresponding value 
of $\alpha$, which had been measured in the previous step 
from the $L^{+-}$ distribution.
We see that the $S^{+-}$ distribution can now further differentiate 
between different spin cases, e.g. it can rule out (in principle)
the $S=4$ and $S=5$ (FVFS and FVFV) models. Interestingly,
now $S=6$ (SFVF) gives a perfect match, but fortunately, 
it has already been eliminated from consideration by the 
analysis of the $L^{+-}$ data at the previous step.
Unfortunately, the ``wrong'' spin scenarios $S=2$ and $S=3$ (FSFS and FSFV)
once again give a perfect match to the data, so that
even after considering both $L^{+-}$ and $S^{+-}$, we are still 
left with 3 distinct possibilities for the spins of the heavy partners.
As we shall see later from the other 5 exercises, the SFSF input ``data''
is somewhat of an unlucky case, since we end up with
several spin models which perfectly fit {\em both} the 
$L^{+-}$ and $S^{+-}$ data. More often than not, 
$L^{+-}$ and $S^{+-}$ by themselves should be sufficient to 
narrow down the spin configuration alternatives to a single one
(or at most two, due to the ``twin'' spin scenarios discussed 
in Sec.~\ref{sec:degenerate}). 

We are therefore forced to consider our third piece of data,
the $D^{+-}$ distribution (\ref{D+-}). This distribution does 
not depend on the previously fitted parameter $\alpha$, 
and instead needs to be fitted with the other two model-dependent
parameters, $\beta$ and $\gamma$. Even though $D^{+-}$ itself
does not explicitly depend on $\alpha$, the fit is nevertheless
impacted by the measured value of $\alpha$, as the latter 
determines the allowed range of values for $\beta$ and $\gamma$
(see Appendix~\ref{app:params} for details).
The results from our fitting exercise to the
$D^{+-}$ SFSF ``data'' are shown in Fig.~\ref{fig:jl_diff}(a).
We see that $D^{+-}$ can now eliminate the remaining two ``wrong'' 
spin scenarios $S=2$ and $S=3$ (FSFS and FSFV) and 
as a result of all three types of fits, we are able to determine
uniquely the spin chain as being $S=1$ (SFSF). In addition, we were also
able to obtain a measurement of the parameter 
$\beta$, which carries information about the
couplings and mixing angles of the heavy partners $D$, $C$ and $B$.
Unfortunately, the parameters $\alpha$ and $\gamma$ are not experimentally 
accessible in this case ($S=1$), since their corresponding basis functions
$\mathcal{F}_{1;\alpha}^{(p)}$ and $\mathcal{F}_{1;\gamma}^{(p)}$
are identically zero for any $p\in \{\ell\ell, j\ell_n, j\ell_f\}$ 
(see Appendix \ref{app:Falpha}).

Having investigated both the $S^{+-}$ and $D^{+-}$ distributions, 
we do not need to consider the lepton charge asymmetry $A^{+-}$,
which is simply the ratio of $D^{+-}$ and $S^{+-}$ (see eq.~(\ref{A+-})).
Numerically the asymmetry $A^{+-}$ and the difference $D^{+-}$ 
show a very similar pattern of their distributions, and thus
provide roughly the same amount of information. However,
as we emphasized in Secs.~\ref{sec:Falphabasis} and \ref{sec:Tevatron},
there are cases where the asymmetry $A^{+-}$ (as well as $D^{+-}$)
does not play any role at all. The cases of S=2 (FSFS) and S=3 (FSFV)
discussed in the next subsection actually provide such an example.

In the remainder of this section, we shall repeat the exercise 
that we just went through, each time taking our ``data'' from 
a different spin configuration, and trying to fit to it the 
remaining\footnote{Obviously the ``correct'' spin configuration
will always give a good fit to its own ``data''.}
5 spin possibilities. 

\subsection{FSFS and FSFV examples ($S=2,3$)}
\label{sec:FSFS}

With the SPS1a parameters (\ref{SPSabg}), eqs.~(\ref{L+-}-\ref{D+-}) 
predict the following observable invariant mass distributions
for the FSFS model
%
\begin{eqnarray}
L_2^{+-} &=& 2 - 2 \hat{m}^2_{\ell\ell} \, ,  
\label{eqn:FSFS_ll}   \\
S_2^{+-} &=&
\left \{\begin{array}{lcccr}
 2.898                                            &&&  ~~~~~~~~~~
\hat{m}^2_{j\ell} \le 0.632   \\[2mm]
 1.316                                            &&& 0.632 \le
\hat{m}^2_{j\ell} \le 0.653   \\[2mm]
-16.583 + 16.583\,\hat{m}^2_{j\ell} - 16.583\,\log
\hat{m}^2_{j\ell}          &&& 0.653 \le \hat{m}^2_{j\ell} \, ,
~~~~~~~~~~~
\end{array}\right.  \label{eqn:FSFS_jl_sum} \\[3mm]
D_2^{+-} &=&  0\ ,
\label{eqn:FSFS_jl_diff}
\end{eqnarray}
which are shown by the solid magenta lines in 
Figs.~\ref{fig:ll}(b), \ref{fig:jl_sum}(b), and \ref{fig:jl_diff}(b),
correspondingly.

Similarly, for the FSFV model we get 
%
\begin{eqnarray}
L_3^{+-} &=& 1.052 - 0.104\,\hat{m}_{\ell\ell}^2 \, ,  \label{eqn:FSFV_ll} \\
S_3^{+-} &=&
\left \{\begin{array}{lcccr}
 2.815                                            &&&  ~~~~~~~~~~
\hat{m}^2_{j\ell} \le 0.632   \\[2mm]
 1.233                                            &&& 0.632 \le
\hat{m}^2_{j\ell} \le 0.653   \\[2mm]
 -0.860 + 0.860\,\hat{m}^2_{j\ell} - 3.590\,\log
\hat{m}^2_{j\ell}         &&& 0.653 \le \hat{m}^2_{j\ell} \, , ~~~~~~~~~~~
\end{array}\right.   \label{eqn:FSFV_jl_sum} \\[3mm]
D_3^{+-} &=&  0\ ,
\label{eqn:FSFV_jl_diff}
\end{eqnarray}
which are shown with solid magenta lines in
Figs.~\ref{fig:ll}(c), \ref{fig:jl_sum}(c), and \ref{fig:jl_diff}(c),
correspondingly.
The distributions (\ref{eqn:FSFS_ll}-\ref{eqn:FSFS_jl_diff}) and
(\ref{eqn:FSFV_ll}-\ref{eqn:FSFV_jl_diff})
will be the input sets of data for our next two exercises.

Perhaps the most striking feature in each of the data sets 
(\ref{eqn:FSFS_ll}-\ref{eqn:FSFS_jl_diff}) and
(\ref{eqn:FSFV_ll}-\ref{eqn:FSFV_jl_diff})
is that the $D^{+-}$ distribution, and consequently, the 
lepton charge asymmetry $A^{+-}$, are both identically zero.
Therefore, they do not convey any information about the spins, 
since any spin configuration can fit those distributions 
with the proper choice of parameters as shown in Figs.~\ref{fig:jl_diff}(b)
and \ref{fig:jl_diff}(c).
This being the case, we should concentrate on the 
$L^{+-}$ and $S^{+-}$ distributions. 

First we shall discuss the case when the data comes from the 
FSFS ($S=2$) model, eqs.~(\ref{eqn:FSFS_ll}-\ref{eqn:FSFS_jl_diff}).
Again, we begin our analysis with $L^{+-}$, which in this case 
shows very good discrimination, and can already rule out {\em all}
of the ``wrong'' spin combinations. As explained in Sec.~\ref{sec:degenerate}, 
FSFS ($S=2$) can sometimes be faked by the FSFV (S=3) model, but this could 
only happen if the $\alpha$ parameter in the data satisfies eq.~(\ref{alpha2_cond}), i.e.
\begin{equation}
|\alpha_2|\le \left|\frac{1-2z}{1+2z}\right|\approx 0.05\ .
\end{equation}
Since for SPS1a $\alpha=1$ (see eq.~(\ref{SPSabg})), this condition 
is not satisfied and the FSFV model cannot fake the FSFS ``data''.
This is confirmed by our result in Fig.~\ref{fig:ll}(b).

Since the $L^{+-}$ distribution alone already singles out the 
correct spin configuration, we do not even need to consider the
$S^{+-}$ distribution. It is worth pointing out, however, that
$S^{+-}$ in this ideal case also can rule out all ``wrong'' spin 
models, although the differences are not so pronounced as for $L^{+-}$,
and in reality are likely to be washed out.
In summary, the FSFS ``data'' can be unambiguously interpreted
in relation to the spin issue, and we can also get a measurement
of the parameter $\alpha$. On the other hand, the parameters
$\beta$ and $\gamma$ will remain undetermined, since their
corresponding basis functions $\mathcal{F}_{2;\beta}^{(p)}$ 
and $\mathcal{F}_{2;\gamma}^{(p)}$ are identically zero for 
any $p\in \{\ell\ell, j\ell_n, j\ell_f\}$ 
(see Appendix \ref{app:Falpha}).

Now we shall discuss the case when the data comes from the 
FSFV ($S=3$) model, eqs.~(\ref{eqn:FSFV_ll}-\ref{eqn:FSFV_jl_diff}).
This will provide our first example where our spin
measurement ends up being inconclusive, yielding two
different, equally plausible, possibilities for the spin chain.
This result should have already been anticipated, 
based on our general discussion in Sec.~\ref{sec:degenerate}.
There we showed that for any given FSFV data, the FSFS model ($S=2$) 
can always provide a perfect fit, and furthermore, the
value of $\alpha_2$ that would be measured for the ``twin''
FSFS model is
\begin{equation}
\alpha_2 = \alpha_3 \ \frac{1-2z}{1+2z}\ \approx 0.05\ ,
\end{equation}
where we used the SPS1a values for $\alpha_3=1$ and $z=0.451$.
Our numerical study explicitly confirms this general expectation
as shown in Figs.~\ref{fig:ll}(c), \ref{fig:jl_sum}(c) and \ref{fig:jl_diff}(c).
In addition, we checked that the $m_{j\ell\ell}$ distributions
for those two ``twin'' spin models are also identical.

\subsection{FVFS and FVFV examples ($S=4,5$)}
\label{sec:FVFS}

In this subsection we discuss the case of the other ``twin'' spin pair
from Sec.~\ref{sec:degenerate}, namely $S=4$ and $S=5$ (FVFS and FVFV). 
Using the SPS1a values for the model-dependent parameters, we 
obtain the following distributions for the FVFS case
%
%
\begin{eqnarray}
L_4^{+-} &=& 0.492 + 1.016\,\hat{m}_{\ell\ell}^2 \, ,  \\
S_4^{+-} &=&
\left \{\begin{array}{lcccr}
 2.307 + 3.455\,\hat{m}^2_{j\ell} - 4.553\,\hat{m}^4_{j\ell}
&&&  ~~~~~~~~~~ \hat{m}^2_{j\ell} \le 0.632   \\[2mm]
 1.028 + 0.577\,\hat{m}^2_{j\ell}
&&& 0.632 \le    \hat{m}^2_{j\ell} \le 0.653   \\[2mm]
- 42.563 - 12.368\,\hat{m}^2_{j\ell} + 54.931\,\hat{m}^4_{j\ell} \\
    ~~~~  - \Big( 7.871 + 90.785\,\hat{m}^2_{j\ell} \Big) \,\log
\hat{m}^2_{j\ell}
&&& 0.653 \le    \hat{m}^2_{j\ell}\, ,  ~~~~~~~~~~~
\end{array}\right.   \\[3mm]
D_4^{+-} &=&
\left \{\begin{array}{lcccr}
 -0.22 + 0.616\,\hat{m}^2_{j\ell} &&& ~~~~~~~~~~\hat{m}^2_{j\ell} \le
0.632   \\[2mm]
 -0.092 + 0.212\,\hat{m}^2_{j\ell} &&& 0.632 \le \hat{m}^2_{j\ell} \le
0.653   \\[2mm]
 - 3.087  + 3.087\,\hat{m}^2_{j\ell} \\ 
 ~~~~  - \Big( 0.874\, +  2.678\,\hat{m}^2_{j\ell}\Big)\log \hat{m}^2_{j\ell}
          &&& 0.653 \le \hat{m}^2_{j\ell}\, , ~~~~~~~~~~~  
\end{array}\right.
\end{eqnarray}
which are shown with the solid (magenta) lines in
Figs.~\ref{fig:ll}(d), \ref{fig:jl_sum}(d), and \ref{fig:jl_diff}(d),
correspondingly.

For the FVFV case we get 
%
\begin{eqnarray}
L_5^{+-} &=& 0.974 + 0.053\,\hat{m}_{\ell\ell}^2   \, ,  \\
S_5^{+-} &=&
\left \{\begin{array}{lcccr}
 2.496 + 2.908\,\hat{m}^2_{j\ell} - 4.553\,\hat{m}^4_{j\ell}
&&&  ~~~~~~~~~~ \hat{m}^2_{j\ell} \le 0.632   \\[2mm]
1.217 + 0.030\,\hat{m}^2_{j\ell}
&&& 0.632 \le    \hat{m}^2_{j\ell} \le 0.653   \\[2mm]
 27.809 - 43.679\,\hat{m}^2_{j\ell} + 15.870\,\hat{m}^4_{j\ell}  \\
~~~~ +\Big( 14.382 - 4.710\,\hat{m}^2_{j\ell} \Big) \,\log
\hat{m}^2_{j\ell}
 &&& 0.653 \le    \hat{m}^2_{j\ell} \, , ~~~~~~~~~~~
\end{array}\right.   \\[3mm]
D_5^{+-} &=&
\left \{\begin{array}{lcccr}
 -0.139 + 0.415\,\hat{m}^2_{j\ell}
&&& ~~~~~~~~~~\hat{m}^2_{j\ell} \le 0.632   \\[2mm]
-0.011 + 0.011\,\hat{m}^2_{j\ell}
&&& 0.632 \le \hat{m}^2_{j\ell} \le 0.653   \\[2mm]
1.109 - 1.109\,\hat{m}^2_{j\ell} \\
~~~~ + \Big( 1.004 - 0.139\,\hat{m}^2_{j\ell}\Big)\log \hat{m}^2_{j\ell}
&&& 0.653 \le \hat{m}^2_{j\ell} \, , ~~~~~~~~~~~ 
\end{array}\right.
\end{eqnarray}
and those are shown by the solid (magenta) lines in
Figs.~\ref{fig:ll}(e), \ref{fig:jl_sum}(e) and \ref{fig:jl_diff}(e).

The end result of the two exercises is very similar to what we obtained 
in the previous subsection for the other ``twin'' model pair (FSFS and FSFV).
It could have also been anticipated from our general discussion in 
Sec.~\ref{sec:degenerate}. When going in the forward direction, i.e. starting 
with the FVFS ``data'' and fitting to it the other 5 models, we do not encounter
any spin ambiguities. As already determined in the previous subsection, this
is because the SPS1a value of the $\alpha$ parameter ($\alpha=1$) does not
satisfy the necessary condition (\ref{alpha4_cond}) for an FVFV model to fake
the FVFS data. As a result, the two $L^{+-}$ and $S^{+-}$ distributions
are already sufficient to pin down the spin case scenario, and the $D^{+-}$ 
distribution can then be used as a cross-check and for a measurement of the
$\beta$ and $\gamma$ parameters.

However, when going in the reverse direction, i.e. starting with the 
FVFV ``data'' and fitting the other 5 models including FVFS to it, 
we do encounter a spin ambiguity, just like in the $S=3$ exercise above. 
Again, the reason for this was already explained in Sec.~\ref{sec:degenerate}.
In agreement with our analytical results, Figs.~\ref{fig:ll}(e), 
\ref{fig:jl_sum}(e) and \ref{fig:jl_diff}(e) show that the FVFS model 
provides an identical match to the FVFV ``data'' for {\em all three} 
observable distributions $L^{+-}$, $S^{+-}$ and $D^{+-}$.
The good news, however, is that while we are left with a two-fold
ambiguity with respect to the spins, for each spin 
scenario the parameters $\alpha$, $\beta$ and $\gamma$ are 
precisely measured, so that we have independent measurements of
{\em three} different combinations of the heavy partner 
couplings and mixing angles. In Sec.~\ref{sec:couplings} 
below we shall show how to interpret those measurements in 
terms of the more fundamental model parameters $a_L$, $a_R$, 
$b_L$, $b_R$, $c_L$, $c_R$ and $f$.

\subsection{SFVF example ($S=6$)}

Our final example is the SFVF spin chain, for which the 
SPS1a model parameters (\ref{SPSabg}) 
predict the following observable distributions
%
\begin{eqnarray}
L_6^{+-} &=&   1.626 - 0.981\,\hat{m}_{\ell\ell}^2 -
0.405\,\hat{m}_{\ell\ell}^4   \, , \label{eqn:SFVF_ll} \\
S_6^{+-} &=&
\left \{\begin{array}{lcccr}
2.87    &&&  ~~~~~~~~~~ \hat{m}^2_{j\ell} \le 0.632   \\[2mm]
1.288   &&& 0.632 \le    \hat{m}^2_{j\ell} \le 0.653   \\[2mm]
 -0.344 - 4.493\,\hat{m}^2_{j\ell} + 4.837\,\hat{m}^4_{j\ell} -
           5.870\,\log \hat{m}^2_{j\ell}
&&& 0.653 \le    \hat{m}^2_{j\ell}, ~~~~~~~~~~~
\end{array}\right.   \\[3mm]
D_6^{+-} &=&
\left \{\begin{array}{lcccr}
-0.322 + 0.786 \,\hat{m}^2_{j\ell}
&&& ~~~~~~~~~~\hat{m}^2_{j\ell} \le 0.632   \\[2mm]
 -0.406 + 1.051\,\hat{m}^2_{j\ell}
&&& 0.632 \le \hat{m}^2_{j\ell} \le 0.653   \\[2mm]
5.870 - 11.674\,\hat{m}^2_{j\ell} + 5.804\,\hat{m}^4_{j\ell} \\
~~~~ +\Big( 3.384 - 3.595\,\hat{m}^2_{j\ell} \Big) \,\log \hat{m}^2_{j\ell}
&&& 0.653 \le \hat{m}^2_{j\ell}, ~~~~~~~~~~~ 
\end{array}\right.
\end{eqnarray}
shown with solid (magenta) lines in Figs.~\ref{fig:ll}(f), 
\ref{fig:jl_sum}(f) and \ref{fig:jl_diff}(f).
                                                                                                                          
The case of $S=6$ (SFVF) is very special, since
in this case the dilepton invariant mass distribution 
(\ref{eqn:SFVF_ll}) exhibits a characteristic $\hat{m}^4$
term which is not present for any of the other 5 spin configurations that
we are considering. Note that the existence of an $\hat{m}^4$ term 
in the dilepton SFVF data is generic, i.e. does not depend on the 
values of the model-dependent parameters such as $\alpha$.
This could be easily understood by realising that the $\hat{m}^4$ 
dependence originates from the ``phase space'' basis function
${\mathcal F}_{6;\delta}^{(\ell\ell)}$, which enters our general 
formula (\ref{L+-}) for the dilepton distribution without any 
model-dependent coefficients. More generally, an inspection of 
Table~\ref{table:Fll2} reveals that the dilepton invariant mass 
distribution is in general given by some polynomial in terms of 
$\hat{m}^2$, whose power is equal to twice the spin of the intermediate 
particle $B$.\footnote{A similar statement can be made about the 
$m_{j\ell_n}$ invariant mass distributions from Table~\ref{table:Fqln2},
relating the power of the $\hat{m}_{j\ell_n}^2$ 
to the spin of the intermediate C particle.} 
Only in the SFVF case ($S$=6) do we have a spin 1 
intermediate particle which brings about an $\hat{m}^4$ term in $L^{+-}$.
If the presence of this term can be observed in the dilepton data, 
it would unambiguously\footnote{This observation is subject to 
our assumption that we do not consider heavy particles of spin 3/2 or higher.
In general, an $\hat{m}^4$ dependence would imply that the spin 
of the mediating particle is {\em at least} 1.} 
signal the presence of a spin 1 mediator.
Of course, the size of the coefficient of the $\hat{m}^4$ term
depends on the mass spectrum in the model, but it cannot be vanishingly 
small -- this would require either $z=1$ or $y=1$, which would 
correspondingly close off the $B\to A \ell$ or the $C\to B \ell$ decay,
and the whole decay chain will become unobservable.
One of our general conclusions, therefore, is that the SFVF model\footnote{Or 
more generally, a spin 1 or higher intermediate particle.}, 
if it exists, should be discernible from the 
dilepton data alone. Our numerical results in Fig.~\ref{fig:ll}(f) 
confirm this conclusion -- we see that none of the other five models 
can reproduce the SFVF dilepton data, due to the presence of the $\hat{m}^4$ 
term. Pictorially this can be seen from the fact that the $L^{+-}$
predictions of the $S=1,2,3,4,5$ models in Fig.~\ref{fig:ll}
are always straight lines, while for the $S=6$ model (SFVF)
the prediction is never a straight line, due to the higher 
order $\hat{m}$ dependence.

%
\TABULAR[ht]{|c||c|c|c|c|c|c||c|}{
Result from                          & \multicolumn{6}{c||}{Data originating from model}    & Total No.  \\
fitting to                           &  SFSF  &  FSFS  &  FSFV  &  FVFS  &  FVFV  &  SFVF  & of fakes   \\  \hline \hline
$L^{+-}$ only                        &   5    &    1   &   3    &   2    &   4    &  1     &  10        \\  \hline
$S^{+-}$ only                        &   4    &    1   &   4    &   1    &   2    &  2     &   8        \\  \hline
$D^{+-}$ only                        &   1    &    6   &   6    &   1    &   3    &  1     &  12        \\  \hline\hline
$L^{+-}\oplus S^{+-}$                &   3    &    1   &   2    &   1    &   2    &  1     &   4        \\  \hline
$L^{+-}\oplus D^{+-}$                &   1    &    1   &   3    &   1    &   2    &  1     &   3        \\  \hline
$S^{+-}\oplus D^{+-}$                &   1    &    1   &   4    &   1    &   2    &  1     &   4        \\  \hline\hline
$L^{+-}\oplus S^{+-}\oplus D^{+-}$   &   1    &    1   &   2    &   1    &   2    &  1     &   2        \\  
\hline\hline
}{\label{table:results2} Summary of the results from our spin discrimination analysis. 
Each entry represents the total number of models $n$ which can perfectly fit the data
sets listed in the first column, i.e. each entry $n$ implies an $n$-fold 
model ambiguity of the corresponding data. The last column lists the total number of 
``wrong'' spin configurations allowed by the corresponding data set, 
which was obtained by summing all the $n$'s from the preceding 6 columns
and subtracting 6 for the correct configurations.}
%

Before we move on to the next subsection, where we shall interpret 
our measurements of the $\alpha$, $\beta$ and $\gamma$ parameters,
we briefly summarise the results from the preceding six exercises
in Table~\ref{table:results2}. The table shows the number of
different spin configurations from Table~\ref{table:spins}
which can fit perfectly a given data set ($L^{+-}$, $S^{+-}$, 
$D^{+-}$, or some combination thereof). Since this number
depends on the spin configuration of the input ``data'', 
we show 6 different columns, one for each different spin 
configuration of the ``data''. The last column lists the total 
number of ``wrong'' spin configurations allowed by the corresponding 
data set in all 6 exercises.
This number was obtained simply by summing all the entries 
from the preceding 6 columns and subtracting 6 to exclude the 
correct configurations among them.

While one should be mindful that the number counts exhibited in 
Table~\ref{table:results2} are only valid for the SPS1a parameter choice,
there are still some interesting conclusions which can be drawn from it.
For example, we do not notice any particular pattern in the horizontal direction.
In particular, the discriminating power of the different data sets,
say $L^{+-}$, $S^{+-}$ and $D^{+-}$, varies greatly from model to model.
There are cases where a single distribution works very well, for example
$L^{+-}$ for FSFS and SFVF,  $S^{+-}$ for FSFS and FVFS and
$D^{+-}$ for SFSF, FVFS and SFVF. In all those cases the spin configuration is 
uniquely fixed by studying a single distribution! On the other hand, there 
are also cases where each one of these individual distributions
performs rather poorly, for example $L^{+-}$ for SFSF,
$S^{+-}$ for SFSF and FSFV and $D^{+-}$ for FSFS and FSFV.
In the end, each one of the $L^{+-}$, $S^{+-}$ and $D^{+-}$ distributions, when
considered in isolation, yields on the order of 10 fake spin configurations.
What this simply means is that no single distribution can be 
universally ``better'' than the others.  

Things begin to get more interesting when we start combining information 
gained from 2 or more different distributions. For example, when we combine 
any 2 out of our three observable distributions $L^{+-}$, $S^{+-}$ and $D^{+-}$,
the total number of fake solutions drops down to 3 or 4. Now again, which particular 
pair works better, is a model-dependent issue: $L^{+-}\oplus S^{+-}$
fails for the SFSF model, while $L^{+-}\oplus D^{+-}$ and 
$S^{+-}\oplus D^{+-}$ both fail for the FSFV model.
Finally, combining the information from all three distributions, 
$L^{+-} \oplus S^{+-} \oplus D^{+-}$, we narrow down the remaining spin 
choices even further, but as we saw in Secs.~\ref{sec:FSFS}
and \ref{sec:FVFS}, there are still two cases of exact duplication, 
which are nothing but the ``twin'' spin scenarios 
of Sec.~\ref{sec:degenerate}. Since this duplication is due to
an exact mathematical identity, it will obviously still persist
if we were to repeat our analysis including all the experimental 
realism (backgrounds, resolution, combinatorics, etc.).
In fact, due to the expected imperfections in the real data,
one may get even more duplicate examples, if anything.

\subsection{Measurements of couplings and mixing angles}
\label{sec:couplings}

\TABULAR[t!]{|c||c|c|c||c|}{
Spin   & \multicolumn{4}{c|}{Parameters measured from distribution}   \\
chain  & $L^{+-}$   & $S^{+-}$     & $D^{+-}$             & $L^{+-} \oplus S^{+-} \oplus D^{+-}$     \\  \hline \hline
SFSF   &   $-$      &  $-$         &  $\beta$             & $\beta$                   \\  \hline
FSFS   & $\alpha$   & $\alpha$     &    $-$               & $\alpha$                  \\  \hline
FSFV   & $\alpha$   & $\alpha$     &     $-$              & $\alpha$                  \\  \hline
FVFS   & $\alpha$   & $\alpha$     &  $\beta, \gamma$     & $\alpha, \beta, \gamma$   \\  \hline
FVFV   & $\alpha$   & $\alpha$     &  $\beta, \gamma$     & $\alpha, \beta, \gamma$   \\  \hline
SFVF   & $\alpha$   & $\alpha$     &  $\beta, \gamma$     & $\alpha, \beta, \gamma$   \\  \hline
}{\label{table:abg} Available measurements of the model-dependent parameters $\alpha$,
$\beta$ and $\gamma$ for each of the six spin configuarions.}

Recall that our general method from Sec.~\ref{sec:method}
yields not only a determination of a possible spin chain 
fitting the data, but also a measurement of the model-dependent
$\alpha$, $\beta$ and $\gamma$ parameters from 
eqs.~(\ref{def_alpha2}-\ref{def_gamma2}). 
Even in the spin duplication scenarios found in Secs.~\ref{sec:FSFS} 
and \ref{sec:FVFS}, we still have a certain measurement 
of the $\alpha$, $\beta$ and $\gamma$ coefficients 
for {\em each} of the two allowed spin chains.
This is illustrated in Table~\ref{table:abg}
where we summarize the available measurements of the
$\alpha$, $\beta$ and $\gamma$ parameters in each individual spin case. 
Notice that only in the last three spin cases (FVFS, FVFV and SFVF)
we are able to measure the complete set of all three parameters 
$\alpha$, $\beta$ and $\gamma$. In contrast, for the SFSF
model chain we can only determine $\beta$, while $\alpha$ and $\gamma$
remain unknown. On the other hand, for the FSFS and FSFV chains we can only
determine $\alpha$, while $\beta$ and $\gamma$ remain arbitrary. 
In the remainder of this section we shall discuss the interpretation
of those measurements in terms of the couplings and mixing 
angles of the heavy partners, i.e. we shall relate the measured 
values of $\alpha$ and/or $\beta$ and/or $\gamma$
to the underlying model parameters $f$, $\varphi_a$, 
$\varphi_b$ and $\varphi_c$.

First, let us consider the case where we determine a spin chain to be
one of the following three: FVFS, FVFV or SFVF. Then, as seen from Table~\ref{table:abg},
we will be able to measure the values of all three parameters
$\alpha$, $\beta$ and $\gamma$. If we have correctly determined the
spin chain, these values will be simply the starting SPS1a inputs (\ref{SPSabg}).
Substituting those in eqs.~(\ref{aL}-\ref{cR}), we obtain the two sets of solutions 
discussed at the beginning of Section~\ref{sec:method}:
\begin{eqnarray}
|a_L| = 0, ~ |a_R| = 1, ~ |b_L| = 0, ~ |b_R| = 1, ~
|c_L| = \sqrt{\frac{1}{2}+\frac{0.2}{2f-1}}, ~ 
|c_R| = \sqrt{\frac{1}{2}-\frac{0.2}{2f-1}},
\label{sol1}
\end{eqnarray}
and
\begin{eqnarray}
|a_L| = 1, ~ |a_R| = 0, ~ |b_L| = 1, ~ |b_R| = 0, ~
|c_L| = \sqrt{\frac{1}{2}-\frac{0.2}{2f-1}}, ~ 
|c_R| = \sqrt{\frac{1}{2}+\frac{0.2}{2f-1}},
\label{sol2}
\end{eqnarray}
where the first (second) 
solution corresponds to choosing the upper (lower) sign in eqs.~(\ref{aL}-\ref{cR}).
As expected, we obtain that each set is a one-parameter family of solutions, 
parameterised by the value of the particle-antiparticle ratio $f$. 
The first solution set (\ref{sol1}) reproduces the SPS1a parameter set for 
$f=0.7$, but of course, we would have no way of knowing that $f=0.7$ 
is the correct value of $f$, since we would have to measure $f$ 
independently by some other means.
However, notice that even though we do not know the exact value of $f$ at this point, 
the solutions (\ref{sol1}-\ref{sol2}) unambiguously restrict 
the allowed range for $f$ from (\ref{frangeLHC}) to be
\begin{equation}
0.7\le f \le 1 \ ,
\label{frangeSPS1a}
\end{equation}
which is by itself already an important and useful experimental determination.

Now let us discuss more specifically the case where the data is due 
to an FVFS or an FVFV spin chain ($S=4$ or $S=5$). 
As already explained in Sec.~\ref{sec:degenerate}
and explicitly seen in our examples in Sec.~\ref{sec:FVFS}, here we may 
encounter a second solution for the spin chain, with its own 
measured $\alpha$, $\beta$ and $\gamma$ parameters. We remind 
the reader that when the data comes from an FVFV chain, there is always 
a duplicate spin solution due to an FVFS chain, while if the data comes 
from an FVFS chain, the duplicate FVFV solution exists only if the
conditions (\ref{alpha4_cond}, \ref{gamma4_cond}) are satisfied.
While the duplicate spin chain prevents us from uniquely resolving the 
spin question, the interpretation of its $\alpha$, $\beta$ and $\gamma$ 
parameters can be done in a very similar fashion. Consider our
duplication example from Sec.~\ref{sec:FVFS} where an FVFS ($S=4$) 
spin chain was able to ``fake'' the FVFV ($S=5$) data. 
All three parameters $\alpha$, $\beta$ and $\gamma$
were still uniquely measured but the obtained values were {\em not} 
the starting SPS1a values. Instead, our fitting procedure found
\begin{equation}
\alpha=0.05 , \quad \beta=-0.4, \quad \gamma=-0.02 \, 
\label{abg_dup}
\end{equation}
as shown in Fig.~\ref{fig:ll}(e), \ref{fig:jl_sum}(e) and \ref{fig:jl_diff}(e).
We see that the $\beta$ parameter for the twin spin chain was found 
to be the same as the true $\beta$ parameter of the data ($\beta=-0.4$),
while both the $\alpha$ and $\gamma$ parameters of the twin spin 
case are a factor of 20 smaller than the original inputs (\ref{SPSabg}).
This fact can be easily understood from our general results 
from Sec.~\ref{sec:degenerate} -- the conditions (\ref{alpha4_cond}, \ref{gamma4_cond})
which guarantee the existence of a duplicate solution, relate the values 
of $\alpha$ and $\gamma$ for the two spin chains, and by the same factor
of $\frac{1-2z}{1+2z}\approx \frac{1}{20}$, where we have used
the SPS1a value of $z=0.451$. Just as before, the measurements
(\ref{abg_dup}) translate into a measurement of the effective couplings 
and mixing angles as a function of $f$, up to a two-fold ambiguity.
Substituting (\ref{abg_dup}) in eqs.~(\ref{aL}-\ref{cR}), we obtain
the two solutions
\begin{equation}
|a_L| = 0.69, ~
|a_R| = 0.72, ~ 
|b_L| = 0, ~
|b_R| = 1, ~
|c_L| = \sqrt{\frac{1}{2}+\frac{0.2}{2f-1}},~ 
|c_R| = \sqrt{\frac{1}{2}-\frac{0.2}{2f-1}}, 
\label{sol1_dup}
\end{equation}
or
\begin{equation}
|a_L| = 0.72,~
|a_R| = 0.69,~ 
|b_L| = 1,~ 
|b_R| = 0,~ 
|c_L| = \sqrt{\frac{1}{2}-\frac{0.2}{2f-1}},~
|c_R| = \sqrt{\frac{1}{2}+\frac{0.2}{2f-1}}.
\label{sol2_dup}
\end{equation}
As expected, these solutions exhibit the same $L\leftrightarrow R$ 
symmetry (\ref{LtoR}) as the solutions (\ref{sol1}) and (\ref{sol2})
for the ``correct'' spin configuration. Comparing 
eqs.~(\ref{sol1_dup}, \ref{sol2_dup}) to
eqs.~(\ref{sol1}, \ref{sol2}), we see that 
we obtain the same result for the 
$|b_L|$, $|b_R|$, $|c_L|$ and $|c_R|$ couplings! 
In other words, although it may not be clear 
what is the correct spin chain -- FVFS or FVFV, 
the chirality of the couplings at the quark and 
at the near lepton vertex will be known 
(up to the inescapable two-fold ambiguity due to (\ref{LtoR})).
This can be simply understood by noticing from eqs.~(\ref{bL}-\ref{cR})
that the couplings $|b_L|$, $|b_R|$, $|c_L|$ and $|c_R|$ only depend
on $\alpha$ and $\gamma$ through their {\em ratio}, which is the same
for the correct and the fake spin solution, since $\alpha$ and $\gamma$
are scaled by the same factor $\frac{1-2z}{1+2z}$ (see 
eqs.~(\ref{alpha4_dup},\ref{gamma4_dup}). Just as before, 
for the ``wrong'' spin chain we also obtain a constraint on the 
allowed range of the particle-antiparticle fraction $f$ at the LHC:
\begin{equation}
0.7 \leq  f \leq 1 \ .
\end{equation}
Notice that this is identical to the result (\ref{frangeSPS1a})
for the ``correct'' spin chain, so that the 
experimental determination of the range of the $f$ parameter 
also does not suffer from the duplicate spin ambiguity.

This concludes our discussion of the spin cases where we can 
measure all three parameters $\alpha$, $\beta$ and $\gamma$.
For the remaining three spin chains, only partial information will be available
(see Table~\ref{table:abg}).
For example, in case of SFSF we can only measure the $\beta$ parameter, 
which gives us one relation among $\varphi_b$ and $\tilde\varphi_c$ 
\begin{equation}
\cos2\varphi_b\,\cos2\tilde\varphi_c = -0.4\ ,
\end{equation}
or alternatively, among $\varphi_b$, $\varphi_c$ and $f$:
\begin{equation}
(2f-1)\,\cos2\varphi_b\,\cos2\varphi_c = -0.4\ .
\end{equation}
Unfortunately, we are unable to pin down further the precise values
of $\varphi_b$, $\varphi_c$ and $f$, and furthermore, $\varphi_a$
remains completely unknown.

Similarly, in case of FSFS and FSFV, we can measure the 
$\alpha$ parameter, which gives us a relation between
$\varphi_a$ and $\varphi_b$:
\begin{equation}
\alpha = \cos2\varphi_b\cos2\varphi_a = 1\ .
\end{equation}
Normally, we would not be able to go any further, but the SPS1a 
parameter set is ``lucky'' in the sense that it yields one of the 
two extreme values of $\alpha$ (see Fig.~\ref{fig:alpha}).
In those circumstances, we can determine the actual values of 
$\varphi_a$ and $\varphi_b$, and subsequently, 
$|a_L|$, $|a_R|$, $|b_L|$ and $|b_R|$, up to the usual $L\leftrightarrow R$
ambiguity:
\begin{equation}
\varphi_a=\varphi_b=\frac{\pi}{2} ~\Longrightarrow ~
|a_L| = 0,~
|a_R| = 1,~ 
|b_L| = 0,~ 
|b_R| = 1\ , 
\end{equation}
or
\begin{equation}
\varphi_a=\varphi_b=0 ~\Longrightarrow ~
|a_L| = 1,~
|a_R| = 0,~ 
|b_L| = 1,~ 
|b_R| = 0\ . 
\end{equation}
Unfortunately, in either case, $|c_L|$, $|c_R|$ and $f$ will remain 
unconstrained.

Finally, we briefly comment on the possibility of spin duplication
between FSFS and FSFV discussed in Sec.~\ref{sec:degenerate}
and Sec.~\ref{sec:FSFS}. Here we will also obtain a measurement of the
$\alpha$ parameter for the ``wrong'' spin chain. The 
two $\alpha$ parameters (for the ``wrong'' and for the ``correct'' 
spin configurations) are related according to (\ref{alpha2_dup})
and the analysis for the couplings in the case of the ``wrong'' 
spin chain can be done in complete analogy.

\section{Summary and conclusions}
\label{sec:conclusions}

We conclude by summarizing the main steps of our method 
for measuring spins, couplings and mixing angles of heavy 
partners in cascade decays with missing energy. We shall then
contrast it to other proposals for spin measurements 
in the literature. 

The method involves the following basic steps.
\begin{enumerate}
\item {\em Data preparation.} 
Identify a decay chain of interest which would yield three 
observable SM fermions. (In this paper we considered the 
example of a quark jet followed by two leptons, which is 
commonly encountered in models of supersymmetry and extra dimensions.)
Then form the three observable invariant mass distributions for
each pair of well-defined objects: $\{\ell^+\ell^-\}$, $\{j\ell^+\}$
and $\{j\ell^-\}$. In order to remove the combinatorial ambiguities, 
perform an opposite-flavor subtraction on the leptons and a mixed-event 
subtraction on the jet. Apply final cuts to possibly suppress
any SM and new physics backgrounds. As the end product from
this step one obtains the three ditributions $L^{+-}$, 
$S^{+-}$ and $D^{+-}$ defined in eqs.~(\ref{L+-}-\ref{D+-}).
\item {\em Mass measurements.} This step is optional, since the mass 
measurements can in principle be performed simultaneously with the
spin fits described below. However, in practice we expect that 
the invariant mass distributions would reveal their kinematic 
endpoints rather early on, so that the mass spectrum can be measured 
in advance of the spin determination. At the end of this step one would
know the mass spectrum, i.e. the values of $x$, $y$ and $z$ which
enter the functions ${\mathcal F}$, as well as the kinematic endpoints
$m_{p}^{max}$ which unit normalise our invariant mass variables
(see eqs.~(\ref{def_mhat}) and (\ref{defmjlhat})).
\item {\em Spin measurements.} This step represents the actual spin 
measurement. One tries to fit\footnote{In general, those are 
three-parameter fits for the floating, a priori unknown, coefficients 
$\alpha$, $\beta$ and $\gamma$. However, as discussed in 
Sec.~\ref{sec:method} and illustrated with our numerical examples
in Sec.~\ref{sec:numerics}, one could make use of the fact that
the $L^{+-}$ and $S^{+-}$ distributions only depend on the 
parameter $\alpha$. Thus one could first extract $\alpha$ from
$L^{+-}$ and/or $S^{+-}$, and then use this value to fit 
$D^{+-}$ for $\beta$ and $\gamma$, 
as shown in Appendix \ref{app:params}.} 
the data for the $L^{+-}$, $S^{+-}$ and $D^{+-}$
distributions obtained in Step 1 with the theoretical predictions
(\ref{L+-}-\ref{D+-}), for {\em each} value of $S$, i.e. for each 
set of allowed spin configurations for particles $D$, $C$, $B$ and $A$
(see Table \ref{table:spins}). If the fit is good, that particular 
spin chain is ruled in, while if the fit is bad, that particular 
spin chain will be ruled out. Our expectations for the generic outcome
of this exercise are summarised in Table~\ref{table:results}.
When using the data from all three distributions
$L^{+-}$, $S^{+-}$ and $D^{+-}$, we expect that the fits will be able 
to rule out all but the correct spin configuration.
The only exceptions are the spin duplication cases discussed in 
Sec.~\ref{sec:degenerate}, when one may end up with at most two
spin chain alternatives. 
\item {\em Measurements of couplings and mixing angles.} In this step 
one uses any available best-fit values for $\alpha$, $\beta$ and $\gamma$
obtained in the previous step, and determines the couplings
$|a_L|$, $|a_R|$, $|b_L|$, $|b_R|$, $|c_L|$ and $|c_R|$ from 
eqs.~(\ref{aL}-\ref{cR}). There will be two different solutions 
due to the $L\leftrightarrow R$ symmetry, as discussed in 
Sec.~\ref{sec:method} and illustrated with some examples in 
Sec.~\ref{sec:couplings}. In addition, eq.~(\ref{frangeLHC})
provides a restriction on the allowed range of values for the 
particle-antiparticle fraction $f$ at the LHC.
\end{enumerate}
\TABULAR[t]{|c||c|c|c|c|c|c|}{
Data   & \multicolumn{6}{c|}{Can this data be fitted by model} \\ 
from   &  SFSF  &  FSFS  &  FSFV  &  FVFS  &  FVFV  &  SFVF  \\  \hline \hline
SFSF   &  yes   &  no    &  no    &  no    &  no    &  no    \\  \hline
FSFS   &  no    &  yes   &  maybe &  no    &  no    &  no    \\  \hline
FSFV   &  no    &  yes   &  yes   &  no    &  no    &  no    \\  \hline
FVFS   &  no    &  no    &  no    &  yes   &  maybe &  no    \\  \hline
FVFV   &  no    &  no    &  no    &  yes   &  yes   &  no    \\  \hline
SFVF   &  no    &  no    &  no    &  no    &  no    &  yes   \\  \hline
}{\label{table:results} Expected outcomes from our spin discrimination analysis,
barring numerical accidents due to very special mass spectra.
The two cases labelled ``maybe'' correspond to the potential 
confusion of an FSFS (FVFS) chain with an FSFV (FVFV) chain,
which occurs only for a certain range of the model-dependent parameters  --
see eqs.~(\ref{alpha2_cond}) and (\ref{alpha4_cond}, \ref{gamma4_cond}).
}

Having summarised the main steps of our method, we are ready to compare
it to other approaches for spin measurements which already exist in 
the literature. In principle, no single method is universally applicable, 
therefore the availability of different and complementary techniques is an
important virtue. Which method ends up being most successful in practice,
will depend on the specific new physics scenario that we may encounter.
With those caveats, we should point out some features of our method which 
are likely to make it relevant and successful, if a missing energy signal of 
new physics is seen at the LHC and/or the Tevatron. 
\begin{itemize}
\item Many of the existing techniques for spin determinations
(see, for example, \cite{Battaglia:2005zf,Battaglia:2005ma,Buckley:2007th,Buckley:2008pp})
have been originally developed in the context of lepton colliders, 
where the total center of mass energy in each event is known. 
Consequently, at hadron colliders, those methods are applicable 
only if the events can be fully reconstructed.
In new physics scenarios with dark matter WIMPs, this appears to be rather
challenging, since there are {\em two} invisible WIMP particles
escaping the detector. In some special circumstances, where two
sufficiently long decay chains can be identified in the event, 
full reconstruction might be possible \cite{Cheng:2007xv,Nojiri:2007pq,Cheng:2008mg}, 
but in any case, this appears to require very large data samples.
In contrast, our method relies on invariant mass distributions, which are 
frame-independent, and we do not need to have the event fully reconstructed.
Furthermore, the event reconstruction techniques currently being discussed
rely on the pair-production of two heavy particles, {\em both of which}
decay visibly to the lightest WIMP. Our method, on the other hand, does not
require the presence of two separate decay chains in the event, and
can be in principle also applied to the associated production
of a WIMP with only one other heavy partner.
\item The invariant mass distributions $L^{+-}$, $S^{+-}$ and $D^{+-}$
that we propose to analyse, are the basic starting point for any
precision study of new physics parameters. In the past they have been
extensively discussed in relation to mass measurements, and we now
simply propose to fully analyse them for the encoded spin information as well.
\item One major advantage of our method in comparison to various 
event counting techniques \cite{Datta:2005vx,Meade:2006dw,Datta:2007xy,Kane:2008kw}
is that we do not need to know anything about a number of additional and 
a priori also unknown quantities such as the production cross-sections
for the different parton-level initial states, the branching fractions, 
the experimental efficiencies, etc. Indeed, our method in essence only uses unit-normalised 
distributions, and is not affected by any of these additional variables.
\item The previous three advantages are common to all studies which have
relied exclusively on invariant mass distributions for spin determinations
\cite{Barr:2004ze,Smillie:2005ar,Athanasiou:2006ef,Athanasiou:2006hv,Datta:2005zs,Alves:2006df,Wang:2006hk,
SA:2006jm,Smillie:2006cd,Kong:2006pi,Kilic:2007zk,Alves:2007xt,Csaki:2007xm}.  
In comparison to those works, the main advantage of our approach is 
that it is completely general and model-independent, in particular 
we make no a priori assumptions about the type of couplings 
in each vertex of Fig.~\ref{fig:ABCD}, or about the particle-antiparticle 
fraction $f$. As a result, we were actually able to come up with
{\em measurements} of certain combinations of those couplings 
and the $f$ parameter (see Secs.~\ref{sec:method} and \ref{sec:couplings}). 
\end{itemize}

In conclusion, we reiterate that 
our goal in this paper was simply to present the basic idea of our method,
and demonstrate that it can work as a matter of principle.
Therefore in our analysis in Sec.~\ref{sec:numerics} 
we did not include any realistic detector simulation,
backgrounds (SM and combinatorial) etc. All of these factors will be 
investigated in a future publication \cite{BKMP}.

\bigskip

\acknowledgments
This work is supported in part by a US Department of Energy 
grant DE-FG02-97ER41029. Fermilab is operated by Fermi Research Alliance, LLC under
Contract No. DE-AC02-07CH11359 with the U.S. Department of Energy.

\newpage
\appendix
\section{Appendix: \ The basis functions ${\mathcal F}_{S;IJ}^{(p)}$}
\label{app:FIJ}
\allowdisplaybreaks
\renewcommand{\theequation}{A.\arabic{equation}}
\setcounter{equation}{0}

The basis functions ${\mathcal F}_{S;IJ}^{(j\ell_n)}(\hat{m}^2;x,y,z)$
are listed in Table~\ref{table:Fqln} and the basis functions
${\mathcal F}_{S;IJ}^{(\ell\ell)}(\hat{m}^2;x,y,z)$
are given in Table~\ref{table:Fll}. Below we explicitly show the 
remaining basis functions ${\mathcal F}_{S;IJ}^{(j\ell_f)}(\hat{m}^2;x,y,z)$:

\vskip 1.0cm
{\bf SFSF} ($S=1$)
\begin{eqnarray}
\mathcal{F}_{1;11}^{(j\ell_{f})}(\hat{m}^{2};x,y,z)
&=&  \mathcal{F}_{1;12}^{(j\ell_{f})}(\hat{m}^{2};x,y,z) 
= \frac{-2}{(1-y)^2} \left \{\begin{array}{ll}
(1-y)+\log{y}  &\textrm{if $\hat{m}^2\le y$}\\[2mm]
1-\hm+\log{\hm}&\textrm{if $y \le \hm \le 1$}\\[2mm]
0 &\textrm{if  $ \hm \ge 1 $}
\end{array}\right.  \qquad \\[3mm]
\mathcal{F}_{1;21}^{(j\ell_{f})}(\hat{m}^{2};x,y,z)
&=&  \mathcal{F}_{1;22}^{(j\ell_{f})}(\hat{m}^{2};x,y,z) 
= \frac{2}{(1-y)^2} \left \{\begin{array}{ll}
(1-y)+y\log{y}  &\textrm{if $\hat{m}^2\le y$}\\[2mm]
1-\hm+y\log{\hm}&\textrm{if $y \le \hm \le 1$}\\[2mm]
0 &\textrm{if  $ \hm \ge 1 $}
\end{array}\right. \qquad 
\end{eqnarray}

{\bf FSFS} ($S=2$)
\begin{eqnarray}
\mathcal{F}_{2;11}^{(j\ell_{f})}(\hat{m}^{2};x,y,z)
&=&  \mathcal{F}_{2;21}^{(j\ell_{f})}(\hat{m}^{2};x,y,z) 
= \frac{-2}{(1-y)^2} \left \{\begin{array}{ll}
(1-y)+\log{y}  &\textrm{if $\hat{m}^2\le y$}\\[2mm]
1-\hm+\log{\hm}&\textrm{if $y \le \hm \le 1$}\\[2mm]
0 &\textrm{if  $ \hm \ge 1 $}
\end{array}\right. \qquad \\[3mm]
\mathcal{F}_{2;12}^{(j\ell_{f})}(\hat{m}^{2};x,y,z)
& =&  \mathcal{F}_{2;22}^{(j\ell_{f})}(\hat{m}^{2};x,y,z) 
= \frac{2}{(1-y)^2} \left \{\begin{array}{ll}
(1-y)+y\log{y}  &\textrm{if $\hat{m}^2\le y$}\\[2mm]
1-\hm+y\log{\hm}&\textrm{if $y \le \hm \le 1$}\\[2mm]
0 &\textrm{if  $ \hm \ge 1 $}
\end{array}\right. \qquad
\end{eqnarray}

{\bf FSFV} ($S=3$)
\begin{eqnarray}
& &  \mathcal{F}_{3;11}^{(j\ell_{f})}(\hat{m}^{2};x,y,z)
 =  \mathcal{F}_{3;21}^{(j\ell_{f})}(\hat{m}^{2};x,y,z) \nonumber \\
&& =\frac{-2}{(1-y)^2(1+2z)} \left \{\begin{array}{ll}
(1-y)(1-2z)+(1-2yz)\log{y}  &\textrm{if $\hat{m}^2\le y$}\\[2mm]
(1-\hm)(1-2z)+(1-2yz)\log{\hm}&\textrm{if $y \le \hm \le 1$}\\[2mm]
0 &\textrm{if  $ \hm \ge 1 $}
\end{array}\right.\\[3mm]
& &  \mathcal{F}_{3;12}^{(j\ell_{f})}(\hat{m}^{2};x,y,z)
 =  \mathcal{F}_{3;22}^{(j\ell_{f})}(\hat{m}^{2};x,y,z) \nonumber \\
&& = \frac{2}{(1-y)^2(1+2z)}\left \{\begin{array}{ll}
(1-y)(1-2z)+(y-2z)\log{y}  &\textrm{if $\hat{m}^2\le y$}\\[2mm]
(1-\hm)(1-2z)+(y-2z)\log{\hm}&\textrm{if $y \le \hm \le 1$}\\[2mm]
0 &\textrm{if  $ \hm \ge 1 $}
\end{array}\right.
\end{eqnarray}

\newpage
{\bf FVFS} ($S=4$)
\begin{eqnarray}
& & \mathcal{F}_{4;11}^{(j\ell_{f})}(\hat{m}^{2};x,y,z) \nonumber \\
&& =\frac{6}{(1+2x)(2+y)(1-y)^2} \left \{\begin{array}{ll}
(1-y)[4x-y-4\hm(2-3x)]\\
+[(-1+4x)y+4\hm\{1-(2+y)(1-x)\}]\log{y} &\textrm{if $\hat{m}^2\le y$}\\[2mm]
(1-\hm)[4x(2y+1)-5y-4\hm(1-x)]\\
+[(-1+4x)y+4\hm\{1-(2+y)(1-x)\}]\log{\hm} &\textrm{if $y \le \hm \le 1$}\\[2mm]
0 &\textrm{if  $ \hm \ge 1 $}
\end{array}\right.\\[3mm]
& & \mathcal{F}_{4;12}^{(j\ell_{f})}(\hat{m}^{2};x,y,z) \nonumber \\
&& =\frac{6}{(1+2x)(2+y)(1-y)^2} \left \{\begin{array}{ll}
(1-y)[2+3y-2x(5+y)+4\hm(2-3x)]\\
+[y(4+y)-4x(1+2y)-4\hm\{1-(2+y)(1-x)\}]\log{y} &\textrm{if $\hat{m}^2\le y$}\\[2mm]
(1-\hm)[2+9y-2x(5+6y)+2\hm(1-x)]\\
+[y(4+y)-4x(1+2y)-4\hm\{1-(2+y)(1-x)\}]\log{\hm} 
&\textrm{if $y \le \hm \le 1$}\\[2mm]
0 &\textrm{if  $ \hm \ge 1 $}
\end{array}\right.\\[3mm]
& & \mathcal{F}_{4;21}^{(j\ell_{f})}(\hat{m}^{2};x,y,z) \nonumber \\
&& =\frac{6}{(1+2x)(2+y)(1-y)^2} \left \{\begin{array}{ll}
(1-y)[-y-4\hm(2-x)]\\
-[y+4\hm\{1+y(1-x)\} ]\log{y} &\textrm{if $\hat{m}^2\le y$}\\[2mm]
(1-\hm)[-5y-4\hm(1-x)]\\
-[y+4\hm\{1+y(1-x)\} ]\log{\hm} 
&\textrm{if $y \le \hm \le 1$}\\[2mm]
0 &\textrm{if  $ \hm \ge 1 $}
\end{array}\right.\\[3mm]
& & \mathcal{F}_{4;22}^{(j\ell_{f})}(\hat{m}^{2};x,y,z) \nonumber \\
&& =\frac{6}{(1+2x)(2+y)(1-y)^2} \left \{\begin{array}{ll}
(1-y)[2+3y+2x(1-y)+4\hm(2-x)]\\
+[y(4+y)+4\hm\{1+y(1-x)\}]\log{y} &\textrm{if $\hat{m}^2\le y$}\\[2mm]
(1-\hm)[2+9y+2x(1-2y)+2\hm(1-x)  ]\\
+[ y(4+y)+4\hm\{1+y(1-x)\}]\log{\hm} 
&\textrm{if $y \le \hm \le 1$}\\[2mm]
0 &\textrm{if  $ \hm \ge 1 $}
\end{array}\right.
\end{eqnarray}

\newpage
\textbf{FVFV} ($S=5$)
\begin{eqnarray}
\mathcal{F}_{5;11}^{(j\ell_{f})}(\hat{m}^{2};x,y,z)
&=&\frac{6}{(1+2x)(2+y)(1-y)^2 (1+2z)}\times\nonumber \\
&\times&\left \{\begin{array}{ll}
(1-y)[4x-y+2z\{2+3y-2x(5+y)\}\\
 -4\hm(2-3x)(1-2z) ]-[y-2yz(4+y)+4x\{2z-y(1-4z)\}\\
+4\hm\{1+y-x(2+y)\}(1-2z)]\log{y} &\textrm{if $\hat{m}^2\le y$}\\[2mm]
(1-\hm)[4x\{1-5z+2y(1-3z)\}-5y+2z(2+9y)\\
-4\hm(1-x)(1-z)]-[y-2yz(4+y)+4x\{2z-y(1-4z)\}\\
+4\hm\{1+y-x(2+y)\}(1-2z)]\log{\hm} 
&\textrm{if $y \le \hm \le 1$}\\[2mm]
0 &\textrm{if  $ \hm \ge 1 $}
\end{array}\right.\\[3mm]
\mathcal{F}_{5;12}^{(j\ell_{f})}(\hat{m}^{2};x,y,z)
&=&\frac{6}{(1+2x)(2+y)(1-y)^2 (1+2z)}\times\nonumber \\
&\times&\left \{\begin{array}{ll}
(1-y)[2+3y-2x(5+y)+2(4x-y)z\\
+4\hm(2-3x)(1-2z)  ]-[4x\{1+2y(1-z)\}-y(4+y-2z)\\
-4\hm\{1+y-x(2+y)\}(1-2z) ]\log{y} &\textrm{if $\hat{m}^2\le y$}\\[2mm]
(1-\hm)[2-2x\{5-4z+2y(3-4z)\}+y(9-10z)\\
+2\hm(1-x)(1-4z)]-[4x\{1+2y(1-z)\}-y(4+y-2z)\\
-4\hm\{1+y-x(2+y)\}(1-2z) ]\log{\hm} 
&\textrm{if $y \le \hm \le 1$}\\[2mm]
0 &\textrm{if  $ \hm \ge 1 $}
\end{array}\right.\\[3mm]
\mathcal{F}_{5;21}^{(j\ell_{f})}(\hat{m}^{2};x,y,z)
&=&\frac{6}{(1+2x)(2+y)(1-y)^2 (1+2z)}\times\nonumber \\
&\times&\left \{\begin{array}{ll}
(1-y)[-y+2\{2+2x(1-y)+3y\}z-4\hm(2-x)(1-2z)  ]\\
-[y\{1-2(4+y)z\}+4\hm(1+y-xy)(1-2z) ]\log{y} &\textrm{if $\hat{m}^2\le y$}\\[2mm]
(1-\hm)[4(1+x)z-y\{5-2(9-4x)z\}-4\hm(1-x)(1-z)  ]\\
-[y\{1-2(4+y)z\}+4\hm(1+y-xy)(1-2z) ]\log{\hm} 
&\textrm{if $y \le \hm \le 1$}\\[2mm]
0 &\textrm{if  $ \hm \ge 1 $}
\end{array}\right.\\[3mm]
\mathcal{F}_{5;22}^{(j\ell_{f})}(\hat{m}^{2};x,y,z)
&=&\frac{6}{(1+2x)(2+y)(1-y)^2 (1+2z)}\times\nonumber \\
&\times& \left \{\begin{array}{ll}
(1-y)[2+2x(1-y)+y(3-2z)+4\hm(2-x)(1-2z)  ]\\
+[y(4+y-2z)+4\hm(1+y-xy)(1-2z) ]\log{y} &\textrm{if $\hat{m}^2\le y$}\\[2mm]
(1-\hm)[ 2+2x(1-2y)+y(9-10z)+2\hm(1-x)(1-4z) ]\\
+[y(4+y-2z)+4\hm(1+y-xy)(1-2z) ]\log{\hm} 
&\textrm{if $y \le \hm \le 1$}\\[2mm]
0 &\textrm{if  $ \hm \ge 1 $}
\end{array}\right.
\end{eqnarray}

\newpage
\textbf{SFVF} ($S=6$)
\begin{eqnarray}
&&\mathcal{F}_{6;11}^{(j\ell_{f})}(\hat{m}^{2};x,y,z)\nonumber \\
&=&\frac{6}{(1+2y)(1-y)^2 (2+z)} 
\left \{\begin{array}{ll}
(1-y)[2-3z-2y(1+z)+4\hm(1-2z)]\\
-[z(1+4y)-4\hm(1-z-yz) ]\log{y} &\textrm{if $\hat{m}^2\le y$}\\[2mm]
(1-\hm)[2-3z-8yz+2\hm(1-z) ]\\
-[z(1+4y)-4\hm(1-z-yz) ]\log{\hm} 
&\textrm{if $y \le \hm \le 1$}\\[2mm]
0 &\textrm{if  $ \hm \ge 1 $}
\end{array}\right.\\[3mm]
&&\mathcal{F}_{6;12}^{(j\ell_{f})}(\hat{m}^{2};x,y,z) \nonumber \\
&=&\frac{6}{(1+2y)(1-y)^2 (2+z)}
\left \{\begin{array}{ll}
(1-y)[2-3z+2y(5-z)+4\hm(3-2z) ]\\
-[z(1+4y)-4y(2+y)-4\hm(1+2y-z-yz) ]\log{y} &\textrm{if $\hat{m}^2\le y$}\\[2mm]
(1-\hm)[2-3z+4y(5-2z)+2\hm(1-z)  ]\\
-[z(1+4y)-4y(2+y)-4\hm(1+2y-z-yz) ]\log{\hm} 
&\textrm{if $y \le \hm \le 1$}\\[2mm]
0 &\textrm{if  $ \hm \ge 1 $}
\end{array}\right.\\[3mm]
&&\mathcal{F}_{6;21}^{(j\ell_{f})}(\hat{m}^{2};x,y,z) \nonumber \\
&=&\frac{6}{(1+2y)(1-y)^2 (2+z)}\left \{\begin{array}{ll}
(1-y)[z-4\hm(1-2z)]\\
+[yz-4\hm(1-z-yz) ]\log{y} &\textrm{if $\hat{m}^2\le y$}\\[2mm]
(1-\hm)[z(1+4y)-4\hm(1-z)]\\
+[yz-4\hm(1-z-yz) ]\log{\hm} 
&\textrm{if $y \le \hm \le 1$}\\[2mm]
0 &\textrm{if  $ \hm \ge 1 $}
\end{array}\right.\\[3mm]
&&\mathcal{F}_{6;22}^{(j\ell_{f})}(\hat{m}^{2};x,y,z) \nonumber \\
&=&\frac{6}{(1+2y)(1-y)^2 (2+z)}\left \{\begin{array}{ll}
(1-y)[z-4y-4\hm(3-2z)]\\
-[y(4-z)+4\hm(1+2y-z-yz)]\log{y} &\textrm{if $\hat{m}^2\le y$}\\[2mm]
(1-\hm)[z-4y(3-z)-4\hm(1-z) ]\\
-[y(4-z)+4\hm(1+2y-z-yz)]\log{\hm} 
&\textrm{if $y \le \hm \le 1$}\\[2mm]
0 &\textrm{if  $ \hm \ge 1 $}
\end{array}\right.
\end{eqnarray}

\begin{landscape}
\TABULAR[ht]{c|c||c|c}{
\hline
$S$ &
Spins  & $\mathcal{F}_{S;11}^{(j\ell_{n})}(\hat{m}^{2};x,y,z)=\mathcal{F}_{S;12}^{(j\ell_{n})}(\hat{m}^{2};x,y,z)$ 
       & $\mathcal{F}_{S;21}^{(j\ell_{n})}(\hat{m}^{2};x,y,z)=\mathcal{F}_{S;22}^{(j\ell_{n})}(\hat{m}^{2};x,y,z)$\\\hline\hline
1& $SFSF$ & $2\hat{m}^{2}$ & $2(1-\hat{m}^{2})$\\\hline
2& $FSFS$ & $1$ & $1$\\\hline
3& $FSFV$ & $1$ & $1$\\\hline
4& $FVFS$ & $\frac{3}{(1+2x)(2+y)}{ \{y   +4(1-y+xy)\hat{m}^{2}   -4(1-x)(1-y)\hat{m}^{4}\}}$ 
          & $\frac{3}{(1+2x)(2+y)}{ \{4x+y+4(1-2x-y+xy)\hat{m}^{2}-4(1-x)(1-y)\hat{m}^{4}\}}$\\\hline
5& $FVFV$ & $\frac{3}{(1+2x)(2+y)}{ \{y   +4(1-y+xy)\hat{m}^{2}   -4(1-x)(1-y)\hat{m}^{4}\}}$ 
          & $\frac{3}{(1+2x)(2+y)}{ \{4x+y+4(1-2x-y+xy)\hat{m}^{2}-4(1-x)(1-y)\hat{m}^{4}\}}$\\\hline
6& $SFVF$ & $\frac{2}{1+2y}\{2y+(1-2y)\hat{m}^{2}\}$ 
          & $\frac{2}{1+2y}\{1 -(1-2y)\hat{m}^{2}\}$\\\hline
}{\label{table:Fqln} Basis functions for the $j\ell_n$ invariant mass distribution. 
}
\TABULAR[ht]{c|c||c|c}{
\hline
$S$ & Spins   
       & $\mathcal{F}_{S;11}^{(\ell\ell)}(\hat{m}^{2};x,y,z)=\mathcal{F}_{S;21}^{(\ell\ell)}(\hat{m}^{2};x,y,z)$ 
       & $\mathcal{F}_{S;12}^{(\ell\ell)}(\hat{m}^{2};x,y,z)=\mathcal{F}_{S;22}^{(\ell\ell)}(\hat{m}^{2};x,y,z)$
\\\hline\hline
1& $SFSF$ & $1$ 
          & $1$ \\ \hline
2& $FSFS$ & $2(1-\hat{m}^{2})$ 
          & $2\hat{m}^{2}$\\\hline
3& $FSFV$ & $\frac{2}{1+2z}\{ 1-(1-2z)\hat{m}^{2}\}$ 
          & $\frac{2}{1+2z}\{2z+(1-2z)\hat{m}^{2}\}$\\\hline
4& $FVFS$ & $\frac{2}{2+y}\{y+(2-y)\hat{m}^{2}\}$ 
          & $\frac{2}{2+y}\{2-(2-y)\hat{m}^{2}\}$\\\hline
5& $FVFV$ & $\frac{2}{(2+y)(1+2z)}\{ y+4z+(2-y)(1-2z)\hat{m}^{2}\}$ 
          & $\frac{2}{(2+y)(1+2z)}\{2+2yz-(2-y)(1-2z)\hat{m}^{2}\}$\\\hline
6& $SFVF$ & $\frac{3}{(1+2y)(2+z)}{\{4y+z+4(1-2y-z+yz)\hat{m}^{2}-4(1-y)(1-z)\hat{m}^{4}\}}$ 
          & $\frac{3}{(1+2y)(2+z)}{\{z   +4(1-z+yz)\hat{m}^{2}   -4(1-y)(1-z)\hat{m}^{4}\}}$\\\hline
}{\label{table:Fll} Basis functions for the dilepton invariant mass distribution. 
}
\end{landscape}

\section{Appendix: \ The basis functions ${\mathcal F}_{S;\alpha}^{(p)}$,
${\mathcal F}_{S;\beta}^{(p)}$, ${\mathcal F}_{S;\gamma}^{(p)}$ and 
${\mathcal F}_{S;\delta}^{(p)}$}
\label{app:Falpha}
\renewcommand{\theequation}{B.\arabic{equation}}
\setcounter{equation}{0}

The basis functions 
${\mathcal F}_{S;\alpha}^{(\ell\ell)}$,
${\mathcal F}_{S;\beta }^{(\ell\ell)}$,
${\mathcal F}_{S;\gamma}^{(\ell\ell)}$ and
${\mathcal F}_{S;\delta}^{(\ell\ell)}$
are listed in Table~\ref{table:Fll2}. The basis functions
${\mathcal F}_{S;\alpha}^{(j\ell_n)}$, 
${\mathcal F}_{S;\beta }^{(j\ell_n)}$, 
${\mathcal F}_{S;\gamma}^{(j\ell_n)}$ and
${\mathcal F}_{S;\delta}^{(j\ell_n)}$
are given in Table~\ref{table:Fqln2}. Below we explicitly show the 
remaining basis functions 
${\mathcal F}_{S;\alpha}^{(j\ell_f)}$,
${\mathcal F}_{S;\beta }^{(j\ell_f)}$,
${\mathcal F}_{S;\gamma}^{(j\ell_f)}$ and
${\mathcal F}_{S;\delta}^{(j\ell_f)}$:

{\bf SFSF} ($S=1$)
\begin{eqnarray}
\mathcal{F}_{1;\alpha}^{(j\ell_{f})}(\hat{m}^{2};x,y,z)
&=&  \mathcal{F}_{1;\gamma}^{(j\ell_{f})}(\hat{m}^{2};x,y,z) =0  \\[3mm] 
\mathcal{F}_{1;\beta}^{(j\ell_{f})}(\hat{m}^{2};x,y,z)
&=&\frac{-1}{(1-y)^2} \left \{\begin{array}{ll}
2(1-y)+(1+y)\log{y}  &\textrm{if $\hat{m}^2\le y$}\\[2mm]
2(1-\hm)+(1+y)\log{\hm}&\textrm{if $y \le \hm \le 1$}\\[2mm]
0 &\textrm{if  $ \hm \ge 1 $}
\end{array}\right.\\[3mm]
\mathcal{F}_{1;\delta}^{(j\ell_{f})}(\hat{m}^{2};x,y,z)
&=&\frac{-1}{(1-y)^2} \left \{\begin{array}{ll}
(1-y)\log{y}  &\textrm{if $\hat{m}^2\le y$}\\[2mm]
(1-y)\log{\hm}&\textrm{if $y \le \hm \le 1$}\\[2mm]
0 &\textrm{if  $ \hm \ge 1 $}
\end{array}\right.
\end{eqnarray}

{\bf FSFS} ($S=2$)
\begin{eqnarray}
\mathcal{F}_{2;\alpha}^{(j\ell_{f})}(\hat{m}^{2};x,y,z)
&=&\frac{-1}{(1-y)^2} \left \{\begin{array}{ll}
2(1-y)+(1+y)\log{y}  &\textrm{if $\hat{m}^2\le y$}\\[2mm]
2(1-\hm)+(1+y)\log{\hm}&\textrm{if $y \le \hm \le 1$}\\[2mm]
0 &\textrm{if  $ \hm \ge 1 $}
\end{array}\right. \\[3mm]
\mathcal{F}_{2;\beta}^{(j\ell_{f})}(\hat{m}^{2};x,y,z)
&=&\mathcal{F}_{2;\gamma}^{(j\ell_{f})}(\hat{m}^{2};x,y,z) =0  \\ [3mm]
\mathcal{F}_{2;\delta}^{(j\ell_{f})}(\hat{m}^{2};x,y,z)
&=&\frac{-1}{(1-y)^2} \left \{\begin{array}{ll}
(1-y)\log{y}  &\textrm{if $\hat{m}^2\le y$}\\[2mm]
(1-y)\log{\hm}&\textrm{if $y \le \hm \le 1$}\\[2mm]
0 &\textrm{if  $ \hm \ge 1 $}
\end{array}\right.
\end{eqnarray}

{\bf FSFV} ($S=3$)
\begin{eqnarray}
\mathcal{F}_{3;\alpha}^{(j\ell_{f})}(\hat{m}^{2};x,y,z)
&=& \frac{-1}{(1-y)^2} \frac{1-2z}{1+2z} 
\left \{\begin{array}{ll}
2(1-y)+(1+y)\log{y}  &\textrm{if $\hat{m}^2\le y$}\\[2mm]
2(1-\hm)+(1+y)\log{\hm}&\textrm{if $y \le \hm \le 1$}\\[2mm]
0 &\textrm{if  $ \hm \ge 1 $}
\end{array}\right. \\[3mm]
\mathcal{F}_{3;\beta}^{(j\ell_{f})}(\hat{m}^{2};x,y,z)
&=&\mathcal{F}_{3;\gamma}^{(j\ell_{f})}(\hat{m}^{2};x,y,z) =0  \\ [3mm]
\mathcal{F}_{3;\delta}^{(j\ell_{f})}(\hat{m}^{2};x,y,z)
&=&\frac{-1}{(1-y)^2} \left \{\begin{array}{ll}
(1-y)\log{y}  &\textrm{if $\hat{m}^2\le y$}\\[2mm]
(1-y)\log{\hm}&\textrm{if $y \le \hm \le 1$}\\[2mm]
0 &\textrm{if  $ \hm \ge 1 $}
\end{array}\right.
\end{eqnarray}

\newpage
{\bf FVFS} ($S=4$)
\begin{eqnarray}
&&\mathcal{F}_{4;\alpha}^{(j\ell_{f})}(\hat{m}^{2};x,y,z) \nonumber \\
&=&\frac{ 3 }{(1+2x)(2+y)(1-y)^2 } \left \{\begin{array}{ll}
(1-y)[-2(1+2y)+2x(3+y)-16\hm(1-x) ] \\ 
-[y(5+y)-2x(1+3y)+8\hm(1-x)(1+y) ]\log{y}  &\textrm{if $\hat{m}^2\le y$}\\[2mm]
(1-\hm)[-2(1+7y)+6x(1+2y)-6\hm(1-x) ]\\ 
-[y(5+y)-2x(1+3y)+8\hm(1-x)(1+y) ]\log{\hm}&\textrm{if $y \le \hm \le 1$}\\[2mm]
0 &\textrm{if  $ \hm \ge 1 $}
\end{array}\right.\\[3mm]
&&\mathcal{F}_{4;\beta}^{(j\ell_{f})}(\hat{m}^{2};x,y,z) \nonumber \\
&=&\frac{-6x}{(1+2x)(2+y)(1-y)^2 } \left \{\begin{array}{ll}
2(1-y)+(1+y)\log{y}  &\textrm{if $\hat{m}^2\le y$}\\[2mm]
2(1-\hm)+(1+y)\log{\hm}&\textrm{if $y \le \hm \le 1$}\\[2mm]
0 &\textrm{if  $ \hm \ge 1 $}
\end{array}\right.\\[3mm]
&&\mathcal{F}_{4;\gamma}^{(j\ell_{f})}(\hat{m}^{2};x,y,z) \nonumber \\
&=&\frac{6x}{(1+2x)(2+y)(1-y)^2 } \left \{\begin{array}{ll}
4(1-y)(1+\hm)+ 
[(1+3y)+4\hm ]\log{y}  &\textrm{if $\hat{m}^2\le y$}\\[2mm]
4(1-\hm)(1+y)+
[(1+3y)+4\hm ]\log{\hm}&\textrm{if $y \le \hm \le 1$}\\[2mm]
0 &\textrm{if  $ \hm \ge 1 $}
\end{array}\right.\\[3mm]
&&\mathcal{F}_{4;\delta}^{(j\ell_{f})}(\hat{m}^{2};x,y,z) \nonumber \\
&=&\frac{3}{(1+2x)(2+y)(1-y)^2 } \left \{\begin{array}{ll}
2(1-y)(1+y)(1-x)\\
+ [ -2x(1+y)+y(3+y)]\log{y}  &\textrm{if $\hat{m}^2\le y$}\\[2mm]
2(1-\hm)(1-x)\{(1+2y)-\hm\}\\
+[ -2x(1+y)+y(3+y)]\log{\hm}&\textrm{if $y \le \hm \le 1$}\\[2mm]
0 &\textrm{if  $ \hm \ge 1 $}
\end{array}\right.
\end{eqnarray}

\newpage
{\bf FVFV} ($S=5$)
\begin{eqnarray}
&&\mathcal{F}_{5;\alpha}^{(j\ell_{f})}(\hat{m}^{2};x,y,z)
\nonumber\\
&=&\frac{ 3}{(1+2x)(2+y)(1-y)^2} \frac{1-2z}{1+2z} 
\left \{\begin{array}{ll}
(1-y)[-2(1+2y)+2x(3+y)-16\hm(1-x) ] \\ 
-[y(5+y)-2x(1+3y)\\
+8\hm(1-x)(1+y) ]\log{y}  &\textrm{if $\hat{m}^2\le y$}\\[2mm]
(1-\hm)[-2(1+7y)+6x(1+2y)-6\hm(1-x) ]\\ 
-[y(5+y)-2x(1+3y)\\
+8\hm(1-x)(1+y) ]\log{\hm}&\textrm{if $y \le \hm \le 1$}\\[2mm]
0 &\textrm{if  $ \hm \ge 1 $}
\end{array}\right.\\[3mm]
&&\mathcal{F}_{5;\beta}^{(j\ell_{f})}(\hat{m}^{2};x,y,z)
\nonumber \\
&=&\frac{-6x}{(1+2x)(2+y)(1-y)^2 } \left \{\begin{array}{ll}
2(1-y)+(1+y)\log{y}  &\textrm{if $\hat{m}^2\le y$}\\[2mm]
2(1-\hm)+(1+y)\log{\hm}&\textrm{if $y \le \hm \le 1$}\\[2mm]
0 &\textrm{if  $ \hm \ge 1 $}
\end{array}\right.\\[3mm]
&&\mathcal{F}_{5;\gamma}^{(j\ell_{f})}(\hat{m}^{2};x,y,z)
\nonumber\\
&=&\frac{6x}{(1+2x)(2+y)(1-y)^2 } \frac{1-2z}{1+2z} 
\left \{\begin{array}{ll}
4(1-y)(1+\hm)\\
+[(1+3y)+4\hm ]\log{y}  &\textrm{if $\hat{m}^2\le y$}\\[2mm]
4(1-\hm)(1+y)\\
+[(1+3y)+4\hm ]\log{\hm}&\textrm{if $y \le \hm \le 1$}\\[2mm]
0 &\textrm{if  $ \hm \ge 1 $}
\end{array}\right.\\[3mm]
&&\mathcal{F}_{5;\delta}^{(j\ell_{f})}(\hat{m}^{2};x,y,z)
\nonumber \\
&=&\frac{3}{(1+2x)(2+y)(1-y)^2 } \left \{\begin{array}{ll}
2(1-y)(1+y)(1-x)\\
+ [ -2x(1+y)+y(3+y)]\log{y}  &\textrm{if $\hat{m}^2\le y$}\\[2mm]
2(1-\hm)(1-x)\{(1+2y)-\hm\}\\
+[ -2x(1+y)+y(3+y)]\log{\hm}&\textrm{if $y \le \hm \le 1$}\\[2mm]
0 &\textrm{if  $ \hm \ge 1 $}
\end{array}\right.
\end{eqnarray}

\newpage
{\bf SFVF} ($S=6$)
\begin{eqnarray}
&&\mathcal{F}_{6;\alpha}^{(j\ell_{f})}(\hat{m}^{2};x,y,z)\nonumber \\
&=&\frac{ -6y}{(1+2y)(1-y)^2 (2+z) } \left \{\begin{array}{ll}
2(1-y)+(1+y)\log{y}  &\textrm{if $\hat{m}^2\le y$}\\[2mm]
2(1-\hm)+(1+y)\log{\hm}&\textrm{if $y \le \hm \le 1$}\\[2mm]
0 &\textrm{if  $ \hm \ge 1 $}
\end{array}\right.\\[3mm]
&&\mathcal{F}_{6;\beta}^{(j\ell_{f})}(\hat{m}^{2};x,y,z)\nonumber \\
&=&\frac{ 3 }{(1+2y)(1-y)^2 (2+z) }\left \{\begin{array}{ll}
(1-y)[ 2(1+3y)-2(2+y)z+16\hm(1-z)]\\
+ [ 2y(3+y)-(1+5y)z+8\hm(1+y)(1-z)]\log{y}  &\textrm{if $\hat{m}^2\le y$}\\[2mm]
(1-\hm)[ 2(1+8y)-4(1+3y)z+6\hm(1-z)]\\
+[ 2y(3+y)-(1+5y)z+8\hm(1+y)(1-z)]\log{\hm}&\textrm{if $y \le \hm \le 1$}\\[2mm]
0 &\textrm{if  $ \hm \ge 1 $}
\end{array}\right.\\[3mm]
&&\mathcal{F}_{6;\gamma}^{(j\ell_{f})}(\hat{m}^{2};x,y,z)\nonumber \\
&=&\frac{ -6 }{(1+2y)(1-y)^2 (2+z) } \left \{\begin{array}{ll}
4(1-y)(y+\hm)+ 
[ y(3+y)+4y\hm]\log{y}  &\textrm{if $\hat{m}^2\le y$}\\[2mm]
8y(1-\hm)+
[ y(3+y)+4y\hm]\log{\hm}&\textrm{if $y \le \hm \le 1$}\\[2mm]
0 &\textrm{if  $ \hm \ge 1 $}
\end{array}\right.\\[3mm]
&&\mathcal{F}_{6;\delta}^{(j\ell_{f})}(\hat{m}^{2};x,y,z)\nonumber \\
&=&\frac{ 3 }{(1+2y)(1-y)^2 (2+z) } \left \{\begin{array}{ll}
2(1-y)(1+y)(1-z)\\
+ [ -(1-y)(1+2y)+(1+3y)(1-z)]\log{y}  &\textrm{if $\hat{m}^2\le y$}\\[2mm]
2(1-\hm)(1-z)\{(1+2y)-\hm \}\\
+[ -(1-y)(1+2y)+(1+3y)(1-z)]\log{\hm}&\textrm{if $y \le \hm \le 1$}\\[2mm]
0 &\textrm{if  $ \hm \ge 1 $}
\end{array}\right.
\end{eqnarray}

\begin{landscape}
\TABULAR[ht]{c|c||c|c|c|c}{
\hline
$S$& Spins
       & $\mathcal{F}_{S;\delta}^{(\ell \ell)}(\hat{m}^2;x,y,z)$  
       & $\mathcal{F}_{S;\alpha}^{(\ell \ell)}(\hat{m}^2;x,y,z)$
       & $\mathcal{F}_{S;\beta }^{(\ell \ell)}(\hat{m}^2;x,y,z)$ 
       & $\mathcal{F}_{S;\gamma}^{(\ell \ell)}(\hat{m}^2;x,y,z)$ \\ \hline\hline
1& $SFSF$ & $1$ 
       & $0$
       & $0$ 
       & $0$  \\ \hline 
2& $FSFS$ & $1$ 
       & $1-2\hat{m}^2$ 
       & $0$ 
       & $0$  \\\hline
3& $FSFV$ & $1$ 
       & $\frac{1-2z}{1+2z} ( 1-2\hat{m}^2 )$
       & $0$ 
       & $0$  \\\hline
4& $FVFS$ & $1$ 
       & $-\frac{2-y}{2+y} ( 1-2\hat{m}^2 )$
       & $0$ 
       & $0$  \\\hline
5& $FVFV$ & $1$ 
       & $-\frac{(2-y)(1-2z)}{(2+y)(1+2z)} ( 1-2\hat{m}^2 )$
       & $0$ 
       & $0$  \\\hline
6& $SFVF$ & $\frac{3}{(1+2y)(2+z)} \{ 2y+z+4(1-y)(1-z)(\hat{m}^2-\hat{m}^4) \} $ 
       & $\frac{6y}{(1+2y)(2+z)} ( 1-2\hat{m}^2 )$
       & $0$ 
       & $0$  \\\hline
}{\label{table:Fll2} Basis functions for the dilepton invariant mass distribution. 
}
\TABULAR[ht]{c|c||c|c|c|c}{
\hline
$S$ & Spins   & $\mathcal{F}_{S;\delta}^{(j \ell_n)}(\hat{m}^2;x,y,z)$
       & $\mathcal{F}_{S;\alpha}^{(j \ell_n)}(\hat{m}^2;x,y,z)$
       & $\mathcal{F}_{S;\beta }^{(j \ell_n)}(\hat{m}^2;x,y,z)$
       & $\mathcal{F}_{S;\gamma}^{(j \ell_n)}(\hat{m}^2;x,y,z)$
\\ \hline\hline
1& $SFSF$ & $1$ 
       & $0$ 
       & $-(1-2\hat{m}^2)$
       & $0$ 
\\ \hline
2& $FSFS$ & $1$ 
       & $0$ 
       & $0$
       & $0$ 
\\\hline
3& $FSFV$ & $1$ 
       & $0$ 
       & $0$
       & $0$ 
\\\hline
4& $FVFS$ & $\frac{3}{(1+2x)(2+y)}\{ 2x+y+4(1-x)(1-y)(\hat{m}^2 -\hat{m}^4  \}$ 
       & $0$ 
       & $-\frac{6x}{(1+2x)(2+y)}(1-2\hat{m}^2 )$
       & $0$ 
\\\hline
5& $FVFV$ & $\frac{3}{(1+2x)(2+y)}\{ 2x+y+4(1-x)(1-y)(\hat{m}^2 -\hat{m}^4  \} $ 
       & $0$ 
       & $-\frac{6x}{(1+2x)(2+y)}(1-2\hat{m}^2 )$
       & $0$ 
\\\hline
6& $SFVF$ & $1$ 
       & $0$ 
       & $-\frac{1-2y}{1+2y} (1-2\hat{m}^2)$
       & $0$ 
\\\hline
}{\label{table:Fqln2} Basis functions for the $j\ell_n$ invariant mass distribution. 
}
\end{landscape}

\section{Appendix: \ Fitting procedure for the parameters $\alpha$, $\beta$ and $\gamma$}
\label{app:params}
\renewcommand{\theequation}{C.\arabic{equation}}
\setcounter{equation}{0}

In the absence of any error bars, we use a rather naive matching criterion, namely
\begin{equation}
\chi^2(\alpha,\beta,\gamma) \equiv \int_0^1 \Big ( f_0(\hat{m}^2,\alpha_0,\beta_0,\gamma_0) - f(\hat{m}^2,\alpha,\beta,\gamma) \Big )^2 d \hat{m}^2 \, ,
\label{eqn:closeness}
\end{equation}
where $f_0(\hat{m}^2,\alpha_0,\beta_0,\gamma_0)$ represents 
the experimental data that needs to be fitted
and $f(\hat{m}^2,\alpha,\beta,\gamma)$ is the theoretical
prediction for it. We then minimize the $\chi^2(\alpha,\beta,\gamma)$
function for $\alpha$, $\beta$ and/or $\gamma$, as appropriate.
$\alpha_0$, $\beta_0$ and $\gamma_0$ are fixed constant values 
of the $\alpha$, $\beta$ and $\gamma$ parameters 
as predicted for the corresponding study point.
A more sophisticated analysis including the expected statistical 
uncertainties is postponed for a future publication \cite{BKMP}.

\FIGURE[h!]{
\epsfig{file=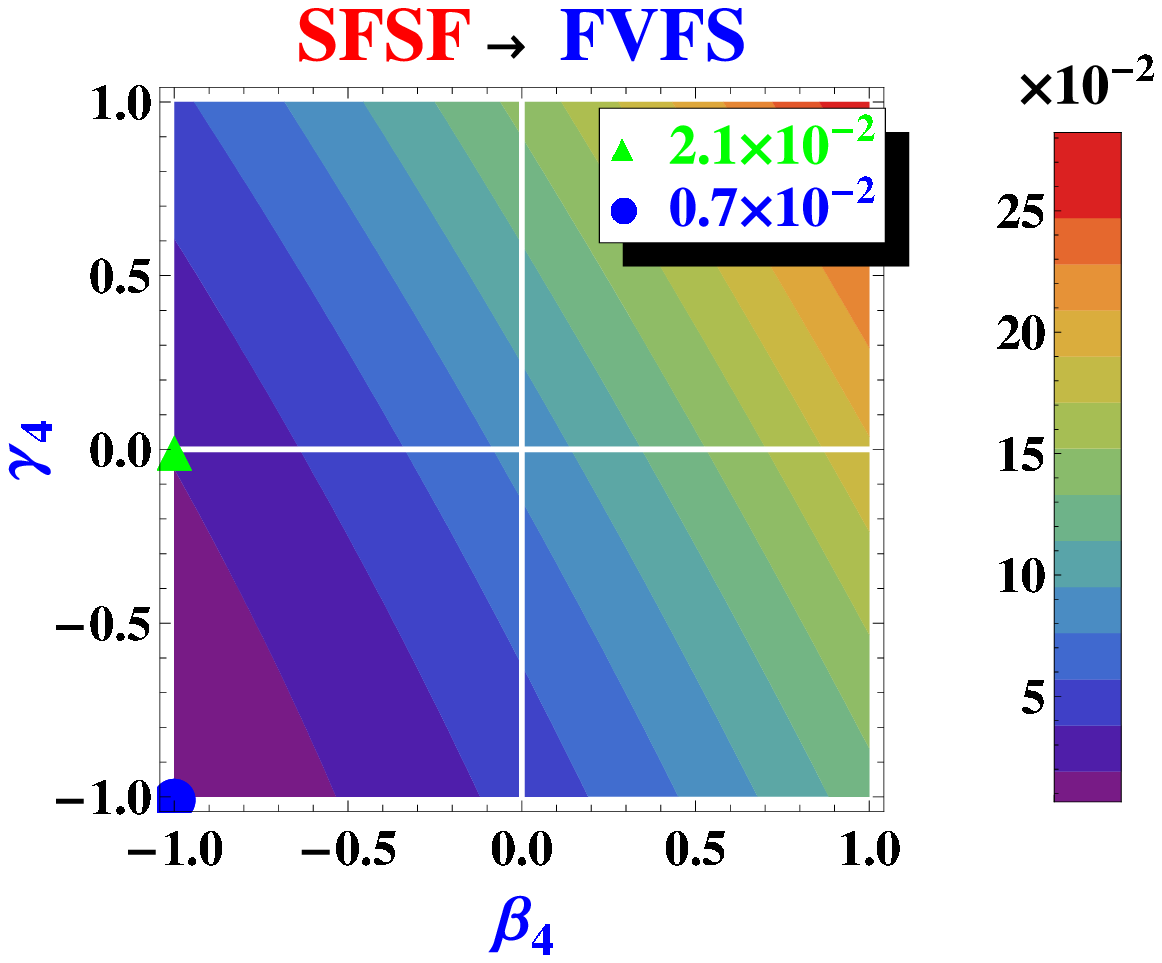,width=5cm} \epsfig{file=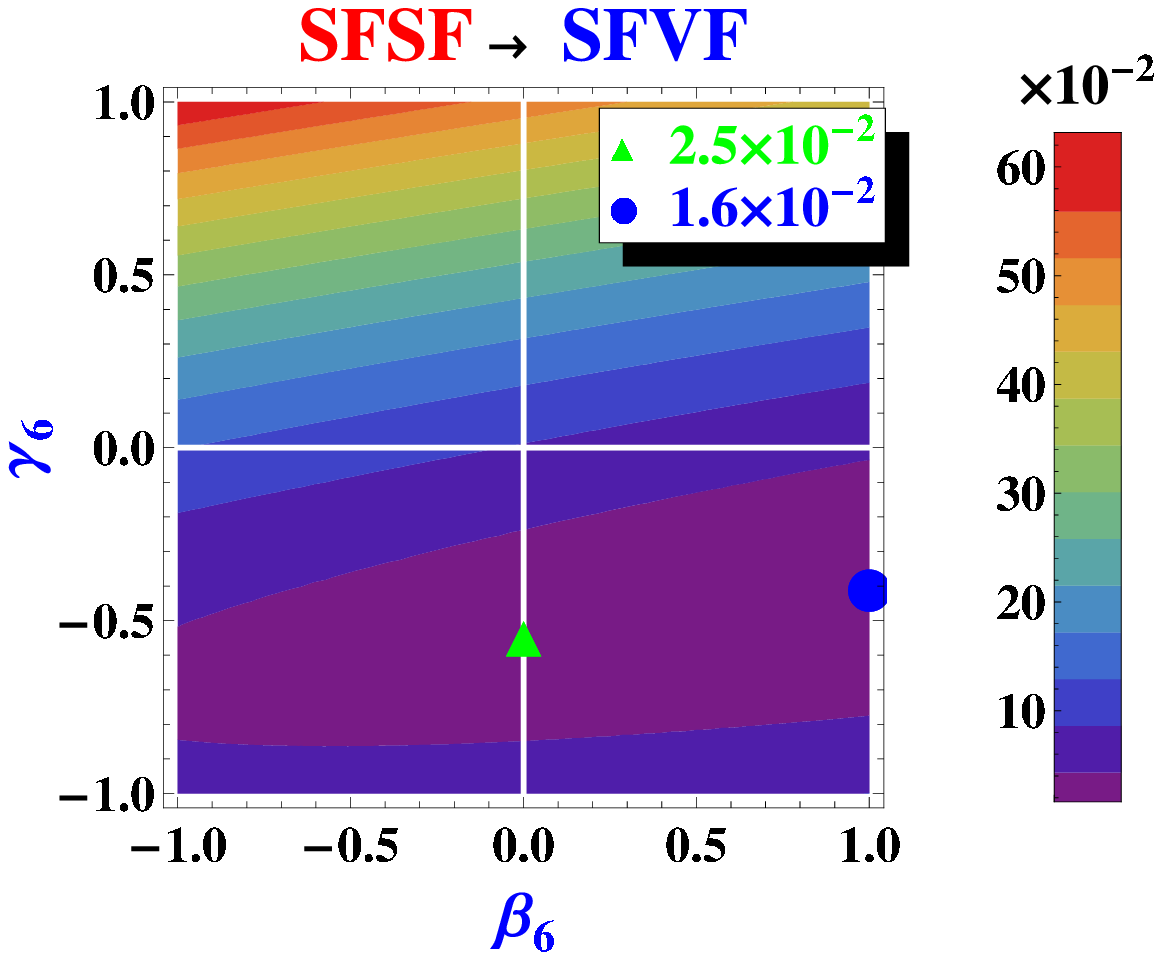,width=5cm} \epsfig{file=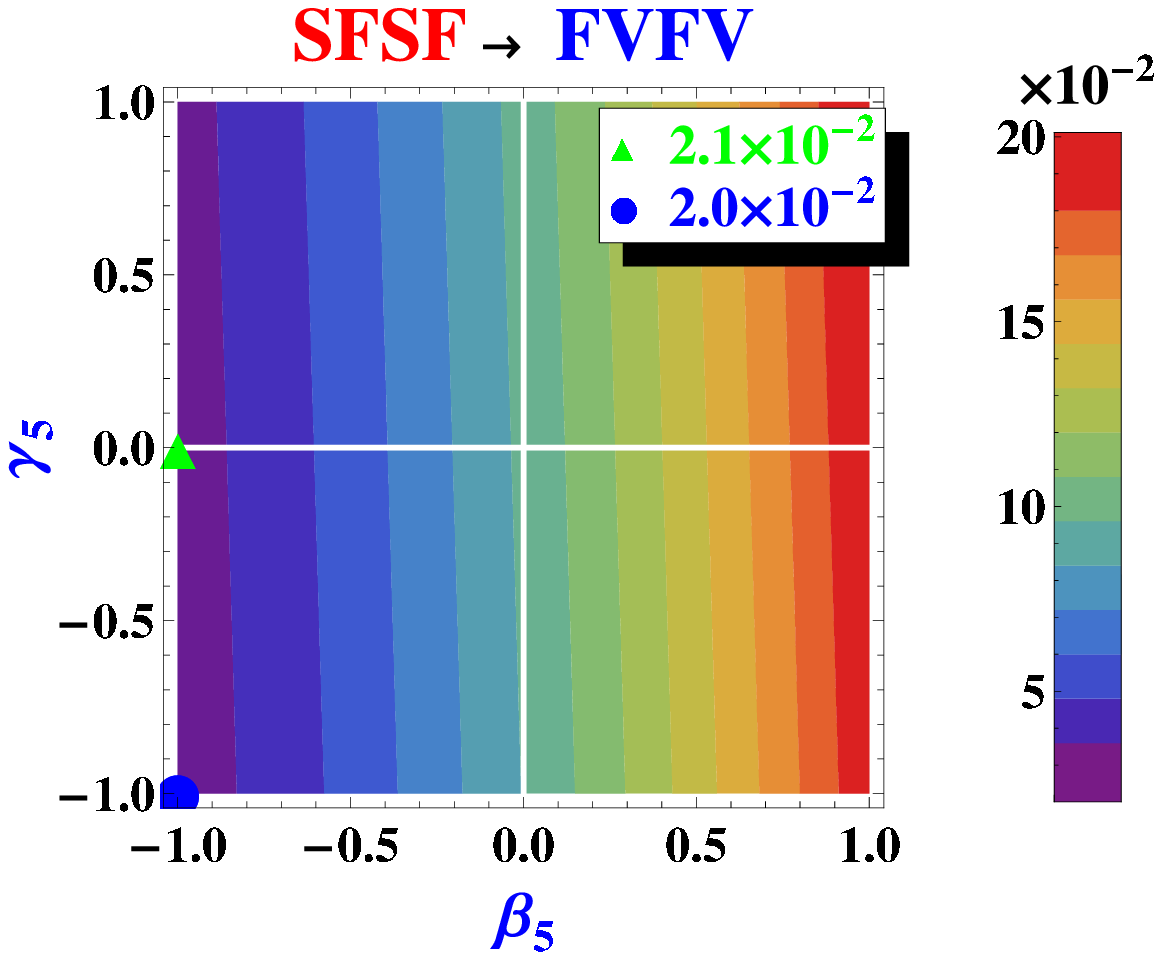,width=5cm}  \\
\epsfig{file=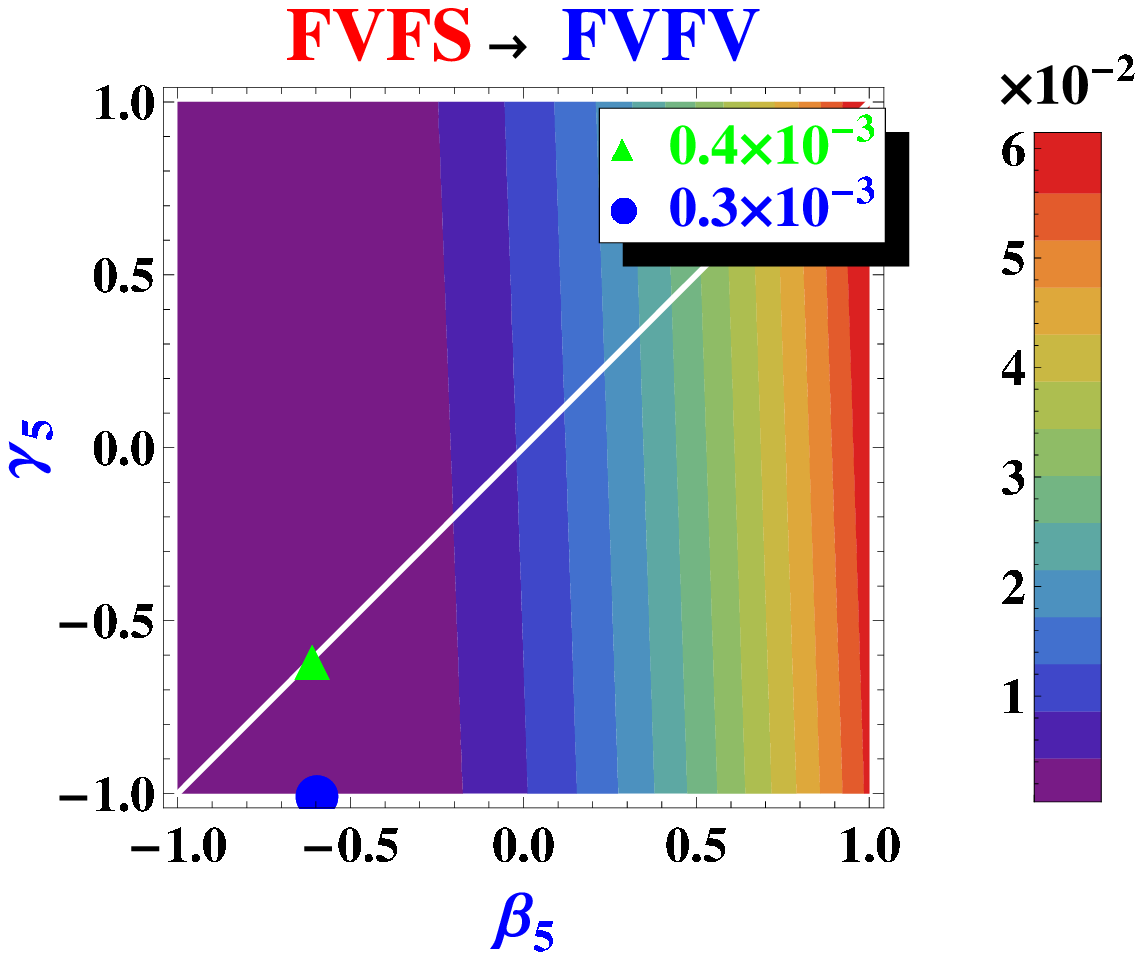,width=5cm} \epsfig{file=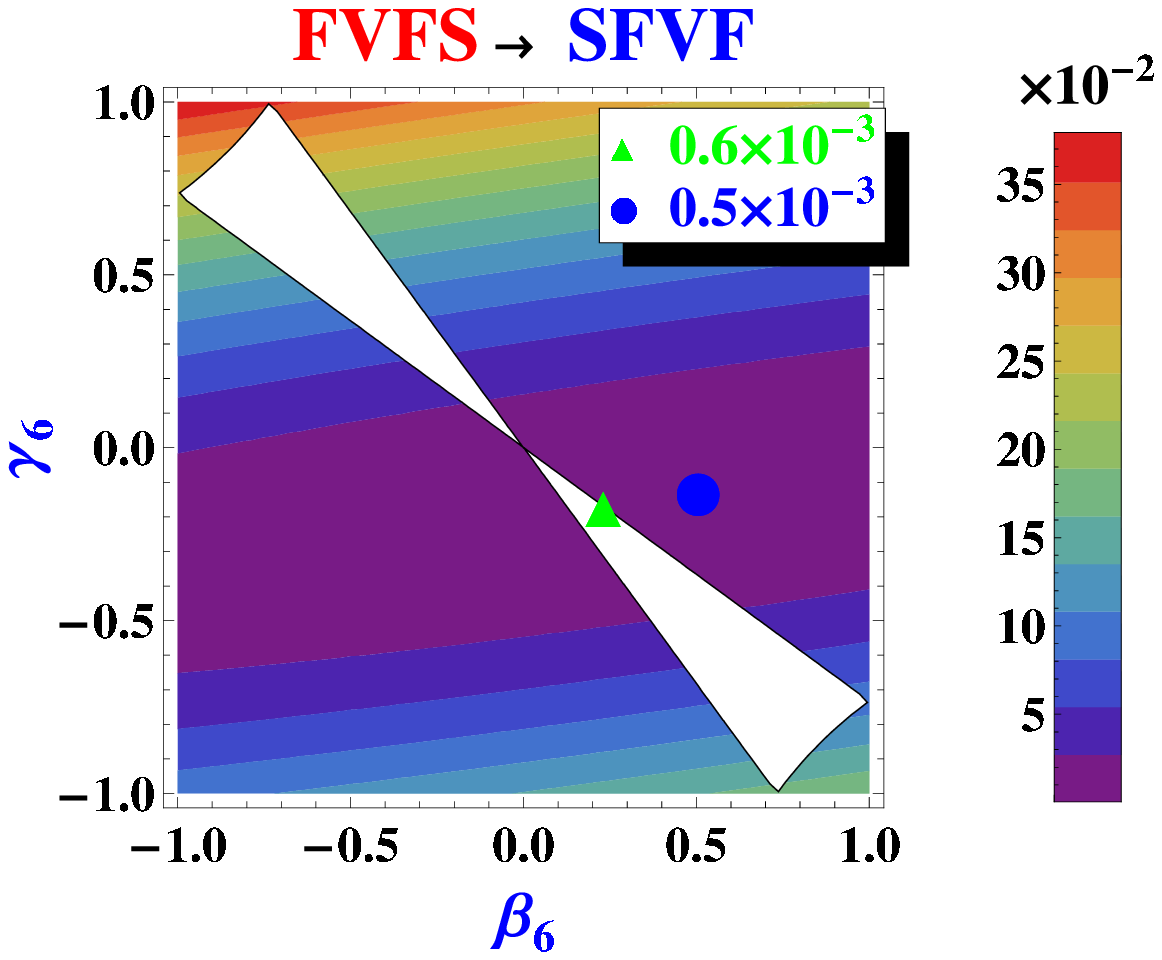,width=5cm} \epsfig{file=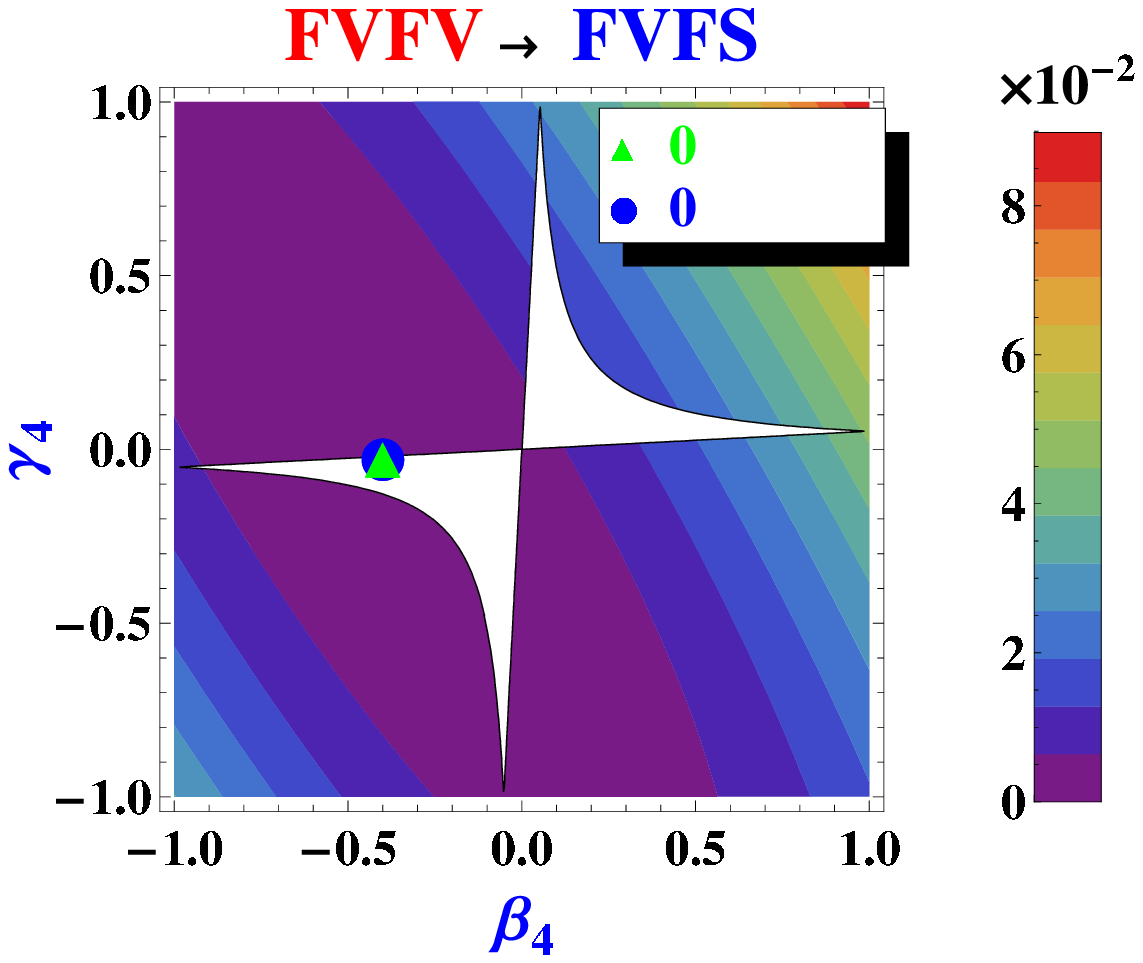,width=5cm}  \\
\epsfig{file=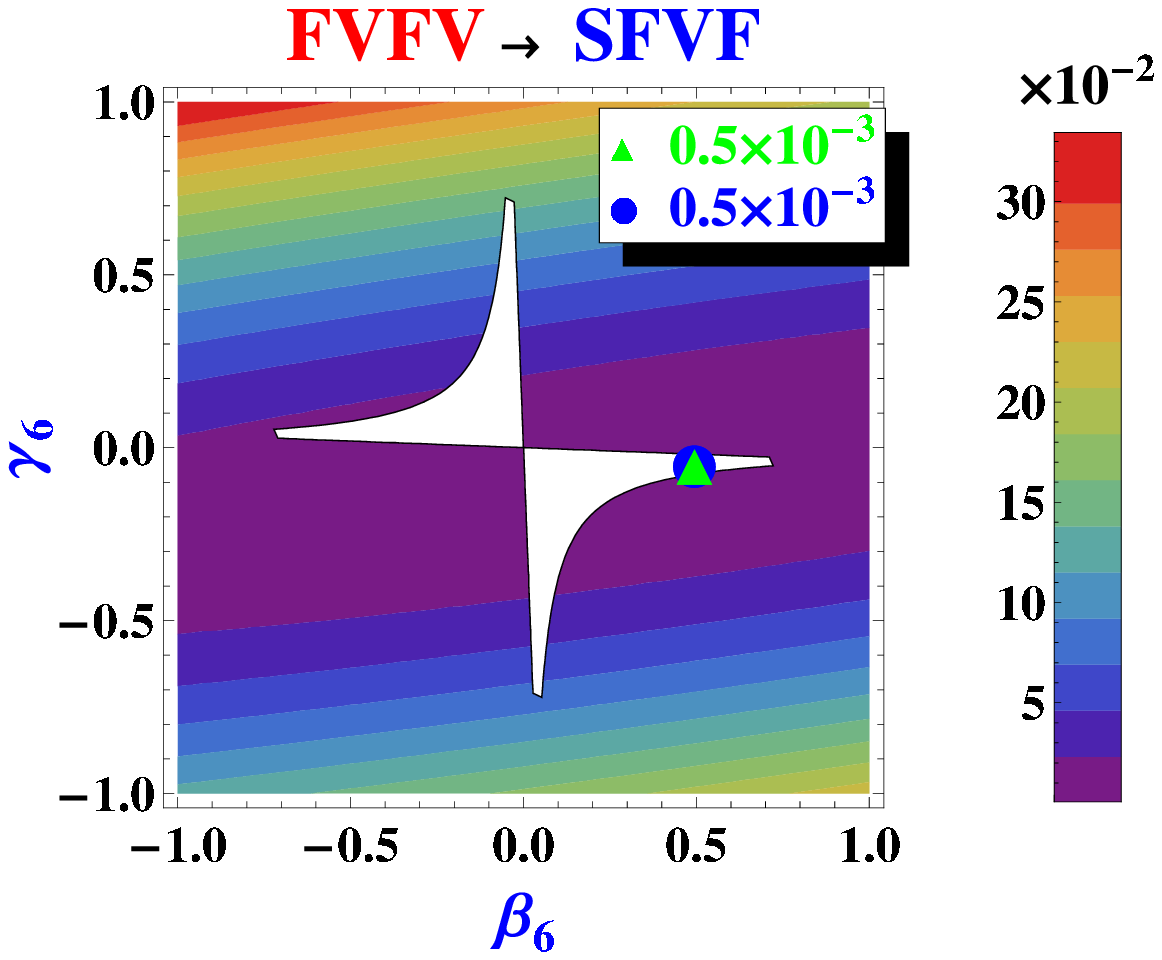,width=5cm} \epsfig{file=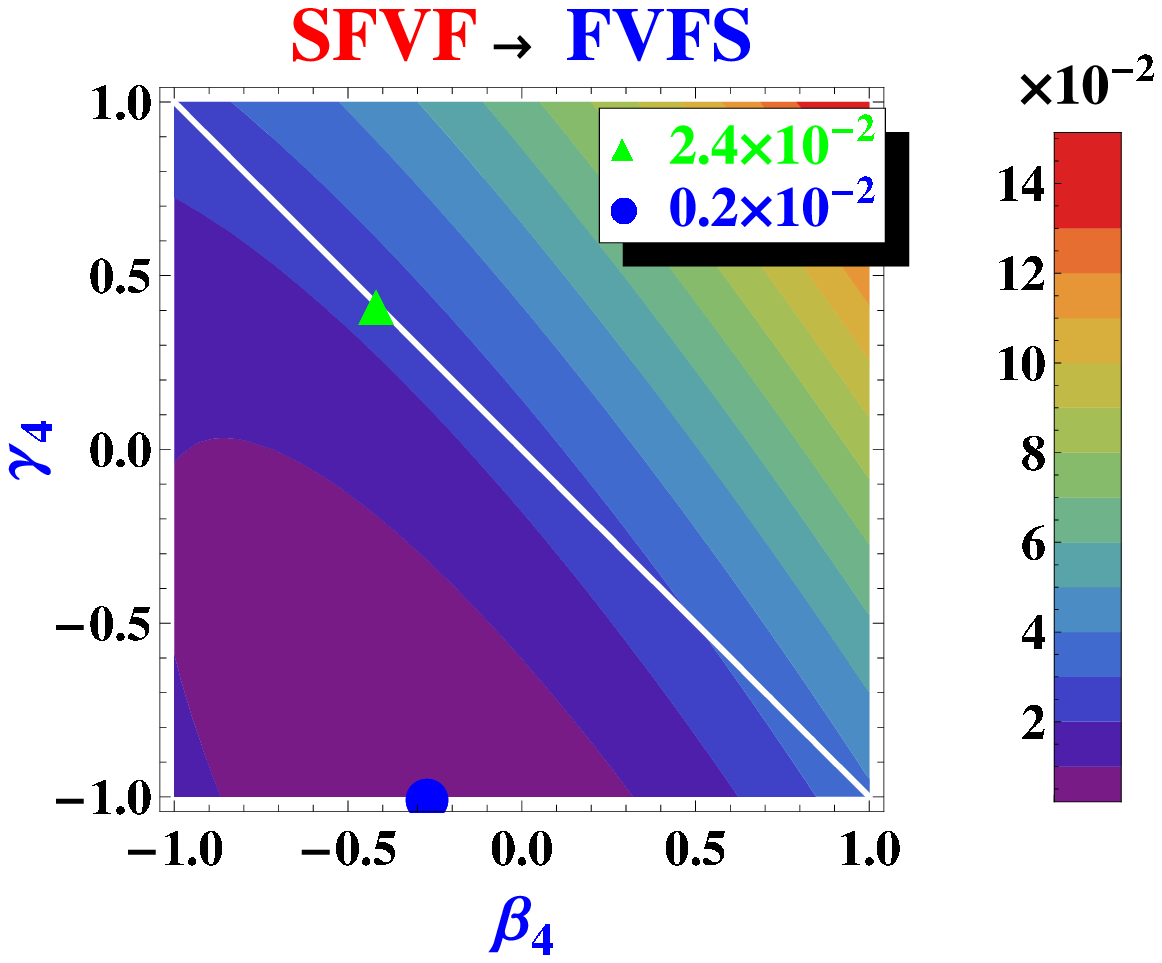,width=5cm} \epsfig{file=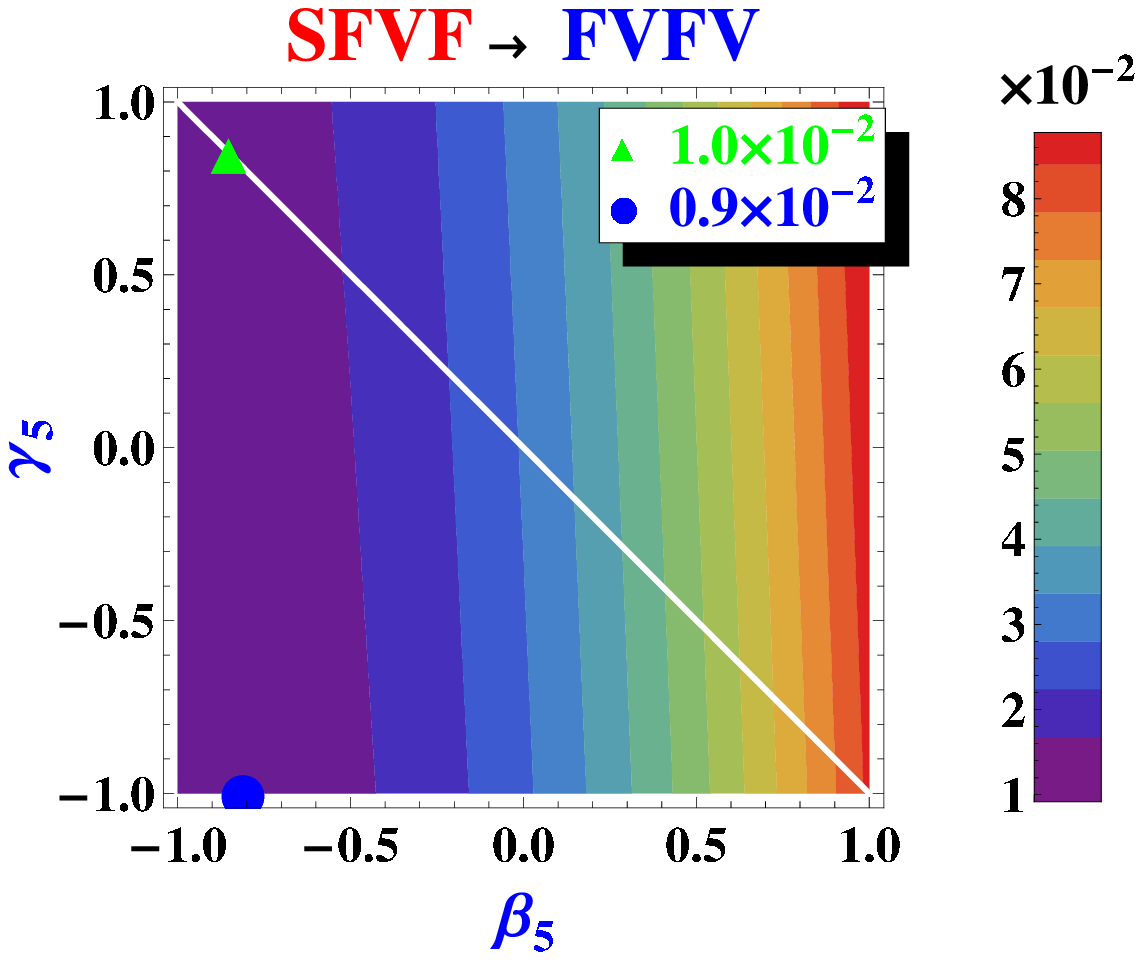,width=5cm}  
\caption{\sl Contour plots of $\chi^2(\alpha,\beta,\gamma)$ as a function of $\beta$ and $\gamma$, 
with $\alpha$ already fixed by the fit to the $L^{+-}$ data.
The physically allowed region satisfying the constraints 
(\ref{eqn:constraints1}-\ref{eqn:constraints4}) is shaded in white.
The blue dot denotes the global $\chi^2$ minimum, while the green triangle
denotes the location of the $\chi^2$ minimum within the physically allowed (white-shaded) region. 
In each plot, the ``data'' $f_0(\hat{m}^2,\alpha_0,\beta_0,\gamma_0)$ 
comes from the first spin chain (shown in red) at the top of each plot, 
which is then fitted with the distribution $f(\hat{m}^2,\alpha,\beta,\gamma)$
predicted by the second spin chain (shown in blue). 
\label{fig:fitting}}
}

As we discussed in Sec.~\ref{sec:method}, fitting to the $L^{+-}$ 
or to the $S^{+-}$ distribution is a simple one-parameter fit for $\alpha$, 
while fitting to the $D^{+-}$ data is a two-parameter fit for $\beta$ and $\gamma$.
Fig.~\ref{fig:fitting} shows sample results from our $D^{+-}$ fits for $\beta$ and $\gamma$
performed in the course of the exercises described in Sec.~\ref{sec:numerics}.
In each plot in Fig.~\ref{fig:fitting}, the ``data'' $f_0(\hat{m}^2,\alpha_0,\beta_0,\gamma_0)$ 
comes from the first spin chain (shown in red) at the top of each plot, 
which is then fitted with the distribution $f(\hat{m}^2,\alpha,\beta,\gamma)$
predicted by the second spin chain (shown in blue). 
The contour lines represent constant values of $\chi^2(\alpha,\beta,\gamma)$,
where $\alpha$ has already been fixed by fitting to $L^{+-}$.
The blue dot corresponds to the absolute minimum
of $\chi^2$, ignoring any restrictions on $\alpha$, $\beta$ and $\gamma$.
However, the parameters $\alpha$, $\beta$ and $\gamma$ 
are not completely independent from each other.
For any given $\alpha$, the physically allowed region in the $(\beta,\gamma)$ 
parameter space is described by an envelope which satisfies
\begin{eqnarray}
&&\alpha\beta \leq \gamma, ~\beta\gamma \leq \alpha, ~\gamma\alpha \leq \beta    \, ,
  \label{eqn:constraints1} \hspace{1cm}{\rm~if~}\alpha > 0, \beta > 0 {\rm~and~} \gamma > 0    \, , \\
&&\alpha\beta \geq \gamma, ~\beta\gamma \leq \alpha, ~\gamma\alpha \geq \beta    \, ,
  \label{eqn:constraints2} \hspace{1cm}{\rm~if~}\alpha > 0, \beta < 0 {\rm~and~} \gamma < 0    \, , \\
&&\alpha\beta \geq \gamma, ~\beta\gamma \geq \alpha, ~\gamma\alpha \leq \beta    \, ,
   \label{eqn:constraints3} \hspace{1cm}{\rm~if~}\alpha < 0, \beta > 0 {\rm~and~} \gamma < 0    \, , \\
&&\alpha\beta \leq \gamma, ~\beta\gamma \geq \alpha, ~\gamma\alpha \geq \beta    \, ,
   \label{eqn:constraints4} \hspace{1cm}{\rm~if~}\alpha < 0, \beta < 0 {\rm~and~} \gamma > 0    \, .
\end{eqnarray}
In Fig.~\ref{fig:fitting} we denote this allowed region in white 
(sometimes it may reduce to a single line). 
The green triangle corresponds to the minimum of the $\chi^2$ function 
within this restricted parameter set. The green triangle solution 
for $\beta$ and $\gamma$ was then used for our plots in Fig.~\ref{fig:jl_diff}.
For the two cases with FVFV (S=5) ``data'', the global minimum happens to 
lie within the (white) allowed region and so the blue dot and the green 
triangle coincide.

For the extreme values of $|\alpha|$, the (white) allowed region collapses to 
one or two lines:
\begin{eqnarray}
\beta&=&0 {\rm ~or~} \gamma=0 \, , \hspace{0.85cm}  {\rm ~if~} \alpha=0  \, , \\
\gamma &=& \pm \beta          \, , \hspace{2cm} {\rm ~if~}   \alpha=\pm 1 \, .
\end{eqnarray}

\listoftables           
\listoffigures          



\begin{thebibliography}{999}

\bibitem{Bertone:2004pz}
  For a recent review, see G.~Bertone, D.~Hooper and J.~Silk,
  ``Particle dark matter: Evidence, candidates and constraints,''
  Phys.\ Rept.\  {\bf 405}, 279 (2005)
  [arXiv:hep-ph/0404175].

\bibitem{Hubisz:2008gg}
See, for example,
  J.~Hubisz, J.~Lykken, M.~Pierini and M.~Spiropulu,
  ``Missing energy look-alikes with 100 pb-1 at the LHC,''
  arXiv:0805.2398 [hep-ph],
and references therein.

\bibitem{Jungman:1995df}
  For a review, see G.~Jungman, M.~Kamionkowski and K.~Griest,
  ``Supersymmetric dark matter,''
  Phys.\ Rept.\  {\bf 267}, 195 (1996)
  [arXiv:hep-ph/9506380].

\bibitem{Hooper:2007qk}
  For a review, see D.~Hooper and S.~Profumo,
  ``Dark matter and collider phenomenology of universal extra dimensions,''
  Phys.\ Rept.\  {\bf 453}, 29 (2007)
  [arXiv:hep-ph/0701197].

\bibitem{Cheng:2003ju}
  H.~C.~Cheng and I.~Low,
  ``TeV symmetry and the little hierarchy problem,''
  JHEP {\bf 0309}, 051 (2003)
  [arXiv:hep-ph/0308199].

\bibitem{Birkedal:2006fz}
  A.~Birkedal, A.~Noble, M.~Perelstein and A.~Spray,
  ``Little Higgs dark matter,''
  Phys.\ Rev.\  D {\bf 74}, 035002 (2006)
  [arXiv:hep-ph/0603077].

\bibitem{Hur:2007ur}
  T.~Hur, H.~S.~Lee and S.~Nasri,
  ``A Supersymmetric U(1)' Model with Multiple Dark Matters,''
  Phys.\ Rev.\  D {\bf 77}, 015008 (2008)
  [arXiv:0710.2653 [hep-ph]].

\bibitem{Lee:2008pc}
  H.~S.~Lee,
  ``Lightest U-parity Particle (LUP) dark matter,''
  Phys.\ Lett.\  B {\bf 663}, 255 (2008)
  [arXiv:0802.0506 [hep-ph]].

\bibitem{Birkedal:2004xn}
  A.~Birkedal, K.~Matchev and M.~Perelstein,
  ``Dark matter at colliders: A model-independent approach,''
  Phys.\ Rev.\  D {\bf 70}, 077701 (2004)
  [arXiv:hep-ph/0403004].

\bibitem{Cheng:2007xv}
  H.~C.~Cheng, J.~F.~Gunion, Z.~Han, G.~Marandella and B.~McElrath,
  ``Mass Determination in SUSY-like Events with Missing Energy,''
  JHEP {\bf 0712}, 076 (2007)
  [arXiv:0707.0030 [hep-ph]].

\bibitem{Nojiri:2007pq}
  M.~M.~Nojiri, G.~Polesello and D.~R.~Tovey,
  ``A hybrid method for determining SUSY particle masses at the LHC with fully
  identified cascade decays,''
  JHEP {\bf 0805}, 014 (2008)
  [arXiv:0712.2718 [hep-ph]].

\bibitem{Cheng:2008mg}
  H.~C.~Cheng, D.~Engelhardt, J.~F.~Gunion, Z.~Han and B.~McElrath,
  ``Accurate Mass Determinations in Decay Chains with Missing Energy,''
  arXiv:0802.4290 [hep-ph].

\bibitem{Hinchliffe:1996iu}
  I.~Hinchliffe, F.~E.~Paige, M.~D.~Shapiro, J.~Soderqvist and W.~Yao,
  ``Precision SUSY measurements at LHC,''
  Phys.\ Rev.\  D {\bf 55}, 5520 (1997)
  [arXiv:hep-ph/9610544].

\bibitem{Allanach:2000kt}
  B.~C.~Allanach, C.~G.~Lester, M.~A.~Parker and B.~R.~Webber,
  ``Measuring sparticle masses in non-universal string inspired models at the LHC,''
  JHEP {\bf 0009}, 004 (2000)
  [arXiv:hep-ph/0007009].

\bibitem{Gjelsten:2004ki}
  B.~K.~Gjelsten, D.~J.~Miller and P.~Osland,
  ``Measurement of SUSY masses via cascade decays for SPS 1a,''
  JHEP {\bf 0412}, 003 (2004)
  [arXiv:hep-ph/0410303].

\bibitem{Gjelsten:2005aw}
  B.~K.~Gjelsten, D.~J.~Miller and P.~Osland,
  ``Measurement of the gluino mass via cascade decays for SPS 1a,''
  JHEP {\bf 0506}, 015 (2005)
  [arXiv:hep-ph/0501033].

\bibitem{Miller:2005zp}
  D.~J.~Miller, P.~Osland and A.~R.~Raklev,
  ``Invariant mass distributions in cascade decays,''
  JHEP {\bf 0603}, 034 (2006)
  [arXiv:hep-ph/0510356].

\bibitem{BKMP}
M.~Burns, K.~Kong, K.~Matchev and M.~Park, in preparation.

\bibitem{Barr:2004ze}
  A.~J.~Barr,
  ``Using lepton charge asymmetry to investigate the spin of supersymmetric particles at the LHC,''
  Phys.\ Lett.\  B {\bf 596}, 205 (2004)
  [arXiv:hep-ph/0405052].

\bibitem{Smillie:2005ar}
  J.~M.~Smillie and B.~R.~Webber,
  ``Distinguishing spins in supersymmetric and universal extra dimension
  models at the Large Hadron Collider,''
  JHEP {\bf 0510}, 069 (2005)
  [arXiv:hep-ph/0507170].

\bibitem{Athanasiou:2006ef}
  C.~Athanasiou, C.~G.~Lester, J.~M.~Smillie and B.~R.~Webber,
  ``Distinguishing spins in decay chains at the Large Hadron Collider,''
  JHEP {\bf 0608}, 055 (2006)
  [arXiv:hep-ph/0605286].

\bibitem{Athanasiou:2006hv}
  C.~Athanasiou, C.~G.~Lester, J.~M.~Smillie and B.~R.~Webber,
  ``Addendum to 'Distinguishing spins in decay chains at the Large Hadron Collider',''
  arXiv:hep-ph/0606212.

\bibitem{Bagger:1996bt}
  J.~A.~Bagger, K.~T.~Matchev, D.~M.~Pierce and R.~j.~Zhang,
  ``Weak-scale phenomenology in models with gauge-mediated supersymmetry breaking,''
  Phys.\ Rev.\  D {\bf 55}, 3188 (1997)
  [arXiv:hep-ph/9609444].

\bibitem{Gherghetta:1999sw}
  T.~Gherghetta, G.~F.~Giudice and J.~D.~Wells,
  ``Phenomenological consequences of supersymmetry with anomaly-induced masses,''
  Nucl.\ Phys.\  B {\bf 559}, 27 (1999)
  [arXiv:hep-ph/9904378].

\bibitem{Feng:1999hg}
  J.~L.~Feng and T.~Moroi,
  ``Supernatural supersymmetry: Phenomenological implications of
  anomaly-mediated supersymmetry breaking,''
  Phys.\ Rev.\  D {\bf 61}, 095004 (2000)
  [arXiv:hep-ph/9907319].

\bibitem{Schmaltz:2000gy}
  M.~Schmaltz and W.~Skiba,
  ``Minimal gaugino mediation,''
  Phys.\ Rev.\  D {\bf 62}, 095005 (2000)
  [arXiv:hep-ph/0001172].

\bibitem{Cheng:2002iz}
  H.~C.~Cheng, K.~T.~Matchev and M.~Schmaltz,
  ``Radiative corrections to Kaluza-Klein masses,''
  Phys.\ Rev.\  D {\bf 66}, 036005 (2002)
  [arXiv:hep-ph/0204342].

\bibitem{Cheng:2002ab}
  H.~C.~Cheng, K.~T.~Matchev and M.~Schmaltz,
  ``Bosonic supersymmetry? Getting fooled at the LHC,''
  Phys.\ Rev.\  D {\bf 66}, 056006 (2002)
  [arXiv:hep-ph/0205314].

\bibitem{Cheng:2005as}
  H.~C.~Cheng, I.~Low and L.~T.~Wang,
  ``Top partners in little Higgs theories with T-parity,''
  Phys.\ Rev.\  D {\bf 74}, 055001 (2006)
  [arXiv:hep-ph/0510225].

\bibitem{Battaglia:2005zf}
  M.~Battaglia, A.~Datta, A.~De Roeck, K.~Kong and K.~T.~Matchev,
  ``Contrasting supersymmetry and universal extra dimensions at the CLIC
  multi-TeV e+ e- collider,''
  JHEP {\bf 0507}, 033 (2005)
  [arXiv:hep-ph/0502041].

\bibitem{Battaglia:2005ma}
  M.~Battaglia, A.~K.~Datta, A.~De Roeck, K.~Kong and K.~T.~Matchev,
  ``Contrasting supersymmetry and universal extra dimensions at colliders,''
{\it In the Proceedings of 2005 International Linear Collider Workshop (LCWS 2005), Stanford, California, 18-22 Mar 2005, pp 0302}
  [arXiv:hep-ph/0507284].

\bibitem{Datta:2005zs}
  A.~Datta, K.~Kong and K.~T.~Matchev,
  ``Discrimination of supersymmetry and universal extra dimensions at  hadron
  colliders,''
  Phys.\ Rev.\  D {\bf 72}, 096006 (2005)
  [Erratum-ibid.\  D {\bf 72}, 119901 (2005)]
  [arXiv:hep-ph/0509246].

\bibitem{Datta:2005vx}
  A.~Datta, G.~L.~Kane and M.~Toharia,
  ``Is it SUSY?,''
  arXiv:hep-ph/0510204.

\bibitem{Barr:2005dz}
  A.~J.~Barr,
  ``Measuring slepton spin at the LHC,''
  JHEP {\bf 0602}, 042 (2006)
  [arXiv:hep-ph/0511115].

\bibitem{Meade:2006dw}
  P.~Meade and M.~Reece,
  ``Top partners at the LHC: Spin and mass measurement,''
  Phys.\ Rev.\  D {\bf 74}, 015010 (2006)
  [arXiv:hep-ph/0601124].

\bibitem{Alves:2006df}
  A.~Alves, O.~Eboli and T.~Plehn,
  ``It's a gluino,''
  Phys.\ Rev.\  D {\bf 74}, 095010 (2006)
  [arXiv:hep-ph/0605067].

\bibitem{Wang:2006hk}
  L.~T.~Wang and I.~Yavin,
  ``Spin Measurements in Cascade Decays at the LHC,''
  JHEP {\bf 0704}, 032 (2007)
  [arXiv:hep-ph/0605296].

\bibitem{SA:2006jm}
  S.~Abdullin {\it et al.}  [TeV4LHC Working Group],
  ``Tevatron-for-LHC report: Preparations for discoveries,''
  arXiv:hep-ph/0608322.

\bibitem{Smillie:2006cd}
  J.~M.~Smillie,
  ``Spin Correlations in Decay Chains Involving W Bosons,''
  Eur.\ Phys.\ J.\  C {\bf 51}, 933 (2007)
  [arXiv:hep-ph/0609296].

\bibitem{Kong:2006pi}
  K.~Kong and K.~T.~Matchev,
  ``Phenomenology of universal extra dimensions,''
  AIP Conf.\ Proc.\  {\bf 903}, 451 (2007)
  [arXiv:hep-ph/0610057].

\bibitem{Kilic:2007zk}
  C.~Kilic, L.~T.~Wang and I.~Yavin,
  ``On the Existence of Angular Correlations in Decays with Heavy Matter Partners,''
  JHEP {\bf 0705}, 052 (2007)
  [arXiv:hep-ph/0703085].

\bibitem{Alves:2007xt}
  A.~Alves and O.~Eboli,
  ``Unravelling the sbottom spin at the CERN LHC,''
  Phys.\ Rev.\  D {\bf 75}, 115013 (2007)
  [arXiv:0704.0254 [hep-ph]].

\bibitem{Csaki:2007xm}
  C.~Csaki, J.~Heinonen and M.~Perelstein,
  ``Testing Gluino Spin with Three-Body Decays,''
  JHEP {\bf 0710}, 107 (2007)
  [arXiv:0707.0014 [hep-ph]].

\bibitem{Datta:2007xy}
  A.~Datta, P.~Dey, S.~K.~Gupta, B.~Mukhopadhyaya and A.~Nyffeler,
  ``Distinguishing the Littlest Higgs model with T-parity from supersymmetry at
  the LHC using trileptons,''
  Phys.\ Lett.\  B {\bf 659}, 308 (2008)
  [arXiv:0708.1912 [hep-ph]].

\bibitem{Buckley:2007th}
  M.~R.~Buckley, H.~Murayama, W.~Klemm and V.~Rentala,
  ``Discriminating spin through quantum interference,''
  arXiv:0711.0364 [hep-ph].

\bibitem{Buckley:2008pp}
  M.~R.~Buckley, B.~Heinemann, W.~Klemm and H.~Murayama,
  ``Quantum Interference Effects Among Helicities at LEP-II and Tevatron,''
  Phys.\ Rev.\  D {\bf 77}, 113017 (2008)
  [arXiv:0804.0476 [hep-ph]].

\bibitem{Kane:2008kw}
  G.~L.~Kane, A.~A.~Petrov, J.~Shao and L.~T.~Wang,
  ``Initial determination of the spins of the gluino and squarks at LHC,''
  arXiv:0805.1397 [hep-ph].

\bibitem{Appelquist:2000nn}
  T.~Appelquist, H.~C.~Cheng and B.~A.~Dobrescu,
  ``Bounds on universal extra dimensions,''
  Phys.\ Rev.\  D {\bf 64}, 035002 (2001)
  [arXiv:hep-ph/0012100].

\bibitem{Dobrescu:2007xf}
  B.~A.~Dobrescu, K.~Kong and R.~Mahbubani,
  ``Leptons and photons at the LHC: Cascades through spinless adjoints,''
  JHEP {\bf 0707}, 006 (2007)
  [arXiv:hep-ph/0703231].

\bibitem{Ozturk:2007ap}
  N.~Ozturk  [ATLAS Collaboration],
  ``SUSY Parameters Determination with ATLAS,''
  arXiv:0710.4546 [hep-ph].

\bibitem{Birkedal:2005cm}
  A.~Birkedal, R.~C.~Group and K.~Matchev,
  ``Slepton mass measurements at the LHC,''
{\it In the Proceedings of 2005 International Linear Collider Workshop (LCWS 2005), Stanford, California, 18-22 Mar 2005, pp 0210}
  [arXiv:hep-ph/0507002].

\bibitem{Gjelsten:2006tg}
  B.~K.~Gjelsten, D.~J.~Miller, P.~Osland and A.~R.~Raklev,
  ``Mass determination in cascade decays using shape formulas,''
  AIP Conf.\ Proc.\  {\bf 903}, 257 (2007)
  [arXiv:hep-ph/0611259].

\bibitem{KM}
K.~Kong and K.~Matchev, in preparation.

\end{thebibliography}
\end{document}